\documentclass[%
 reprint,
showpacs,
 amsmath,
 amssymb,
 aps,
prl,
nobalancelastpage]{revtex4-1}

\usepackage{color}
\usepackage{graphicx}% Include figure files
\usepackage{dcolumn}% Align table columns on decimal point
\usepackage{bm}% bold math
\usepackage{esint}
\usepackage{todo}
\usepackage[colorlinks,allcolors=black]{hyperref}
\usepackage[capitalize]{cleveref}
\usepackage{csquotes}
\usepackage{verbatim}
\usepackage{hhline}
\usepackage[normalem]{ulem}
\usepackage[english]{babel} 
\usepackage{microtype}
\usepackage{siunitx}
\usepackage{dsfont}
\usepackage[dvipsnames]{xcolor}
\usepackage{eucal}

{}

\crefname{appsec}{appendix}{appendices}

\begin{document}

%\title{Learning to control nanometric islands under electromigration}
%\title{Nanomanipulation with reinforcement learning}
%\title{Macroscopic control of nanostructures via reinforcement learning}
%\title{Control of nanostructures with a macroscopic parameter and reinforcement learning}
%\title{Learning to manipulate nanostructures with a macroscopic control parameter}

%^these ones above all have "learning" in the title, which maybe is not the best since we finally use DP?

\title{Controlling the shape of small clusters with and without macroscopic fields}

\author{Francesco Boccardo}
\email{francesco.boccardo@univ-lyon1.fr}
\author{Olivier Pierre-Louis}
\email{olivier.pierre-louis@univ-lyon1.fr}
\affiliation{Institut Lumi\`ere Mati\`ere, UMR5306 Universit\'e Lyon 1 - CNRS, 69622 Villeurbanne, France}
\date{\today}
\begin{abstract}
{
%Major challenges in the control of nanostructure morphology lie in the use of macroscopic fields that do not address one atom or particle at a time, and in handling inherent stochasticity caused by thermal fluctuations. 
%We report on theoretical investigations to 
%tackle both difficulties 
Despite major advances in the understanding of the formation and dynamics of nano-clusters in the past decades,
theoretical bases for the control of their shape are still lacking. 
We investigate strategies for driving fluctuating few-particle clusters to an arbitrary target shape in minimum time
with or without an external field.
This question is recast into a first passage problem, solved numerically, and discussed within 
a high temperature expansion. 
Without field, large-enough low-energy target shapes exhibit an optimal temperature at which they are reached in minimum time. 
%Using Dynamic Programming, we obtain 
We then compute the optimal way to set an external field to minimize the time to reach the target, 
leading to a gain of time that grows 
when increasing cluster size or decreasing temperature.  
This gain can shift the optimal temperature or even create one.  Our results could apply to 
clusters of atoms at equilibrium, and colloidal or nanoparticle clusters under thermo- or electrophoresis.
}
\end{abstract}

\maketitle
% \tableofcontents

{

Less than a decade after its discovery~\cite{Binnig1983}, scanning tunneling microscope (STM)
was used to position atoms on a surface with \AA nsgtrom precision~\cite{Eigler1990}, reaching
atomic-scale control on the organization of matter.
Following this seminal work, many examples of organization 
of atoms~\cite{Cooper2018,Barredo2016}, molecules or nanoparticles~\cite{Martinez2014,Erickson2011,Junno1995,Baur1998,RubioSierra2005} 
and colloids~\cite{Korda2002,Grier2003} were obtained
with tools like STM, atomic force microscope, or optical tweezers.
%that manipulate single particles~\cite{Grier2003} or atoms~\cite{Barredo2016} 
%were realized.
However, important challenges are still open in the control of few-particle clusters.

The first one is to control matter at the nanoscale
with an external macroscopic field that does not
act on one single particle or atom at a time, but on the whole cluster. 
External fields such as  light acting on metal nanoparticle clusters~\cite{McCormack2018}
or electromigration acting on atomic monolayer clusters~\cite{Kuhn2005,Mahadevan1999,PierreLouis2000,Curiotto2019}
are known to lead to complex equilibrium or non-equilibrium cluster shapes.
However, these shapes are only a very small fraction of all possible shapes,
that are dictated by the physics of the interaction of the driving force with the system.

Another challenge lies in the ability to obtain refined
control of nanostructure shapes in the presence of thermal
fluctuations that activate the random diffusion of particles and atoms,
leading to shape fluctuations~\cite{Khare1995,PierreLouis2000,Lai2017}.
Some progress in this direction has been achieved with the control of the formation and order
of colloidal clusters~\cite{Juarez2012,Juarez2012a,Xue2014} in finite-temperature experiments. 
However, the control of the cluster shape is still an open issue.

In order to address these challenges, we investigate strategies 
%that allow one
to reach arbitrary cluster shapes in minimum time in the presence of fluctuations.
We focus on the control of few-particles two-dimensional clusters
%. More precisely, we 
and find
how a given target shape
%of the cluster 
can be reached in minimum time
%in the presence of thermal fluctuations and 
with and without macroscopic external field.
This problem which is formulated as the minimization of a first passage time on the graph of cluster configurations,
is solved numerically and studied analytically
with the help of a high temperature expansion.

In the absence of field, we find that large compact target shapes 
% that are large enough 
exhibit an optimal temperature
at which they can be reached in minimum time. 
In the presence of an external field we use dynamic programming~\cite{Sutton1998,Bellman1957}
to find the optimal way to set the external field as a function of 
the 
%observed 
cluster shape to reach the target in minimum time. 
%\com{The requirement of looking at the shape in order to set the field makes our method a feedback control approach.}
The gain in time 
%to reach the target 
due to the forces increases
with decreasing temperature and with increasing clusters size. 
This gain can 
shift the optimal temperature, or create one when it does not exist in the absence of forces.

We  focus on clusters 
with edge diffusion dynamics.  
Edge diffusion was observed in metal atomic monolayer islands~\cite{Giesen2001,Tao2010}, 
and for colloids~\cite{Hubartt2015}.
However, our strategy can
readily be extended to any type of dynamics that preserve the number of particles
such as surface-diffusion dynamics inside vacancy clusters~\cite{Plass2001,Heinonen1999,Leroy2020}, or
dislocation-mediated cluster rearrangements in colloids~\cite{VanSaders2021}
and metal nanoclusters~\cite{Huang2018,Trushin2001}.
%and diffusion of adsorbed molecules and polymers~\cite{Sukhishvili2000}.
We discuss possible experiments
with clusters of atoms
%, nanoparticles, 
or colloids. 

%In the absence of force,
%atomic and colloidal clusters can both 
%present an optimal temperature.
%In the presence of forces, control can be achieved
%with colloids and nanoparticles under thermo or electrophoresis.
%However, the forces are not strong enough to achieve
%control of few-atoms clusters under electromigration.

%%%%%%%%%%%%%%%%%%%%%%%%%%%%%%%%%%%%%%%%%%%%%%%%%%%%%%%%%
\paragraph{\bf Model.}
%%%%%%%%%%%%%%%%%%%%%%%%%%%%%%%%%%%%%%%%%%%%%%%%%%%%%%%%%
We consider a small cluster on a square lattice with lattice parameter $a$
and nearest-neighbor bonds $J$ under an external macroscopic force $\bm{F}$. 
We assume that the current configuration of the cluster, hereafter denoted
as the state $s$, can be observed at all times.
The force is chosen as a function of $s$.
This choice is encoded in the policy ${\boldsymbol \phi}$,
so that $\bm{F}={\boldsymbol \phi}(s)$.
%A state $s$ is associated to each cluster configuration on the lattice.
%Our goal is to find the value $\bm{F}$ in each state $s$ that minimizes the first-passage time to reach a given target state $\bar s$. 
%This choice is encoded in the policy ${\boldsymbol \phi}$, which sets a value of the force $\bm{F}(s)$ in each state $s$.
%The choice of the force $\bm{F}$ in each state $s$ is encoded in the policy ${\boldsymbol \phi}$,
%so that ${\boldsymbol \phi}(s)=\bm{F}$.
The state $s$ can change to another state $s'$ via the motion
of a single particle to one of its nearest or next-nearest neighbor sites along the cluster edge. Moves that break the cluster are forbidden.
Following usual models for biased diffusion~\cite{Glasstone1941,Liu1998,PierreLouis2000},
the hopping rate is assumed to take an Arrhenius form
\begin{align}\label{eq:TST_hopping}
    \gamma_{\boldsymbol \phi}(s,s') = \nu \exp[-(n_{ss'}J - {\boldsymbol \phi}(s)\cdot\bm{u}_{ss'})/k_BT],
\end{align}  
where $ k_B T $ is the thermal energy and $\nu$ is an attempt frequency,  
$n_{ss'}$ is the number of in-plane nearest neighbors in state $s$ before hopping.
To gain computation time, we freeze atoms with $n_{ss'}=4$.
In addition, we assume that the displacement vector to the diffusion saddle point $\bm{u}_{ss'}$ is half the displacement 
vector between the initial and final positions of the moving atom~\cite{PierreLouis2000}.
%~\footnote{ More generally,
%$\bm{u}_{ss'}$ is the displacement vector between the equilibrium position of the atom
%before the move and the saddle point of the diffusion energy potential.}.
In the following, we choose units where $k_B=1$, $J=1$, $\nu=1$ and $a=1$.

%Our goal is to study the time to reach a given target cluster configuration $\bar{s}$.
%This time can be seen as a first passage time in a random walk on the space of cluster configurations~\cite{Sanchez1999,Combe2000}.
%This space of configurations is represented by a graph in Fig.~\ref{fig:graph_dynamics}.
%%Under a given policy ${\boldsymbol \phi}$, 
%The expected first passage time $\tau_{\boldsymbol \phi}(s,{\bar{s}})$ from state $s\neq\bar{s}$ 
%obeys a recursion relation. Indeed, since the dynamics is Markovian, 
%%the first passage time 
%$\tau_{\boldsymbol \phi}(s,{\bar{s}})$
%is equal to the expected residence time $t_{\boldsymbol \phi}(s)$ in state $s$
%plus the first passage time from the neighboring state $s'$ after the move~\cite{VanKampen1992}

Our goal is to study the time to reach a target cluster configuration $\bar{s}$ from
an initial state $s$.
This time can be seen as a first passage time $\tau_{\boldsymbol \phi}(s,{\bar{s}})$
in a random walk on the graph of cluster configurations~\cite{Sanchez1999,Combe2000},
as represented in Fig.~\ref{fig:graph_dynamics}.
Since the dynamics is Markovian, 
$\tau_{\boldsymbol \phi}(s,{\bar{s}})$
is equal to the expected residence time $t_{\boldsymbol \phi}(s)$ in state $s$
plus the first passage time from the new state $s'$ after the move~\cite{VanKampen1992}.
Considering all possible moves, we obtain a recursion relation
\begin{align}
\label{eq:recursion}
\tau_{\boldsymbol \phi}(s,{\bar{s}}) = t_{\boldsymbol \phi}(s) +\! \sum_{s'\in {\cal B}_s}\!p_{\boldsymbol \phi}(s,s')\tau_{{\boldsymbol \phi}}(s',{\bar{s}})\,, %\\
%\label{eq:mean_res_time}
%\!\!\!\!\! \frac{1}{t_{\boldsymbol \phi}(s)}= \!\sum_{s'\in {\cal B}_s}\!\gamma_{\boldsymbol \phi}(s,s') \,, \quad\,\,\,
%p_{\boldsymbol \phi}(s,s') = \gamma_{\boldsymbol \phi}(s,s')\,t_{\boldsymbol \phi}(s) \,,
\end{align}
where $p_{\boldsymbol \phi}(s,s')= \gamma_{\boldsymbol \phi}(s,s')\,t_{\boldsymbol \phi}(s)$ 
is the transition probability  from $s$ to $s'$,
 $t_{\boldsymbol \phi}(s)=1/\!\sum_{s'\in {\cal B}_s}\!\gamma_{\boldsymbol \phi}(s,s')$,
%under force ${\boldsymbol \phi}(s)$.
%\com{, when the force is set according to the policy ${\boldsymbol \phi}$}.
and ${\cal B}_s$  
%\com{neighbors $\rightarrow$ bordering states (or ${\cal V}_s$ for vicinal states? or ${\cal C}_s$ for contiguous states?)}  
the set of states that can be reached from $s$ via a single move.
Relation Eq.~(\ref{eq:recursion}) 
is supplemented with the boundary condition $\tau_{\boldsymbol \phi}(\bar{s},{\bar{s}})=0$.

\begin{figure}[b]
	\includegraphics[width=\linewidth]{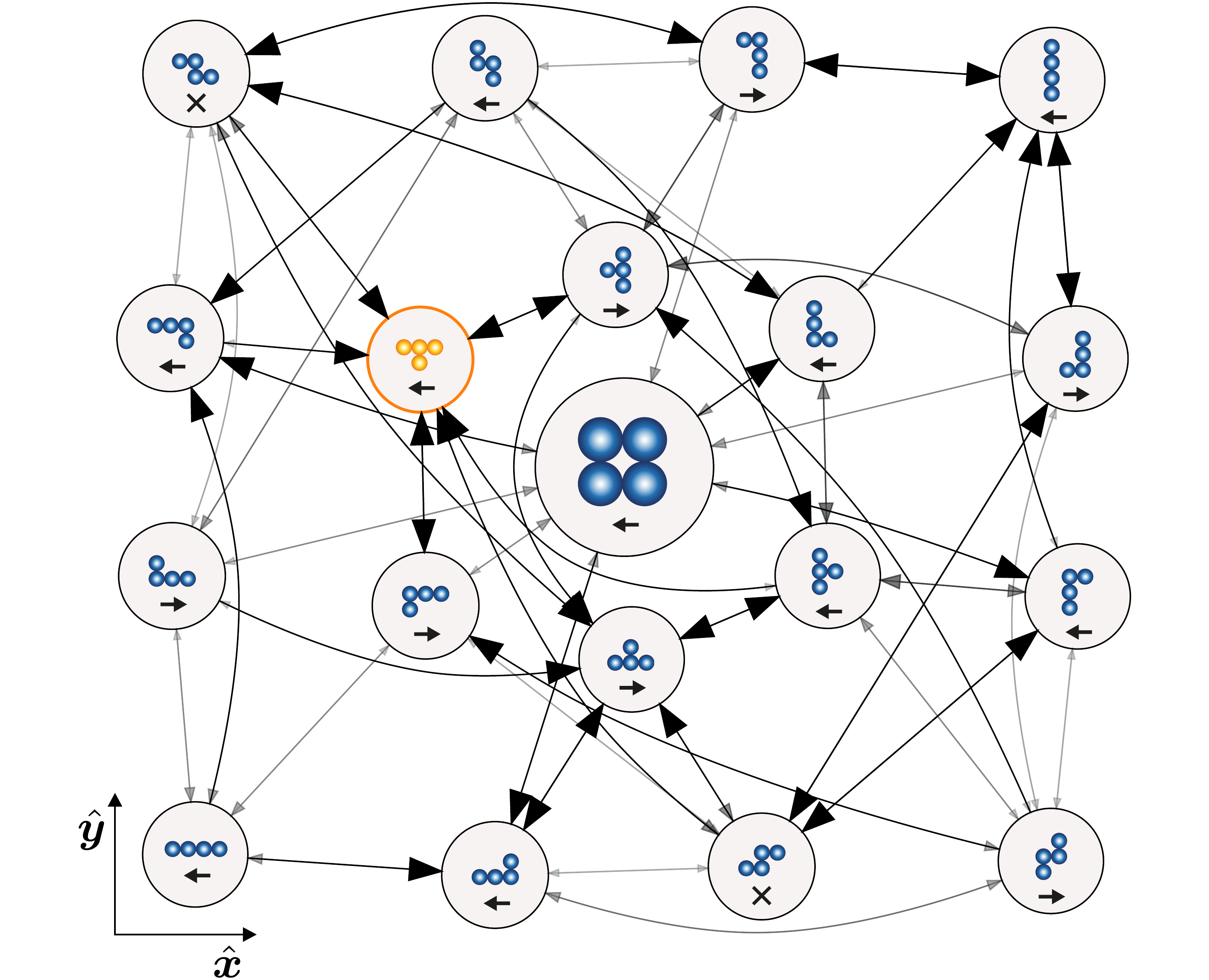}
	\caption{Graph of configurations for a tetramer cluster ($N=4$) at $T=0.24$.
	The node size is proportional to the expected residence time $t_{\boldsymbol \phi}(s)$.
	The thickness and shade of the edges are proportional to the transition probability $p_{\boldsymbol \phi}(s,s')$. 
	Arrows in the nodes represent an optimal policy ${\boldsymbol \phi}_*(s)$ to reach the orange target shape
	(crosses correspond to a zero force). 
	}
	\label{fig:graph_dynamics}
\end{figure}

We also define the expected return time to target,
i.e. spent outside the target before returning to it when starting from the target itself~\footnote{
As a technical remark, this definition requires to extend the policy and define a force on the target state itself.
However, due to the Markovian character of the dynamics, 
this does not affect the mean first passage time to target and the optimal policy in the other states outside the target.}
%\com{In Supp. Mat. we show that $\tau_{\boldsymbol \phi}^\mathrm{r}({\bar{s}})$ behaves in a similar way as  $\tau_{\boldsymbol \phi}(s,{\bar{s}})$ from all the other states.}
\begin{equation}
\label{eq:mean_return_time}
\tau_{\boldsymbol \phi}^\mathrm{r}({\bar{s}}) = \sum_{s\in{\cal B}_{\bar{s}}} p_{\boldsymbol \phi}(\bar s,s)\, \tau_{\boldsymbol \phi}(s,{\bar{s}})
\, . 
\end{equation}
%where the sum is performed on the states $s$ that can be reached in one move from the target
For the sake of concision, we mainly focus on the analysis
of $\tau_{\boldsymbol \phi}^\mathrm{r}({\bar{s}})$
instead of $\tau_{\boldsymbol \phi}(s,{\bar{s}})$ which is different for each $s$. 

%%%%%%%%%%%%%%%%%%%%%%%%%%%%%%%%%%%%%%%%%%%%%%%%%%%%%%%%%
\paragraph{\bf Zero force.}
%%%%%%%%%%%%%%%%%%%%%%%%%%%%%%%%%%%%%%%%%%%%%%%%%%%%%%%%%
Let us first set the force to zero 
 in all states, ${\boldsymbol \phi}(s)={\bm 0}$. This leads to standard equilibrium fluctuation
dynamics that have been investigated thoroughly in the case of edge diffusion~\cite{Khare1995,Giesen2001,Lai2017}.
Although some quantities related to first passage processes
have been discussed within the frame of persistence of fluctuations~\cite{Dougherty2002,Constantin2003},
there is to our knowledge no study of the first passage times
to cluster configurations. We have evaluated $\tau_{\boldsymbol \phi}(s,{\bar{s}})$ numerically using the method of
iterative evaluation~\cite{Sutton1998}:
for a given $\bar{s}$, we iterate the evaluation of  $\tau_{{\boldsymbol \phi}}(s,{\bar{s}})$
by substitution of its value in the right hand side of Eq.~(\ref{eq:recursion}).
Since it requires to list all states, such a method is suitable for
small clusters, which corresponds to our focus in this Letter.
Indeed, the total number $S_N$ of configurations for a cluster with $N$ particles, often called polyominoes 
or lattice animals~\cite{Guttmann2009}, grows exponentially
with $N$: 
$ S_N \sim c\lambda^N/N $, with $ \lambda \approx 4.0626$ and $ c \approx 0.3169$~\cite{Jensen2000}.
We have performed simulations with $N\leq12$, with $S_{12}\approx 5\times 10^{5}$ states.

The resulting expected return time to target with zero force $\tau_0^\mathrm{r}({\bar{s}})$ is shown in Fig.~\ref{fig:returntime}
as a function of $1/T$.
For small clusters, $\tau_0^\mathrm{r}({\bar{s}})$ increases monotonously as
the temperature is decreased. This is expected because thermally activated 
hopping diffusion events become slower at low temperatures.
However, $\tau_0^\mathrm{r}({\bar{s}})$ exhibits a minimum
as a function of temperature for clusters that are larger and more compact.
As shown in {\color{red} Supp. Mat.}, a similar minimum is found in the time $\tau_0(s,{\bar{s}})$ to reach the target
starting from any state $s$.
This striking result implies that some targets exhibit an optimal
temperature at which the target can be reached in minimum time.

\begin{figure}[t]
	\includegraphics[width=\linewidth]{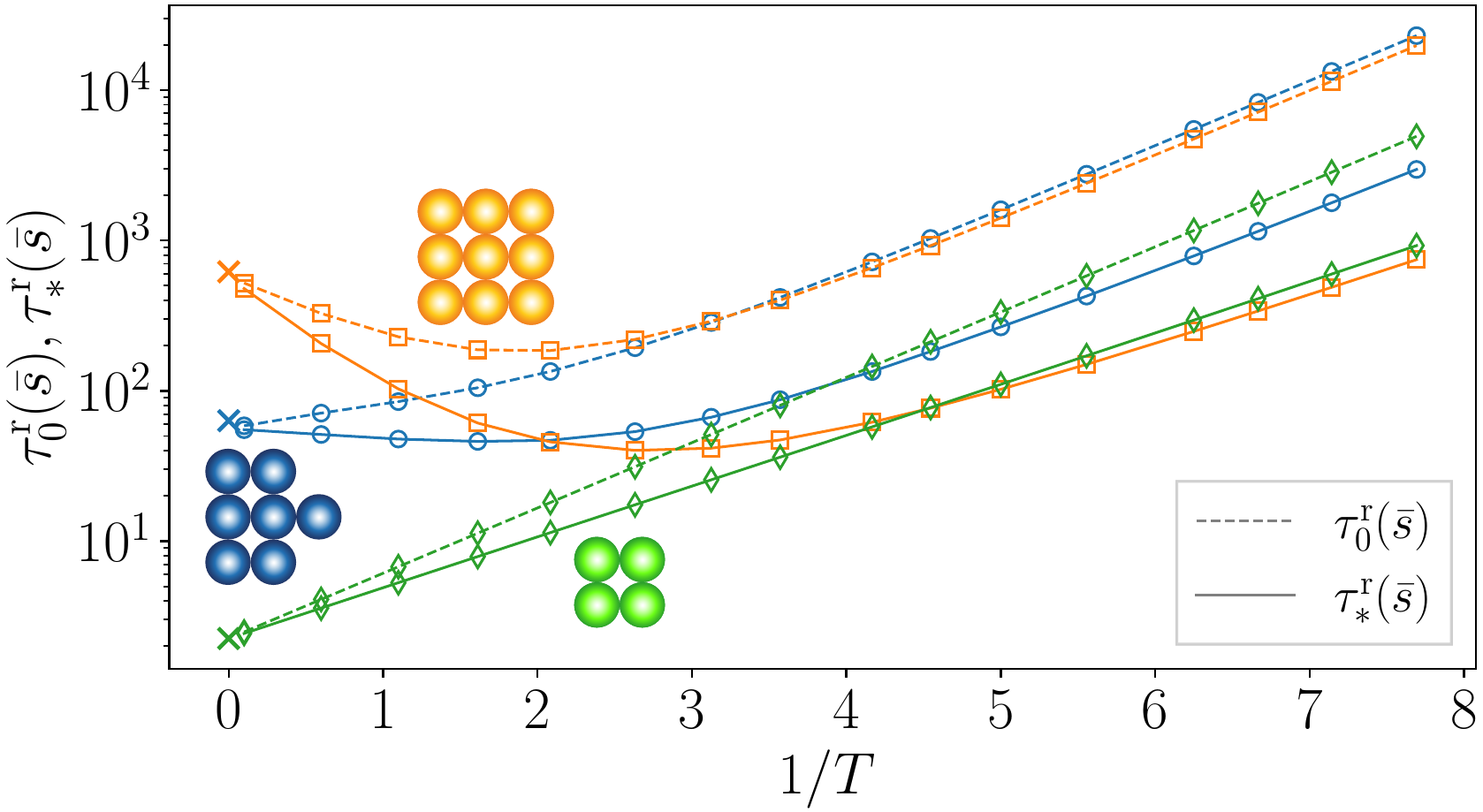}
	\caption{Expected return time to target as a function of $1/T$. Zero-force case $\tau^{\mathrm r}_{0}({\bar{s}})$ 
	and under the optimal policy $\tau^{\mathrm r}_{*}({\bar{s}})$ with $F_0=0.4$.  
	The \mbox{$\times$ correspond} to $\tau_{\infty}^\mathrm{r}(\bar{s})$ for $T \rightarrow \infty$.}
	\label{fig:returntime}
\end{figure}

The presence of a minimum is associated to a change of slope of $\tau_0^\mathrm{r}({\bar{s}})$ as 
a function of $1/T$ at high temperatures. 
We therefore study the high temperature behavior in more details.
In the limit 
%of infinitely high temperatures 
$T\rightarrow\infty$,
the rates (\ref{eq:TST_hopping}) are independent of the initial and final state and of the force: $\gamma_{\boldsymbol \phi}(s,s')\rightarrow 1$.
As a consequence, $\tau_{\boldsymbol \phi}^\mathrm{r}({\bar{s}})$ is independent of the policy ${\boldsymbol \phi}$ 
at infinite temperature $\tau_{\boldsymbol \phi}^\mathrm{r}({\bar{s}})\rightarrow {\tau}^\mathrm{r}_\infty({\bar{s}})$.
A simple result is known from the literature~\cite{Lovasz1993} (see also {\color{red} Supp. Mat.})
when all rates are equal: ${\tau}^\mathrm{r}_\infty({\bar{s}})=(S_N-1)/d_{\bar{s}}$,
where the degree $d_{\bar{s}}$ of the target is the number of states that can be reached
from the target $\bar{s}$ in one move. This quantity is similar to the 
lower bound of the mean first passage time (averaged over all initial states $s$),
which is often used to characterize first passages in
random graphs~\cite{Lin2012,Tejedor2009,Baronchelli2006,Noh2004}.

%Let us first provide some heuristic analysis of
%the variation of the first passage times to the target at high temperature. At infinite temperature,
%all moves have the same rate. Then, 
When the temperature is decreased,
the moves become sensitive to the energy. From detailed balance,
a move that leads to a decrease of energy is faster than the reverse move.
As a consequence, the cluster goes faster towards states with lower energy.
Thus, the time to reach the target decreases if the target has a lower energy.
However, this trend is only describing relative variations
of the time to reach different targets. When decreasing the temperature,
there is also a global slowing-down of the dynamics because of 
the Arrhenius dependence of the rates on temperature in Eq.~(\ref{eq:TST_hopping}).
The decrease or increase of first passage times to the target---or equivalently of $\tau_{\boldsymbol \phi}^\mathrm{r}(\bar{s})$---depends on the competition between these two effects: relative energy effect vs global slowing down.
%Since the expected return time is always growing at very low temperatures, 
%a decrease of $\tau_0^{\mathrm r}({\bar{s}})$ as a function of the inverse temperature $1/T$ at high temperatures
%is enforces the presence of a minimum of the expected return time to the target
%at some finite temperature.

This competition can be analyzed from a high temperature expansion to first order in $1/T$ 
%In order to investigate the high temperature regime for small $1/T$ more quantitatively, 
%we performed a high temperature expansion 
(details are reported in {\color{red} Supp. Mat.}),
leading to 
\begin{align}
\label{eq:tau^r_F=0}
\tau^{\mathrm r}_{0}({\bar{s}}) &= \left(1+ \frac{M_0({\bar{s}})}{T}\right) \tau^{\mathrm r}_{\infty}({\bar{s}})\,,
\\
%\end{align}
%The normalized high temperature slope reads
%\begin{align}
\label{eq:M_0}
%\left(1-\frac{1}{S_N}\right)
M_0({\bar{s}})&=\frac{1}{1-S_N^{-1}}\Bigl\langle 
% (1-\delta_{s\bar{s}})  
\tilde\delta_{s\bar{s}}  \langle n_{ss'}\rangle_{\prime} -d_s g_n(s,{\bar{s}})
\Bigr\rangle_{s\in {\cal S}}\,, 
\end{align}
where $\tilde\delta_{ss'}=1-\delta_{ss'}$ with $\delta$ the Kronecker symbol, $\langle\, \cdot\, \rangle_{s\in {\cal Z}}$ indicates
the average over the states in the set of states ${\cal Z}$, and
${\cal S}$ is the set of all states. We also use the notation 
$\langle\, \cdot\, \rangle_{\prime}=\langle\, \cdot\, \rangle_{s'\in{\cal B}_s}$. In addition, the local covariance of any quantity $q_{ss'}$ with $\tau_{\infty}(s,{\bar{s}})$ is defined as
\begin{align*}
%&g_x(s,{\bar{s}})=
%\label{eq:gn_def}%\\
g_q(s,{\bar{s}})=\Bigl\langle 
(q_{ss'}-\langle q_{ss''}\rangle_{\prime\prime} ) 
(\tau_\infty(s',{\bar{s}})-\langle \tau_\infty(s'',{\bar{s}})\rangle_{{\prime\prime}} )
\Bigr\rangle_{\prime} ,
%g_x(s,{\bar{s}})=\Bigl\langle 
%(x_{ss'}-\langle x_{ss''}\rangle_{s''} ) 
%(\tau_\infty(s',{\bar{s}})-\langle \tau_\infty(s'',{\bar{s}})\rangle_{s''} )
%\Bigr\rangle_{s'}
%\Bigl(x_{ss'}-\langle x_{ss''}\rangle_{s''\in {\cal B}_{s}} \Bigr) 
%\Bigl(\tau_\infty(s',{\bar{s}})-\langle \tau_\infty(s'',{\bar{s}})\rangle_{s''\in {\cal B}_{s}} \Bigr)
%\Bigr\rangle_{s'\in {\cal B}_{s}}
%\nonumber
\end{align*}
where 
%$\langle\, \cdot\, \rangle_{\prime}=\langle\, \cdot\, \rangle_{s'\in{\cal B}_s}$ and 
$\langle\, \cdot\, \rangle_{\prime\prime}=\langle\, \cdot\, \rangle_{s''\in{\cal B}_s}$. 
In Fig.~\ref{fig:R0_R*}(a), we see that Eq.~(\ref{eq:M_0})
is in good agreement with the value $M_0^{\mathrm{sim}}(\bar s)$ obtained from a 
high temperature fit 
%(parabolic fit at $T=5, 10, 20$)  
of the numerical solution from iterative evaluation (small deviations are caused by the 
freezing of 4-neighbors particles).

%%%%%%%%%%%%%%%%%%%%%%%%%%%%%%%%%%%%%%%%%%%%%%%%%%%%%%%%
\begin{figure}[b]
	\includegraphics[width=\linewidth]{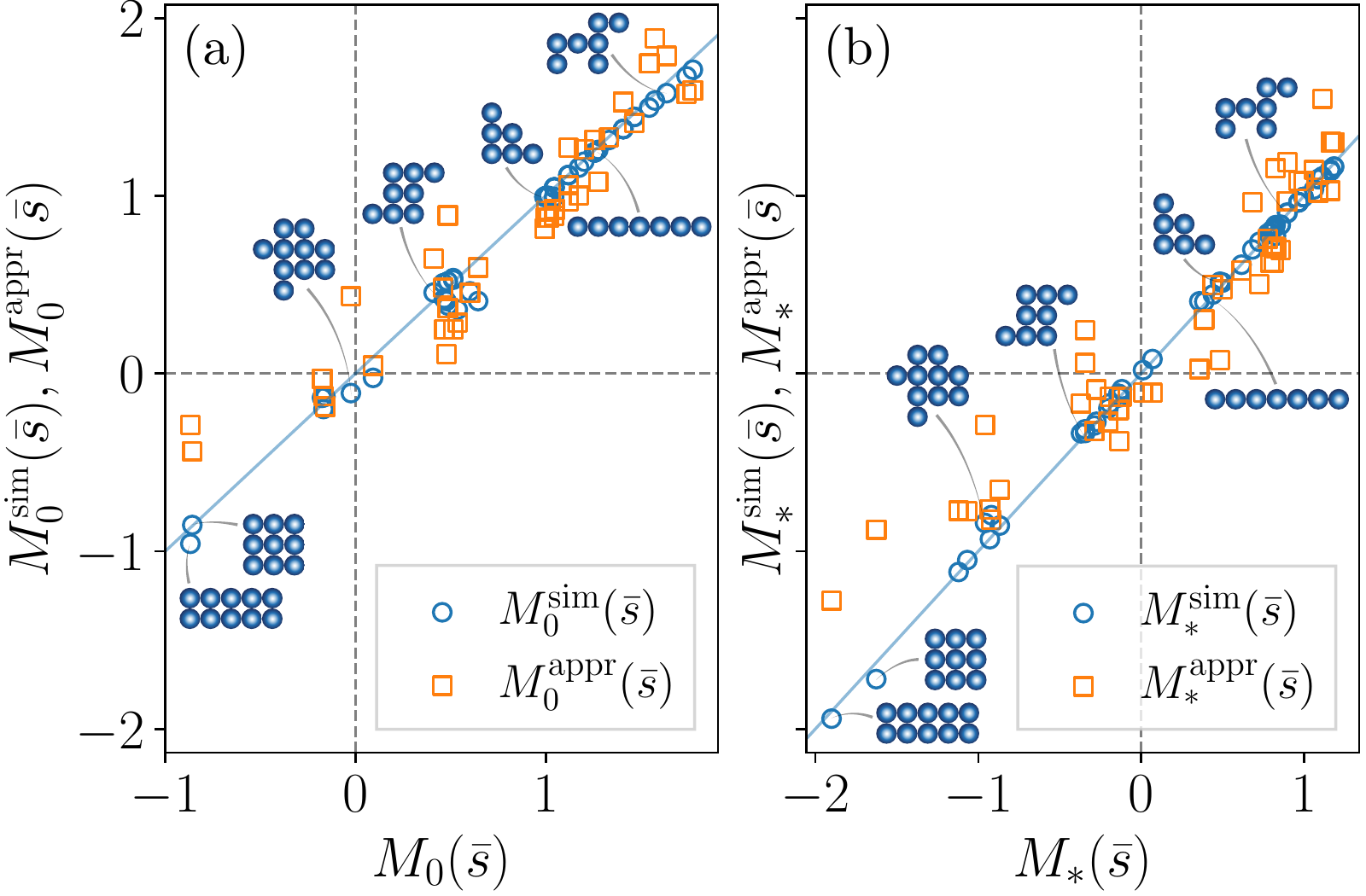}
	\caption{Estimates of the high temperature slope. 
%	The approximations $M_0^{\textrm{appr}}(\bar s)$ and $M_*^{\textrm{appr}}(\bar s)$ 
%	defined in Eqs.~(\ref{eq:approximate_R0},\ref{aeq:M_HT_*_approx}) 
%	are plotted against the analytical expressions of Eqs.~(\ref{eq:M_0},\ref{aeq:M_HT_*}). 
%	$M_0^{\textrm{sim}}(\bar s)$ and $M_*^{\textrm{sim}}(\bar s)$ refer instead to the slopes extracted 
%	%from a parabolic fit at high temperatures ($T=5, 10, 20$) of the expected return time to target obtained 
%	high temperature fit from iterative numerical methods.
    $M_0(\bar s)$ and $M_*(\bar s)$ correspond to the analytical 
    expressions of Eqs.~(\ref{eq:M_0},\ref{eq:tau^r_F}).
    $M_0^{\textrm{sim}}(\bar s)$ and $M_*^{\textrm{sim}}(\bar s)$ are the slopes extracted 
	from a high temperature fit using iterative numerical methods.
	$M_0^{\textrm{appr}}(\bar s)$ and $M_*^{\textrm{appr}}(\bar s)$ 
	are approximate expressions from Eqs.~(\ref{eq:approximate_R0},\ref{aeq:M_HT_*_approx}).
	}
	\label{fig:R0_R*}
\end{figure}
%%%%%%%%%%%%%%%%%%%%%%%%%%%%%%%%%%%%%%%%%%%%%%%%%%%%%%%%

In Eq.~(\ref{eq:M_0}), the first term proportional
to $\langle n_{ss'}\rangle_{\prime}$ accounts for the global slowing down of the 
dynamics, while the second term proportional to $d_sg_n(s,\bar{s})$
accounts for the relative energy effect.
%An approximate expression of $M_0(\bar{s})$ can be obtained 
%in the limit of large $N$. Indeed, 
The  global slowing down contribution
can be approximated by $\rho_0(N)=\langle\langle n_{ss'}\rangle_{\prime}\rangle_{s\in {\cal S}}$,
which converges exponentially to $\rho_0(\infty)\approx 1.64$ for large $N$
(see {\color{red} Supp. Mat.} for details).
The relative energy effect is dominated by moves
from ${\cal B}_{\bar{s}}$ to the target.
%In addition, the relative energy effect contribution 
It is approximately proportional to a measure of cluster compactness defined as
the difference in the number of bonds to break between moves leading to and moves not leading to the target 
$\langle n_{s\bar{s}}-\langle n_{ss''}\rangle_{\prime\prime}  \rangle_{-}$,
where $\langle\, \cdot\, \rangle_{-}=\langle\, \cdot\, \rangle_{s\in{\cal B}_{\bar{s}}}$.
This relation is derived and checked in {\color{red} Supp. Mat.}
%For $N$ large enough, the global slowing down contribution leads to a constant contribution $\rho_0\approx 1.64$,
%while the relative energy effect is proportional to the number of  bonds that are broken to perform
%moves leading directly to the target. We therefore obtain an approximation as
We therefore obtain
\begin{align}
\label{eq:approximate_R0}
M_0^\mathrm{appr}({\bar{s}}) = \rho_0(N) +\rho_1\langle 
%n_{s\bar{s}}-\langle n_{ss''}\rangle_{s''\in {\cal B}_{s}}  \rangle_{s\in {\cal B}_{\bar{s}}},
% n_{s\bar{s}}-\langle n_{ss''}\rangle_{s''\in {\cal B}_{s}}  \rangle_{-}\,,
n_{s\bar{s}}-\langle n_{ss''}\rangle_{\prime\prime}  \rangle_{-}\,,
\end{align}
where $\rho_1\approx 1.60$. 
As shown in Fig.~\ref{fig:R0_R*}(a), $M_0^\mathrm{appr}({\bar{s}})$ provides a fair approximation
to $M_0({\bar{s}})$.
The dispersion originates from assumptions of 
uncorrelation of $\tau_\infty$ with $n_{ss'}$,
and of variations of $\tau_\infty$ dominated by 
the difference between $\tau_\infty(\bar{s},\bar{s})=0$ on the target and $\tau_\infty(s,\bar{s})$ in
its neighborhood ${\cal B}_{\bar{s}}$.
The sign of $M_0^\mathrm{appr}({\bar{s}})$ can serve as a simple guide to
the presence of a minimum as a function of $T$, i.e. an optimal temperature,
and also makes explicit the link between the minimum and the compactness
of the target.
For example, 
in a linear one-atom-thick target, 
only the two atoms at the tips can move, so that  $\langle n_{s\bar{s}} \rangle_{-}= 1$ 
and an inspection of the possible moves shows that 
% $\langle\langle n_{ss''}\rangle_{s''\in {\cal B}_{s}}  \rangle_{-}=4/3$. 
$\langle\langle n_{ss''}\rangle_{\prime\prime}  \rangle_{-}=4/3$. 
This leads to $M_0^\mathrm{appr}({\bar{s}})=\rho_0(N)-\rho_1/3\approx 0.9>0$ for $N=7$,
in agreement with $M_0^\mathrm{sim}({\bar{s}})\approx 1.04$ found by iterative evaluation.
In contrast, in the limit of large compact (square, rectangular, etc) islands, for which
$\langle n_{s\bar{s}} \rangle_{-}=1$ 
and 
% $\langle\langle n_{ss''}\rangle_{s''\in {\cal B}_{s}}  \rangle_{-}\rightarrow 3$,
$\langle\langle n_{ss''}\rangle_{\prime\prime}  \rangle_{-}\rightarrow 3$,
we obtain $M_0^\mathrm{appr}({\bar{s}})= -1.57<0$ leading to a minimum. 
%The sign of this approximate expression provides an approximate
%criterion for the presence of an optimal temperature: $\langle 
%n_{s\bar{s}}-\langle n_{ss''}\rangle_{s''\in {\cal B}_{s}}  \rangle_{-}\geq\rho_0/\rho_1\approx 1$.
%{\color{blue} More generally, in the limit of large $N$, we have
%$\langle n_{s\bar{s}}-\langle n_{ss''}\rangle_{\prime\prime}  \rangle_{-}
%\approx H(\bar{s})-\langle H(\bar{s})\rangle_{-}$
%where $H(\bar s)$ is the cluster energy (see {\color{blue} Supp. Mat.}).  
%Hence, this quantity also describes the difference between the 
%energy of the target and its neighboring states.}

Note however that the convergence of value iteration 
is difficult not only for large clusters, but
also for high-energy (i.e., non-compact) target shapes when the temperature is decreased.
Indeed, disparate times-scales have to be resolved (fast relaxation
towards low-energy shapes vs large time to reach high-energy shapes).

%%%%%%%%%%%%%%%%%%%%%%%%%%%%%%%%%%%%%%%%%%%%%%%%%%%%%%%%%
\paragraph{\bf Optimal policy in the presence of forces.}
%%%%%%%%%%%%%%%%%%%%%%%%%%%%%%%%%%%%%%%%%%%%%%%%%%%%%%%%%
Our goal now is to determine the optimal policy ${\boldsymbol \phi}_*$ that minimizes $\tau_{\boldsymbol \phi}(s)$,
and the resulting optimal first passage time $\tau_*(s, \bar s) = \min_{{\boldsymbol \phi}}\tau_{{\boldsymbol \phi}}(s,\bar{s})$ for non-zero forces.
Such a problem, called a Markov decision process, can be solved
using well-known dynamic programming methods~\cite{Sutton1998,Bellman1957}. 
We substitute the optimal policy in Eq.~(\ref{eq:recursion}) to obtain
the so-called Bellmann optimality equation
\begin{equation}\label{eq:Bellman}
\tau_*(s,{\bar{s}}) = \min_{{\boldsymbol \phi}(s)}\Bigl[  
t_{\boldsymbol \phi}(s)  + \sum_{s'\in{\cal B}_s}p_{\boldsymbol \phi}(s,s')\tau_*(s',{\bar{s}})
\Bigr] \,.
\end{equation}
As in the zero-force case, we iterate Eq.(\ref{eq:Bellman}). 
However minimization over the force in $s$ is taken at each iteration.
This method, called value iteration, provides both $\tau_*(s,{\bar{s}})$ 
and the optimal policy ${\boldsymbol \phi}_*$.
Due to the fast increase of $S_N$ with $N$,
its computational cost grows exponentially with $N$
(see {\color{red}Supp. Mat.}).

%In order to prove the efficiency of our method
%when the degrees of freedom and the symmetry of the external field are strongly restricted,
%we choose a field along a fixed direction that can be switched to
%three values only, positive, negative and zero.

We choose a force $\bm{F}$ that is always oriented in $\hat{\bm{x}}$ direction ((10) lattice direction), with 3 possible values: 
$\{-F_0\hat{\bm{x}}$, $\bm 0$, $F_0 \hat{\bm{x}}\}$, with $F_0>0$.
An example of optimal policy is shown in Fig.~\ref{fig:graph_dynamics}.
As an important remark, the force can drive the cluster towards
any target shape even if the symmetries of the target are not compatible
with those of the force because observation itself (i.e. the knowledge of $s$) breaks
the symmetry.
%\com{what does this mean exactly?}.
%{\color{blue} [[It means that since we know $s$, we can
%take two different and arbitrary values $\boldsymbol\phi(s_1)$
%and $\boldsymbol\phi(s_2)$ for two states $s_1$ and $s_2$ that are
%obtained from each other using a given symmetry.]]}

The gain due to the optimal policy is reported
in Fig.~\ref{fig:gain}, using the zero-force policy as a reference.
In {\color{red} Supp. Mat.}, we show that using a random-force policy as a reference leads to similar results.
As seen from Fig.~\ref{fig:gain}(a), the gain increases not only when $F_0$ is increased, 
but also when $T$ is decreased. This is intuitively expected since
%the forces have no effect in the limit of infinitely high temperatures, and 
the relative change between different rates
due to a change of the force increases when $F_0/T$ is increased.
In addition, the gain increases
when the size of the cluster increases, as shown in Fig.~\ref{fig:gain}(b). A naive explanation for this
trend is that an increase of $N$ leads to an increase of the number of states $S_N$,
and therefore to an increase of the number of ways to tune the policy ${\boldsymbol \phi}$
in order to minimize $\tau_{\boldsymbol \phi}(s,\bar{s})$.

%Again, a high temperature expansion provides
%further hints on the dependence in $N$.
Again, a high temperature expansion leads to
(derivation reported in {\color{red} Supp. Mat.})
\begin{align}
\label{eq:tau^r_F}
\tau^{\mathrm r}_{*}({\bar{s}}) &= \left(1+ \frac{M_*({\bar{s}})}{T}\right) \tau^{\mathrm r}_{\infty}({\bar{s}})\,,
\\
%\end{align}
%with the normalized slope for the optimal policy 
%\begin{align}
% \label{aeq:M_HT_*}
M_*({\bar{s}}) &= M_0({\bar{s}})
%\nonumber \\&
-\frac{F_0}{1-S_N^{-1}}\Bigl\langle 
 \Bigl|
 %(1-\delta_{s\bar{s}}) 
 \tilde \delta_{s\bar{s}}\langle u_{ss'}\rangle_{\prime} -d_s g_u(s,{\bar{s}})\Bigr|
\Bigr\rangle_{s\in {\cal S}},
\nonumber
\end{align}
where $u_{ss'}={\bm{u}_{ss'}}\cdot \hat{\bm x}$. 
The numerical solution of \cref{eq:Bellman}
is in agreement with \cref{eq:tau^r_F} (up to
small deviations due to 4-neighbors particle freezing, see {\color{red} Supp. Mat.}).

\begin{figure}[t]
	\includegraphics[width=\linewidth]{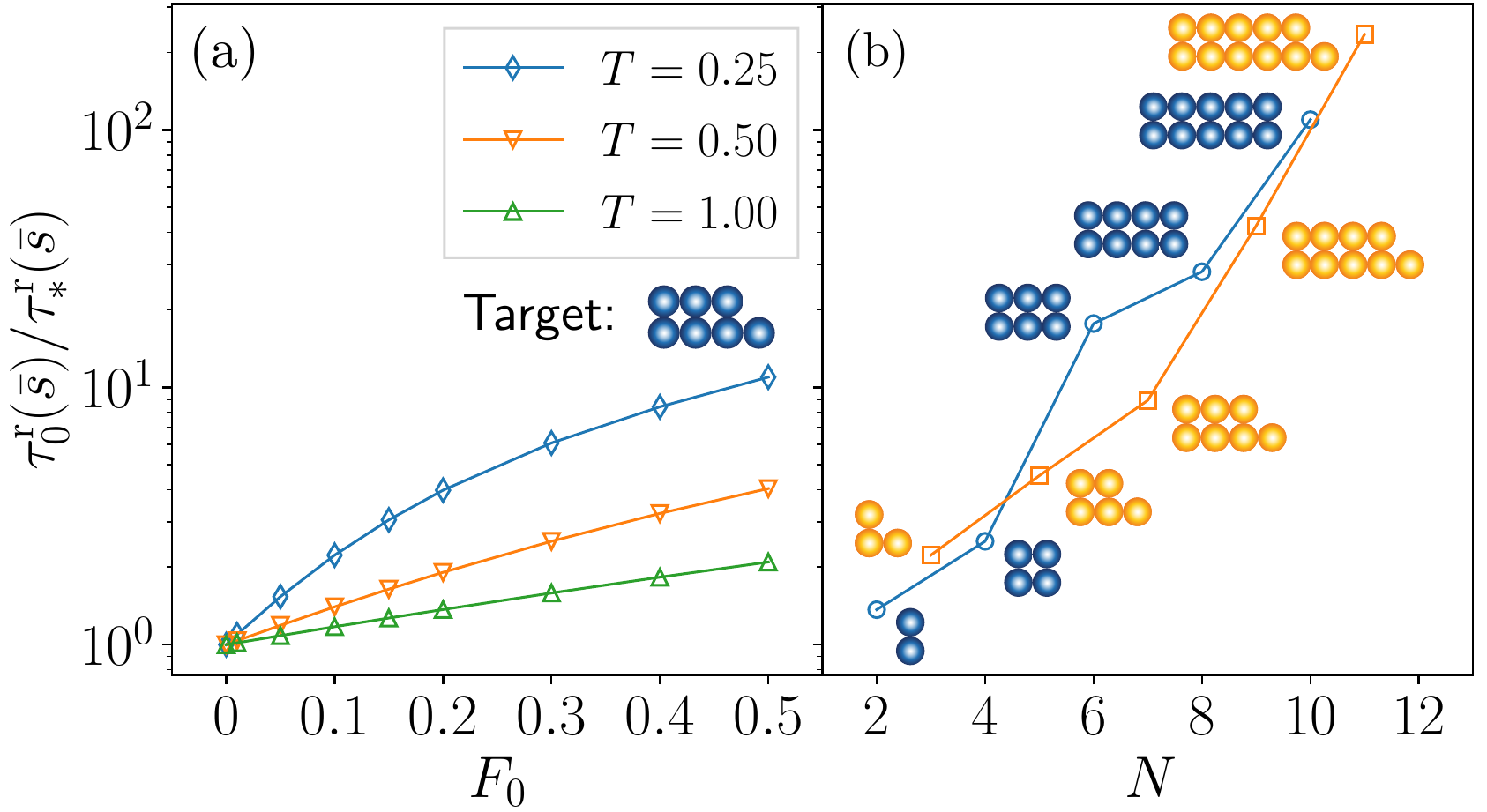}
	\caption{Gain $\tau^{\mathrm r}_0({\bar{s}})/\tau^{\mathrm r}_*({\bar{s}})$ in the return time to target 
	due to the optimization of the forces. 
	(a) As a function of the force magnitude $F_0$, for a fixed target at different $T$. (b) For similar targets of increasing size, with $T=0.24$ and $F_0=0.4$.}
	\label{fig:gain}
\end{figure}

Two remarks are in order. 
First,  $\langle u_{ss'}\rangle_{\prime}$ 
is small and its contribution
to the term proportional to $F_0$ in Eq.~(\ref{eq:tau^r_F})
is negligible.
Second, the absolute value  forbids the cancellation 
of contributions with randomly different signs, leading to a 
behavior which is qualitatively different from that of $M_0(\bar s)$.
Indeed, the average is not dominated by the largest terms coming 
from the strong change of $\tau_\infty(s,{\bar{s}})$ between the target
and its first neighbors, but by the typical values
of $|d_s g_u(s,{\bar{s}})|$ in all states.
Based on this observation, a detailed analysis reported in {\color{red} Supp. Mat.} leads to
the approximation
\begin{align}
\label{aeq:M_HT_*_approx}
M_*^\mathrm{appr}({\bar{s}})= M_0({\bar{s}}) - 
\frac{2^{1/2}}{\pi^{1/2}}
F_0\;\sigma_u\;\sigma_{\tau_\infty}\!({\bar{s}})\;
\langle d_s^{1/2}\rangle_{s\in {\cal S}},
\end{align}
where we have defined the standard deviations
\begin{align}
\sigma_u &= \left\langle \left\langle 
(u_{ss'}-\langle u_{ss''}\rangle_{\prime\prime})^2
\right\rangle_{\prime}^{1/2}\right\rangle_{s\in {\cal S}} \,,
\\
\sigma_{\tau_\infty}({\bar s}) &= \left\langle \left\langle 
(\tau_\infty(s',{\bar{s}})-\langle \tau_\infty(s'',{\bar{s}})\rangle_{\prime\prime})^2
\right\rangle_{\prime}^{1/2}\right\rangle_{s\in {\cal S}}.
\end{align}
In \cref{fig:R0_R*}, the approximation Eq.(\ref{aeq:M_HT_*_approx})
is seen to be valid up to some dispersion originating
mainly in the assumption of uncorrelation between $u_{ss'}$ and $\tau_\infty$.

While $\langle n_{s\bar{s}}-\langle n_{ss''}\rangle_{\prime\prime}\rangle_{-}$
and $\sigma_u$ are bounded
because $1\leq n_{ss'}\leq 4$ and $-1/2\leq -u_{ss'}\leq 1/2$,
the quantities $\sigma_{\tau_\infty}({\bar{s}})$ and  $\langle d_s^{1/2}\rangle_{s\in{\cal S}}$ grow with $N$
(see {\color{red} Supp. Mat.}).
Hence, from Eq.~(\ref{eq:approximate_R0}), $M_0^{\mathrm{appr}}(\bar s)$ is bounded and the contribution proportional to $F_0$ in Eq.~(\ref{aeq:M_HT_*_approx})
usually dominates over the term $M_0(\bar{s})$ for large $N$. Thus, when $N$ is large enough,
$M_*(\bar s)$ should be negative and an optimal temperature should be generically present.
This trend is confirmed by Fig.~\ref{fig:R0_R*}(b).
Simulations with $N=12$ and $F_0=0.4$ (reported in Fig.~10
% \ref{fig:returntime_big_targets} 
of {\color{red} Supp. Mat.})
also confirm the generic presence of a minimum for larger targets.

%%%%%%%%%%%%%%%%%%%%%%%%%%%%%%%%%%%%%%%%%%%%%%%%
\paragraph{\bf Discussion.}
%%%%%%%%%%%%%%%%%%%%%%%%%%%%%%%%%%%%%%%%%%%%%%%%

For atomic metals clusters where edge diffusion is observed (Ag, Cu, etc.), 
estimates of the edge diffusion barrier $\sim J$ or kink energies $\sim J/2$ suggest 
that $J\approx$ 0.2 to 0.3 \si{\electronvolt}~\cite{Ferrando1994,Yu1997,Mehl1999,Nelson1993,Giesen2001}.
For the square 9-atom target depicted in Fig.\ref{fig:returntime} 
the optimal temperature corresponds to $J/k_BT\approx 2$. Choosing $J=0.2\, \si{\electronvolt}$ we obtain
an optimal temperature \mbox{$\approx 10^3$ K} which is too high to be observed in usual experiments.
Thus, $\tau_0(s,\bar{s})$ should decrease with temperature
in usual experimental conditions.
If needed, a quench can also be used to freeze the cluster once the target shape is reached.
However, using electromigration as an external force
leads to~\cite{Tao2010} $ F_0a/J \approx 10^{-4} $,
which is too small to allow for the control of few-atoms clusters.

Edge diffusion can also be observed with colloids~\cite{Hubartt2015}.
Using colloids with depletants, $J\sim$ few $k_B T$~\cite{Nozawa2018}.
The optimal temperature should then be observable in the absence of force.
Thermophoretic forces~\cite{Helden2015,Braibanti2008,Wrger2010}
for polystyrene beads of radius $ 2.5 \, \si{\micro\meter} $
are \mbox{$\sim 10 \, k_BT/ \si{\micro\meter}$}~\cite{Helden2015}. 
Hence, micron-size colloids can lead to $F_0a/J\sim 1$,
which allows for shape control by a macroscopic force.

However, most experiments on colloid clusters report mass transport
dominated by attachment-detachment at the edges~\cite{Ganapathy2010}. 
Our analysis can be adapted to 
vacancy clusters~\cite{Martinez2014,Garcia2007,Pariente2020} with volume-preserving 
detachment-diffusion-reattachment events.
%within vacancy clusters,
%which preserve the total volume
 %As a final remark, other two-dimensional 
% lattices can be
% used depending on the cluster structure (e.g., most colloid clusters have hexagonal symmetry). 
% Three-dimensional nano-clusters could also be handled. Their number of configurations
% (known as polycubes) grows again exponentially, 
% but faster than in two dimensions~\cite{Guttmann2009}.
Moreover, other two or three-dimensional lattices also could be analyzed.
Furthermore, multi-particle  and off-lattice processes
(such as those involved in dislocation-mediated dynamics)
can be included as long as they are pre-determined
(using e.g. energy-exploration methods) and their number is finite,
to allow for the numerical solution 
of \cref{eq:recursion,eq:Bellman}.
Since the presence of the minimum
depends only on the generic competition between 
the relative energy effect and global slowing down,
we speculate that it should not depend on the details
of mass transport kinetics.

%%%%%%%%%%%%%%%%%%%%%%%%%%%%%%%%%%%%%%%%%%%%%%%%
%%%%%%%%%%%%%%%%%%%%CONCLUSION%%%%%%%%%%%%%%%%%%
%%%%%%%%%%%%%%%%%%%%%%%%%%%%%%%%%%%%%%%%%%%%%%%%

In conclusion, thermal fluctuations
% ---which are usually considered to be detrimental---
can be used to reach desired nano-cluster shapes.
% When the target shape is large-enough and compact, 
There is a temperature
that minimizes the time to reach large-enough and compact shapes.
Furthermore, macroscopic fields can help gaining orders
of magnitude in the time to reach arbitrary shapes.
%In conclusion, we have investigated ways to minimize the time 
%for fluctuating few-particles clusters to reach a target shape,
%choosing the best 
%temperature without force or using the optimal policy with force.
%%In the absence of external field, there is an optimal temperature where low-energy target shapes are reached in minimum time.
%%This phenomena should be observable in atomic and colloidal clusters.
%%In the presence of a macroscopic field, the time to reach the target shape
%%can be decreased. This gain increases with decreasing temperature. 
%%For large-enough clusters, an optimal temperature
%%should be present for most target shapes. 
%%%In a companion paper, we also show that the policy exhibits discrete transitions
%%%as a function of temperature, and can be degenerate in the sense that
%%%different actions can lead to the same value of the first passage time~\cite{Boccardo2021b}. 
We hope that our work will motivate experimental investigations
for the control of atomic and colloidal clusters, and will open theoretical directions
for the  optimization of first passage times on graphs.
}

\bibliography{references}

\begin{thebibliography}{58}%
\makeatletter
\providecommand \@ifxundefined [1]{%
 \@ifx{#1\undefined}
}%
\providecommand \@ifnum [1]{%
 \ifnum #1\expandafter \@firstoftwo
 \else \expandafter \@secondoftwo
 \fi
}%
\providecommand \@ifx [1]{%
 \ifx #1\expandafter \@firstoftwo
 \else \expandafter \@secondoftwo
 \fi
}%
\providecommand \natexlab [1]{#1}%
\providecommand \enquote  [1]{``#1''}%
\providecommand \bibnamefont  [1]{#1}%
\providecommand \bibfnamefont [1]{#1}%
\providecommand \citenamefont [1]{#1}%
\providecommand \href@noop [0]{\@secondoftwo}%
\providecommand \href [0]{\begingroup \@sanitize@url \@href}%
\providecommand \@href[1]{\@@startlink{#1}\@@href}%
\providecommand \@@href[1]{\endgroup#1\@@endlink}%
\providecommand \@sanitize@url [0]{\catcode `\\12\catcode `\$12\catcode
  `\&12\catcode `\#12\catcode `\^12\catcode `\_12\catcode `\%12\relax}%
\providecommand \@@startlink[1]{}%
\providecommand \@@endlink[0]{}%
\providecommand \url  [0]{\begingroup\@sanitize@url \@url }%
\providecommand \@url [1]{\endgroup\@href {#1}{\urlprefix }}%
\providecommand \urlprefix  [0]{URL }%
\providecommand \Eprint [0]{\href }%
\providecommand \doibase [0]{http://dx.doi.org/}%
\providecommand \selectlanguage [0]{\@gobble}%
\providecommand \bibinfo  [0]{\@secondoftwo}%
\providecommand \bibfield  [0]{\@secondoftwo}%
\providecommand \translation [1]{[#1]}%
\providecommand \BibitemOpen [0]{}%
\providecommand \bibitemStop [0]{}%
\providecommand \bibitemNoStop [0]{.\EOS\space}%
\providecommand \EOS [0]{\spacefactor3000\relax}%
\providecommand \BibitemShut  [1]{\csname bibitem#1\endcsname}%
\let\auto@bib@innerbib\@empty
%</preamble>
\bibitem [{\citenamefont {Binnig}\ and\ \citenamefont
  {Rohrer}(1983)}]{Binnig1983}%
  \BibitemOpen
  \bibfield  {author} {\bibinfo {author} {\bibfnamefont {G.}~\bibnamefont
  {Binnig}}\ and\ \bibinfo {author} {\bibfnamefont {H.}~\bibnamefont
  {Rohrer}},\ }\href
  {https://www.sciencedirect.com/science/article/pii/0039602883907161}
  {\bibfield  {journal} {\bibinfo  {journal} {Surf. Sci.}\ }\textbf {\bibinfo
  {volume} {126}},\ \bibinfo {pages} {236} (\bibinfo {year}
  {1983})}\BibitemShut {NoStop}%
\bibitem [{\citenamefont {Eigler}\ and\ \citenamefont
  {Schweizer}(1990)}]{Eigler1990}%
  \BibitemOpen
  \bibfield  {author} {\bibinfo {author} {\bibfnamefont {D.~M.}\ \bibnamefont
  {Eigler}}\ and\ \bibinfo {author} {\bibfnamefont {E.~K.}\ \bibnamefont
  {Schweizer}},\ }\href {https://doi.org/10.1038/344524a0} {\bibfield
  {journal} {\bibinfo  {journal} {Nature}\ }\textbf {\bibinfo {volume} {344}},\
  \bibinfo {pages} {524} (\bibinfo {year} {1990})}\BibitemShut {NoStop}%
\bibitem [{\citenamefont {Cooper}\ \emph {et~al.}(2018)\citenamefont {Cooper},
  \citenamefont {Covey}, \citenamefont {Madjarov}, \citenamefont {Porsev},
  \citenamefont {Safronova},\ and\ \citenamefont {Endres}}]{Cooper2018}%
  \BibitemOpen
  \bibfield  {author} {\bibinfo {author} {\bibfnamefont {A.}~\bibnamefont
  {Cooper}}, \bibinfo {author} {\bibfnamefont {J.~P.}\ \bibnamefont {Covey}},
  \bibinfo {author} {\bibfnamefont {I.~S.}\ \bibnamefont {Madjarov}}, \bibinfo
  {author} {\bibfnamefont {S.~G.}\ \bibnamefont {Porsev}}, \bibinfo {author}
  {\bibfnamefont {M.~S.}\ \bibnamefont {Safronova}}, \ and\ \bibinfo {author}
  {\bibfnamefont {M.}~\bibnamefont {Endres}},\ }\href
  {https://link.aps.org/doi/10.1103/PhysRevX.8.041055} {\bibfield  {journal}
  {\bibinfo  {journal} {Phys. Rev. X}\ }\textbf {\bibinfo {volume} {8}},\
  \bibinfo {pages} {041055} (\bibinfo {year} {2018})}\BibitemShut {NoStop}%
\bibitem [{\citenamefont {Barredo}\ \emph {et~al.}(2016)\citenamefont
  {Barredo}, \citenamefont {de~L{\'{e}}s{\'{e}}leuc}, \citenamefont {Lienhard},
  \citenamefont {Lahaye},\ and\ \citenamefont {Browaeys}}]{Barredo2016}%
  \BibitemOpen
  \bibfield  {author} {\bibinfo {author} {\bibfnamefont {D.}~\bibnamefont
  {Barredo}}, \bibinfo {author} {\bibfnamefont {S.}~\bibnamefont
  {de~L{\'{e}}s{\'{e}}leuc}}, \bibinfo {author} {\bibfnamefont
  {V.}~\bibnamefont {Lienhard}}, \bibinfo {author} {\bibfnamefont
  {T.}~\bibnamefont {Lahaye}}, \ and\ \bibinfo {author} {\bibfnamefont
  {A.}~\bibnamefont {Browaeys}},\ }\href
  {https://doi.org/10.1126/science.aah3778} {\bibfield  {journal} {\bibinfo
  {journal} {Science}\ }\textbf {\bibinfo {volume} {354}},\ \bibinfo {pages}
  {1021} (\bibinfo {year} {2016})}\BibitemShut {NoStop}%
\bibitem [{\citenamefont {Mart{\'\i}nez-Galera}\ \emph
  {et~al.}(2014)\citenamefont {Mart{\'\i}nez-Galera}, \citenamefont {Brihuega},
  \citenamefont {Guti{\'e}rrez-Rubio}, \citenamefont {Stauber},\ and\
  \citenamefont {G{\'o}mez-Rodr{\'\i}guez}}]{Martinez2014}%
  \BibitemOpen
  \bibfield  {author} {\bibinfo {author} {\bibfnamefont {A.}~\bibnamefont
  {Mart{\'\i}nez-Galera}}, \bibinfo {author} {\bibfnamefont {I.}~\bibnamefont
  {Brihuega}}, \bibinfo {author} {\bibfnamefont {A.}~\bibnamefont
  {Guti{\'e}rrez-Rubio}}, \bibinfo {author} {\bibfnamefont {T.}~\bibnamefont
  {Stauber}}, \ and\ \bibinfo {author} {\bibfnamefont {J.}~\bibnamefont
  {G{\'o}mez-Rodr{\'\i}guez}},\ }\href@noop {} {\bibfield  {journal} {\bibinfo
  {journal} {Sci. Rep.}\ }\textbf {\bibinfo {volume} {4}},\ \bibinfo {pages}
  {1} (\bibinfo {year} {2014})}\BibitemShut {NoStop}%
\bibitem [{\citenamefont {Erickson}\ \emph {et~al.}(2011)\citenamefont
  {Erickson}, \citenamefont {Serey}, \citenamefont {Chen},\ and\ \citenamefont
  {Mandal}}]{Erickson2011}%
  \BibitemOpen
  \bibfield  {author} {\bibinfo {author} {\bibfnamefont {D.}~\bibnamefont
  {Erickson}}, \bibinfo {author} {\bibfnamefont {X.}~\bibnamefont {Serey}},
  \bibinfo {author} {\bibfnamefont {Y.-F.}\ \bibnamefont {Chen}}, \ and\
  \bibinfo {author} {\bibfnamefont {S.}~\bibnamefont {Mandal}},\ }\href
  {https://doi.org/10.1039/c0lc00482k} {\bibfield  {journal} {\bibinfo
  {journal} {Lab Chip}\ }\textbf {\bibinfo {volume} {11}},\ \bibinfo {pages}
  {995} (\bibinfo {year} {2011})}\BibitemShut {NoStop}%
\bibitem [{\citenamefont {Junno}\ \emph {et~al.}(1995)\citenamefont {Junno},
  \citenamefont {Deppert}, \citenamefont {Montelius},\ and\ \citenamefont
  {Samuelson}}]{Junno1995}%
  \BibitemOpen
  \bibfield  {author} {\bibinfo {author} {\bibfnamefont {T.}~\bibnamefont
  {Junno}}, \bibinfo {author} {\bibfnamefont {K.}~\bibnamefont {Deppert}},
  \bibinfo {author} {\bibfnamefont {L.}~\bibnamefont {Montelius}}, \ and\
  \bibinfo {author} {\bibfnamefont {L.}~\bibnamefont {Samuelson}},\ }\href
  {https://doi.org/10.1063/1.113809} {\bibfield  {journal} {\bibinfo  {journal}
  {Appl. Phys. Lett.}\ }\textbf {\bibinfo {volume} {66}},\ \bibinfo {pages}
  {3627} (\bibinfo {year} {1995})}\BibitemShut {NoStop}%
\bibitem [{\citenamefont {Baur}\ \emph {et~al.}(1998)\citenamefont {Baur},
  \citenamefont {Bugacov}, \citenamefont {Koel}, \citenamefont {Madhukar},
  \citenamefont {Montoya}, \citenamefont {Ramachandran}, \citenamefont
  {Requicha}, \citenamefont {Resch},\ and\ \citenamefont {Will}}]{Baur1998}%
  \BibitemOpen
  \bibfield  {author} {\bibinfo {author} {\bibfnamefont {C.}~\bibnamefont
  {Baur}}, \bibinfo {author} {\bibfnamefont {A.}~\bibnamefont {Bugacov}},
  \bibinfo {author} {\bibfnamefont {B.~E.}\ \bibnamefont {Koel}}, \bibinfo
  {author} {\bibfnamefont {A.}~\bibnamefont {Madhukar}}, \bibinfo {author}
  {\bibfnamefont {N.}~\bibnamefont {Montoya}}, \bibinfo {author} {\bibfnamefont
  {T.~R.}\ \bibnamefont {Ramachandran}}, \bibinfo {author} {\bibfnamefont
  {A.~A.~G.}\ \bibnamefont {Requicha}}, \bibinfo {author} {\bibfnamefont
  {R.}~\bibnamefont {Resch}}, \ and\ \bibinfo {author} {\bibfnamefont
  {P.}~\bibnamefont {Will}},\ }\href
  {https://doi.org/10.1088/0957-4484/9/4/011} {\bibfield  {journal} {\bibinfo
  {journal} {Nanotechnology}\ }\textbf {\bibinfo {volume} {9}},\ \bibinfo
  {pages} {360} (\bibinfo {year} {1998})}\BibitemShut {NoStop}%
\bibitem [{\citenamefont {Rubio-Sierra}\ \emph {et~al.}(2005)\citenamefont
  {Rubio-Sierra}, \citenamefont {Heckl},\ and\ \citenamefont
  {Stark}}]{RubioSierra2005}%
  \BibitemOpen
  \bibfield  {author} {\bibinfo {author} {\bibfnamefont {F.~J.}\ \bibnamefont
  {Rubio-Sierra}}, \bibinfo {author} {\bibfnamefont {W.~M.}\ \bibnamefont
  {Heckl}}, \ and\ \bibinfo {author} {\bibfnamefont {R.~W.}\ \bibnamefont
  {Stark}},\ }\href {https://doi.org/10.1002/adem.200400174} {\bibfield
  {journal} {\bibinfo  {journal} {Adv. Eng. Mater.}\ }\textbf {\bibinfo
  {volume} {7}},\ \bibinfo {pages} {193} (\bibinfo {year} {2005})}\BibitemShut
  {NoStop}%
\bibitem [{\citenamefont {Korda}\ \emph {et~al.}(2002)\citenamefont {Korda},
  \citenamefont {Taylor},\ and\ \citenamefont {Grier}}]{Korda2002}%
  \BibitemOpen
  \bibfield  {author} {\bibinfo {author} {\bibfnamefont {P.~T.}\ \bibnamefont
  {Korda}}, \bibinfo {author} {\bibfnamefont {M.~B.}\ \bibnamefont {Taylor}}, \
  and\ \bibinfo {author} {\bibfnamefont {D.~G.}\ \bibnamefont {Grier}},\ }\href
  {https://link.aps.org/doi/10.1103/PhysRevLett.89.128301} {\bibfield
  {journal} {\bibinfo  {journal} {Phys. Rev. Lett.}\ }\textbf {\bibinfo
  {volume} {89}},\ \bibinfo {pages} {128301} (\bibinfo {year}
  {2002})}\BibitemShut {NoStop}%
\bibitem [{\citenamefont {Grier}(2003)}]{Grier2003}%
  \BibitemOpen
  \bibfield  {author} {\bibinfo {author} {\bibfnamefont {D.~G.}\ \bibnamefont
  {Grier}},\ }\href {https://www.nature.com/articles/nature01935} {\bibfield
  {journal} {\bibinfo  {journal} {Nature}\ }\textbf {\bibinfo {volume} {424}},\
  \bibinfo {pages} {810} (\bibinfo {year} {2003})}\BibitemShut {NoStop}%
\bibitem [{\citenamefont {McCormack}\ \emph {et~al.}(2018)\citenamefont
  {McCormack}, \citenamefont {Han},\ and\ \citenamefont {Yan}}]{McCormack2018}%
  \BibitemOpen
  \bibfield  {author} {\bibinfo {author} {\bibfnamefont {P.}~\bibnamefont
  {McCormack}}, \bibinfo {author} {\bibfnamefont {F.}~\bibnamefont {Han}}, \
  and\ \bibinfo {author} {\bibfnamefont {Z.}~\bibnamefont {Yan}},\ }\href
  {https://doi.org/10.1021/acs.jpclett.7b03188} {\bibfield  {journal} {\bibinfo
   {journal} {J. Phys. Chem. Lett.}\ }\textbf {\bibinfo {volume} {9}},\
  \bibinfo {pages} {545} (\bibinfo {year} {2018})}\BibitemShut {NoStop}%
\bibitem [{\citenamefont {Kuhn}\ \emph {et~al.}(2005)\citenamefont {Kuhn},
  \citenamefont {Krug}, \citenamefont {Hausser},\ and\ \citenamefont
  {Voigt}}]{Kuhn2005}%
  \BibitemOpen
  \bibfield  {author} {\bibinfo {author} {\bibfnamefont {P.}~\bibnamefont
  {Kuhn}}, \bibinfo {author} {\bibfnamefont {J.}~\bibnamefont {Krug}}, \bibinfo
  {author} {\bibfnamefont {F.}~\bibnamefont {Hausser}}, \ and\ \bibinfo
  {author} {\bibfnamefont {A.}~\bibnamefont {Voigt}},\ }\href
  {https://link.aps.org/doi/10.1103/PhysRevLett.94.166105} {\bibfield
  {journal} {\bibinfo  {journal} {Phys. Rev. Lett.}\ }\textbf {\bibinfo
  {volume} {94}},\ \bibinfo {pages} {166105} (\bibinfo {year}
  {2005})}\BibitemShut {NoStop}%
\bibitem [{\citenamefont {Mahadevan}\ and\ \citenamefont
  {Bradley}(1999)}]{Mahadevan1999}%
  \BibitemOpen
  \bibfield  {author} {\bibinfo {author} {\bibfnamefont {M.}~\bibnamefont
  {Mahadevan}}\ and\ \bibinfo {author} {\bibfnamefont {R.~M.}\ \bibnamefont
  {Bradley}},\ }\href {https://link.aps.org/doi/10.1103/PhysRevB.59.11037}
  {\bibfield  {journal} {\bibinfo  {journal} {Phys. Rev. B}\ }\textbf {\bibinfo
  {volume} {59}},\ \bibinfo {pages} {11037} (\bibinfo {year}
  {1999})}\BibitemShut {NoStop}%
\bibitem [{\citenamefont {Pierre-Louis}\ and\ \citenamefont
  {Einstein}(2000)}]{PierreLouis2000}%
  \BibitemOpen
  \bibfield  {author} {\bibinfo {author} {\bibfnamefont {O.}~\bibnamefont
  {Pierre-Louis}}\ and\ \bibinfo {author} {\bibfnamefont {T.~L.}\ \bibnamefont
  {Einstein}},\ }\href {https://doi.org/10.1103/physrevb.62.13697} {\bibfield
  {journal} {\bibinfo  {journal} {Phys. Rev. B}\ }\textbf {\bibinfo {volume}
  {62}},\ \bibinfo {pages} {13697} (\bibinfo {year} {2000})}\BibitemShut
  {NoStop}%
\bibitem [{\citenamefont {Curiotto}\ \emph {et~al.}(2019)\citenamefont
  {Curiotto}, \citenamefont {Leroy}, \citenamefont {M\"{u}ller}, \citenamefont
  {Cheynis}, \citenamefont {Michailov}, \citenamefont {El-Barraj},\ and\
  \citenamefont {Ranguelov}}]{Curiotto2019}%
  \BibitemOpen
  \bibfield  {author} {\bibinfo {author} {\bibfnamefont {S.}~\bibnamefont
  {Curiotto}}, \bibinfo {author} {\bibfnamefont {F.}~\bibnamefont {Leroy}},
  \bibinfo {author} {\bibfnamefont {P.}~\bibnamefont {M\"{u}ller}}, \bibinfo
  {author} {\bibfnamefont {F.}~\bibnamefont {Cheynis}}, \bibinfo {author}
  {\bibfnamefont {M.}~\bibnamefont {Michailov}}, \bibinfo {author}
  {\bibfnamefont {A.}~\bibnamefont {El-Barraj}}, \ and\ \bibinfo {author}
  {\bibfnamefont {B.}~\bibnamefont {Ranguelov}},\ }\href
  {https://doi.org/10.1016/j.jcrysgro.2019.05.016} {\bibfield  {journal}
  {\bibinfo  {journal} {J. Cryst. Growth}\ }\textbf {\bibinfo {volume} {520}},\
  \bibinfo {pages} {42} (\bibinfo {year} {2019})}\BibitemShut {NoStop}%
\bibitem [{\citenamefont {Khare}\ \emph {et~al.}(1995)\citenamefont {Khare},
  \citenamefont {Bartelt},\ and\ \citenamefont {Einstein}}]{Khare1995}%
  \BibitemOpen
  \bibfield  {author} {\bibinfo {author} {\bibfnamefont {S.~V.}\ \bibnamefont
  {Khare}}, \bibinfo {author} {\bibfnamefont {N.~C.}\ \bibnamefont {Bartelt}},
  \ and\ \bibinfo {author} {\bibfnamefont {T.~L.}\ \bibnamefont {Einstein}},\
  }\href {https://link.aps.org/doi/10.1103/PhysRevLett.75.2148} {\bibfield
  {journal} {\bibinfo  {journal} {Phys. Rev. Lett.}\ }\textbf {\bibinfo
  {volume} {75}},\ \bibinfo {pages} {2148} (\bibinfo {year}
  {1995})}\BibitemShut {NoStop}%
\bibitem [{\citenamefont {Lai}\ \emph {et~al.}(2017)\citenamefont {Lai},
  \citenamefont {Liu},\ and\ \citenamefont {Evans}}]{Lai2017}%
  \BibitemOpen
  \bibfield  {author} {\bibinfo {author} {\bibfnamefont {K.~C.}\ \bibnamefont
  {Lai}}, \bibinfo {author} {\bibfnamefont {D.-J.}\ \bibnamefont {Liu}}, \ and\
  \bibinfo {author} {\bibfnamefont {J.~W.}\ \bibnamefont {Evans}},\ }\href
  {https://link.aps.org/doi/10.1103/PhysRevB.96.235406} {\bibfield  {journal}
  {\bibinfo  {journal} {Phys. Rev. B}\ }\textbf {\bibinfo {volume} {96}},\
  \bibinfo {pages} {235406} (\bibinfo {year} {2017})}\BibitemShut {NoStop}%
\bibitem [{\citenamefont {Ju{\'{a}}rez}\ and\ \citenamefont
  {Bevan}(2012)}]{Juarez2012}%
  \BibitemOpen
  \bibfield  {author} {\bibinfo {author} {\bibfnamefont {J.~J.}\ \bibnamefont
  {Ju{\'{a}}rez}}\ and\ \bibinfo {author} {\bibfnamefont {M.~A.}\ \bibnamefont
  {Bevan}},\ }\href {https://doi.org/10.1002/adfm.201200400} {\bibfield
  {journal} {\bibinfo  {journal} {Adv. Funct. Mater.}\ }\textbf {\bibinfo
  {volume} {22}},\ \bibinfo {pages} {3833} (\bibinfo {year}
  {2012})}\BibitemShut {NoStop}%
\bibitem [{\citenamefont {Ju{\'{a}}rez}\ \emph {et~al.}(2012)\citenamefont
  {Ju{\'{a}}rez}, \citenamefont {Mathai}, \citenamefont {Liddle},\ and\
  \citenamefont {Bevan}}]{Juarez2012a}%
  \BibitemOpen
  \bibfield  {author} {\bibinfo {author} {\bibfnamefont {J.~J.}\ \bibnamefont
  {Ju{\'{a}}rez}}, \bibinfo {author} {\bibfnamefont {P.~P.}\ \bibnamefont
  {Mathai}}, \bibinfo {author} {\bibfnamefont {J.~A.}\ \bibnamefont {Liddle}},
  \ and\ \bibinfo {author} {\bibfnamefont {M.~A.}\ \bibnamefont {Bevan}},\
  }\href {https://doi.org/10.1039/c2lc40692f} {\bibfield  {journal} {\bibinfo
  {journal} {Lab Chip}\ }\textbf {\bibinfo {volume} {12}},\ \bibinfo {pages}
  {4063} (\bibinfo {year} {2012})}\BibitemShut {NoStop}%
\bibitem [{\citenamefont {Xue}\ \emph {et~al.}(2014)\citenamefont {Xue},
  \citenamefont {Beltran-Villegas}, \citenamefont {Tang}, \citenamefont
  {Bevan},\ and\ \citenamefont {Grover}}]{Xue2014}%
  \BibitemOpen
  \bibfield  {author} {\bibinfo {author} {\bibfnamefont {Y.}~\bibnamefont
  {Xue}}, \bibinfo {author} {\bibfnamefont {D.~J.}\ \bibnamefont
  {Beltran-Villegas}}, \bibinfo {author} {\bibfnamefont {X.}~\bibnamefont
  {Tang}}, \bibinfo {author} {\bibfnamefont {M.~A.}\ \bibnamefont {Bevan}}, \
  and\ \bibinfo {author} {\bibfnamefont {M.~A.}\ \bibnamefont {Grover}},\
  }\href {https://doi.org/10.1109/tcst.2013.2296700} {\bibfield  {journal}
  {\bibinfo  {journal} {IEEE Trans. Control Syst. Technol.}\ }\textbf {\bibinfo
  {volume} {22}},\ \bibinfo {pages} {1956} (\bibinfo {year}
  {2014})}\BibitemShut {NoStop}%
\bibitem [{\citenamefont {Sutton}\ and\ \citenamefont
  {Barto}(2018)}]{Sutton1998}%
  \BibitemOpen
  \bibfield  {author} {\bibinfo {author} {\bibfnamefont {R.~S.}\ \bibnamefont
  {Sutton}}\ and\ \bibinfo {author} {\bibfnamefont {A.~G.}\ \bibnamefont
  {Barto}},\ }\href {http://incompleteideas.net/book/the-book.html} {\emph
  {\bibinfo {title} {Reinforcement Learning: An Introduction}}},\ \bibinfo
  {edition} {2nd}\ ed.\ (\bibinfo  {publisher} {The MIT Press},\ \bibinfo
  {year} {2018})\BibitemShut {NoStop}%
\bibitem [{\citenamefont {Bellman}(2003)}]{Bellman1957}%
  \BibitemOpen
  \bibfield  {author} {\bibinfo {author} {\bibfnamefont {R.~E.}\ \bibnamefont
  {Bellman}},\ }\href@noop {} {\emph {\bibinfo {title} {Dynamic Programming}}}\
  (\bibinfo  {publisher} {Dover Publications, Inc.},\ \bibinfo {address}
  {USA},\ \bibinfo {year} {2003})\BibitemShut {NoStop}%
\bibitem [{\citenamefont {Giesen}(2001)}]{Giesen2001}%
  \BibitemOpen
  \bibfield  {author} {\bibinfo {author} {\bibfnamefont {M.}~\bibnamefont
  {Giesen}},\ }\href {https://doi.org/10.1016/s0079-6816(00)00021-6} {\bibfield
   {journal} {\bibinfo  {journal} {Prog. Surf. Sci.}\ }\textbf {\bibinfo
  {volume} {68}},\ \bibinfo {pages} {1} (\bibinfo {year} {2001})}\BibitemShut
  {NoStop}%
\bibitem [{\citenamefont {Tao}\ \emph {et~al.}(2010)\citenamefont {Tao},
  \citenamefont {Cullen},\ and\ \citenamefont {Williams}}]{Tao2010}%
  \BibitemOpen
  \bibfield  {author} {\bibinfo {author} {\bibfnamefont {C.}~\bibnamefont
  {Tao}}, \bibinfo {author} {\bibfnamefont {W.~G.}\ \bibnamefont {Cullen}}, \
  and\ \bibinfo {author} {\bibfnamefont {E.~D.}\ \bibnamefont {Williams}},\
  }\href {https://doi.org/10.1126/science.1186648} {\bibfield  {journal}
  {\bibinfo  {journal} {Science}\ }\textbf {\bibinfo {volume} {328}},\ \bibinfo
  {pages} {736} (\bibinfo {year} {2010})}\BibitemShut {NoStop}%
\bibitem [{\citenamefont {Hubartt}\ and\ \citenamefont
  {Amar}(2015)}]{Hubartt2015}%
  \BibitemOpen
  \bibfield  {author} {\bibinfo {author} {\bibfnamefont {B.~C.}\ \bibnamefont
  {Hubartt}}\ and\ \bibinfo {author} {\bibfnamefont {J.~G.}\ \bibnamefont
  {Amar}},\ }\href@noop {} {\bibfield  {journal} {\bibinfo  {journal} {J. Chem.
  Phys.}\ }\textbf {\bibinfo {volume} {142}},\ \bibinfo {pages} {024709}
  (\bibinfo {year} {2015})}\BibitemShut {NoStop}%
\bibitem [{\citenamefont {Plass}\ \emph {et~al.}(2001)\citenamefont {Plass},
  \citenamefont {Last}, \citenamefont {Bartelt},\ and\ \citenamefont
  {Kellogg}}]{Plass2001}%
  \BibitemOpen
  \bibfield  {author} {\bibinfo {author} {\bibfnamefont {R.}~\bibnamefont
  {Plass}}, \bibinfo {author} {\bibfnamefont {J.~A.}\ \bibnamefont {Last}},
  \bibinfo {author} {\bibfnamefont {N.}~\bibnamefont {Bartelt}}, \ and\
  \bibinfo {author} {\bibfnamefont {G.}~\bibnamefont {Kellogg}},\ }\href@noop
  {} {\bibfield  {journal} {\bibinfo  {journal} {Nature}\ }\textbf {\bibinfo
  {volume} {412}},\ \bibinfo {pages} {875} (\bibinfo {year}
  {2001})}\BibitemShut {NoStop}%
\bibitem [{\citenamefont {Heinonen}\ \emph {et~al.}(1999)\citenamefont
  {Heinonen}, \citenamefont {Koponen}, \citenamefont {Merikoski},\ and\
  \citenamefont {Ala-Nissila}}]{Heinonen1999}%
  \BibitemOpen
  \bibfield  {author} {\bibinfo {author} {\bibfnamefont {J.}~\bibnamefont
  {Heinonen}}, \bibinfo {author} {\bibfnamefont {I.}~\bibnamefont {Koponen}},
  \bibinfo {author} {\bibfnamefont {J.}~\bibnamefont {Merikoski}}, \ and\
  \bibinfo {author} {\bibfnamefont {T.}~\bibnamefont {Ala-Nissila}},\ }\href
  {https://link.aps.org/doi/10.1103/PhysRevLett.82.2733} {\bibfield  {journal}
  {\bibinfo  {journal} {Phys. Rev. Lett.}\ }\textbf {\bibinfo {volume} {82}},\
  \bibinfo {pages} {2733} (\bibinfo {year} {1999})}\BibitemShut {NoStop}%
\bibitem [{\citenamefont {Leroy}\ \emph {et~al.}(2020)\citenamefont {Leroy},
  \citenamefont {El~Barraj}, \citenamefont {Cheynis}, \citenamefont
  {M\"uller},\ and\ \citenamefont {Curiotto}}]{Leroy2020}%
  \BibitemOpen
  \bibfield  {author} {\bibinfo {author} {\bibfnamefont {F.}~\bibnamefont
  {Leroy}}, \bibinfo {author} {\bibfnamefont {A.}~\bibnamefont {El~Barraj}},
  \bibinfo {author} {\bibfnamefont {F.}~\bibnamefont {Cheynis}}, \bibinfo
  {author} {\bibfnamefont {P.}~\bibnamefont {M\"uller}}, \ and\ \bibinfo
  {author} {\bibfnamefont {S.}~\bibnamefont {Curiotto}},\ }\href
  {https://link.aps.org/doi/10.1103/PhysRevB.102.235412} {\bibfield  {journal}
  {\bibinfo  {journal} {Phys. Rev. B}\ }\textbf {\bibinfo {volume} {102}},\
  \bibinfo {pages} {235412} (\bibinfo {year} {2020})}\BibitemShut {NoStop}%
\bibitem [{\citenamefont {VanSaders}\ and\ \citenamefont
  {Glotzer}(2021)}]{VanSaders2021}%
  \BibitemOpen
  \bibfield  {author} {\bibinfo {author} {\bibfnamefont {B.}~\bibnamefont
  {VanSaders}}\ and\ \bibinfo {author} {\bibfnamefont {S.~C.}\ \bibnamefont
  {Glotzer}},\ }\href {https://doi.org/10.1073/pnas.2017377118} {\bibfield
  {journal} {\bibinfo  {journal} {Proc. Natl. Acad. Sci. U.S.A}\ }\textbf
  {\bibinfo {volume} {118}},\ \bibinfo {pages} {e2017377118} (\bibinfo {year}
  {2021})}\BibitemShut {NoStop}%
\bibitem [{\citenamefont {Huang}\ \emph {et~al.}(2018)\citenamefont {Huang},
  \citenamefont {Wen}, \citenamefont {Voter},\ and\ \citenamefont
  {Perez}}]{Huang2018}%
  \BibitemOpen
  \bibfield  {author} {\bibinfo {author} {\bibfnamefont {R.}~\bibnamefont
  {Huang}}, \bibinfo {author} {\bibfnamefont {Y.}~\bibnamefont {Wen}}, \bibinfo
  {author} {\bibfnamefont {A.~F.}\ \bibnamefont {Voter}}, \ and\ \bibinfo
  {author} {\bibfnamefont {D.}~\bibnamefont {Perez}},\ }\href
  {https://link.aps.org/doi/10.1103/PhysRevMaterials.2.126002} {\bibfield
  {journal} {\bibinfo  {journal} {Phys. Rev. Materials}\ }\textbf {\bibinfo
  {volume} {2}},\ \bibinfo {pages} {126002} (\bibinfo {year}
  {2018})}\BibitemShut {NoStop}%
\bibitem [{\citenamefont {Trushin}\ \emph {et~al.}(2001)\citenamefont
  {Trushin}, \citenamefont {Salo}, \citenamefont {Alatalo},\ and\ \citenamefont
  {Ala-Nissila}}]{Trushin2001}%
  \BibitemOpen
  \bibfield  {author} {\bibinfo {author} {\bibfnamefont {O.}~\bibnamefont
  {Trushin}}, \bibinfo {author} {\bibfnamefont {P.}~\bibnamefont {Salo}},
  \bibinfo {author} {\bibfnamefont {M.}~\bibnamefont {Alatalo}}, \ and\
  \bibinfo {author} {\bibfnamefont {T.}~\bibnamefont {Ala-Nissila}},\ }\href
  {https://doi.org/10.1016/s0039-6028(00)01013-x} {\bibfield  {journal}
  {\bibinfo  {journal} {Surf. Sci.}\ }\textbf {\bibinfo {volume} {482-485}},\
  \bibinfo {pages} {365} (\bibinfo {year} {2001})}\BibitemShut {NoStop}%
\bibitem [{\citenamefont {Glasstone}\ \emph {et~al.}(1941)\citenamefont
  {Glasstone}, \citenamefont {Laidler},\ and\ \citenamefont
  {Eyring}}]{Glasstone1941}%
  \BibitemOpen
  \bibfield  {author} {\bibinfo {author} {\bibfnamefont {S.}~\bibnamefont
  {Glasstone}}, \bibinfo {author} {\bibfnamefont {K.~J.}\ \bibnamefont
  {Laidler}}, \ and\ \bibinfo {author} {\bibfnamefont {H.}~\bibnamefont
  {Eyring}},\ }\href@noop {} {\emph {\bibinfo {title} {The Theory of Rate
  Processes: The Kinetics of Chemical Reactions, Viscosity, Diffusion and
  Electrochemical Phenomena}}},\ International chemical series\ (\bibinfo
  {publisher} {McGraw-Hill Book Company, Inc.},\ \bibinfo {year}
  {1941})\BibitemShut {NoStop}%
\bibitem [{\citenamefont {Liu}\ and\ \citenamefont {Weeks}(1998)}]{Liu1998}%
  \BibitemOpen
  \bibfield  {author} {\bibinfo {author} {\bibfnamefont {D.-J.}\ \bibnamefont
  {Liu}}\ and\ \bibinfo {author} {\bibfnamefont {J.~D.}\ \bibnamefont
  {Weeks}},\ }\href {https://link.aps.org/doi/10.1103/PhysRevB.57.14891}
  {\bibfield  {journal} {\bibinfo  {journal} {Phys. Rev. B}\ }\textbf {\bibinfo
  {volume} {57}},\ \bibinfo {pages} {14891} (\bibinfo {year}
  {1998})}\BibitemShut {NoStop}%
\bibitem [{\citenamefont {Sanchez}\ and\ \citenamefont
  {Evans}(1999)}]{Sanchez1999}%
  \BibitemOpen
  \bibfield  {author} {\bibinfo {author} {\bibfnamefont {J.~R.}\ \bibnamefont
  {Sanchez}}\ and\ \bibinfo {author} {\bibfnamefont {J.~W.}\ \bibnamefont
  {Evans}},\ }\href {https://doi.org/10.1103/physrevb.59.3224} {\bibfield
  {journal} {\bibinfo  {journal} {Phys. Rev. B}\ }\textbf {\bibinfo {volume}
  {59}},\ \bibinfo {pages} {3224} (\bibinfo {year} {1999})}\BibitemShut
  {NoStop}%
\bibitem [{\citenamefont {Combe}\ and\ \citenamefont
  {Larralde}(2000)}]{Combe2000}%
  \BibitemOpen
  \bibfield  {author} {\bibinfo {author} {\bibfnamefont {N.}~\bibnamefont
  {Combe}}\ and\ \bibinfo {author} {\bibfnamefont {H.}~\bibnamefont
  {Larralde}},\ }\href {https://link.aps.org/doi/10.1103/PhysRevB.62.16074}
  {\bibfield  {journal} {\bibinfo  {journal} {Phys. Rev. B}\ }\textbf {\bibinfo
  {volume} {62}},\ \bibinfo {pages} {16074} (\bibinfo {year}
  {2000})}\BibitemShut {NoStop}%
\bibitem [{\citenamefont {Van~Kampen}(1992)}]{VanKampen1992}%
  \BibitemOpen
  \bibfield  {author} {\bibinfo {author} {\bibfnamefont {N.~G.}\ \bibnamefont
  {Van~Kampen}},\ }\href@noop {} {\emph {\bibinfo {title} {Stochastic processes
  in physics and chemistry}}}\ (\bibinfo  {publisher} {Elsevier},\ \bibinfo
  {year} {1992})\BibitemShut {NoStop}%
\bibitem [{Note1()}]{Note1}%
  \BibitemOpen
  \bibinfo {note} {As a technical remark, this definition requires to extend
  the policy and define a force on the target state itself. However, due to the
  Markovian character of the dynamics, this does not affect the mean first
  passage time to target and the optimal policy in the other states outside the
  target.}\BibitemShut {Stop}%
\bibitem [{\citenamefont {Dougherty}\ \emph {et~al.}(2002)\citenamefont
  {Dougherty}, \citenamefont {Lyubinetsky}, \citenamefont {Williams},
  \citenamefont {Constantin}, \citenamefont {Dasgupta},\ and\ \citenamefont
  {Sarma}}]{Dougherty2002}%
  \BibitemOpen
  \bibfield  {author} {\bibinfo {author} {\bibfnamefont {D.~B.}\ \bibnamefont
  {Dougherty}}, \bibinfo {author} {\bibfnamefont {I.}~\bibnamefont
  {Lyubinetsky}}, \bibinfo {author} {\bibfnamefont {E.~D.}\ \bibnamefont
  {Williams}}, \bibinfo {author} {\bibfnamefont {M.}~\bibnamefont
  {Constantin}}, \bibinfo {author} {\bibfnamefont {C.}~\bibnamefont
  {Dasgupta}}, \ and\ \bibinfo {author} {\bibfnamefont {S.~D.}\ \bibnamefont
  {Sarma}},\ }\href {https://link.aps.org/doi/10.1103/PhysRevLett.89.136102}
  {\bibfield  {journal} {\bibinfo  {journal} {Phys. Rev. Lett.}\ }\textbf
  {\bibinfo {volume} {89}},\ \bibinfo {pages} {136102} (\bibinfo {year}
  {2002})}\BibitemShut {NoStop}%
\bibitem [{\citenamefont {Constantin}\ \emph {et~al.}(2003)\citenamefont
  {Constantin}, \citenamefont {Das~Sarma}, \citenamefont {Dasgupta},
  \citenamefont {Bondarchuk}, \citenamefont {Dougherty},\ and\ \citenamefont
  {Williams}}]{Constantin2003}%
  \BibitemOpen
  \bibfield  {author} {\bibinfo {author} {\bibfnamefont {M.}~\bibnamefont
  {Constantin}}, \bibinfo {author} {\bibfnamefont {S.}~\bibnamefont
  {Das~Sarma}}, \bibinfo {author} {\bibfnamefont {C.}~\bibnamefont {Dasgupta}},
  \bibinfo {author} {\bibfnamefont {O.}~\bibnamefont {Bondarchuk}}, \bibinfo
  {author} {\bibfnamefont {D.~B.}\ \bibnamefont {Dougherty}}, \ and\ \bibinfo
  {author} {\bibfnamefont {E.~D.}\ \bibnamefont {Williams}},\ }\href
  {https://link.aps.org/doi/10.1103/PhysRevLett.91.086103} {\bibfield
  {journal} {\bibinfo  {journal} {Phys. Rev. Lett.}\ }\textbf {\bibinfo
  {volume} {91}},\ \bibinfo {pages} {086103} (\bibinfo {year}
  {2003})}\BibitemShut {NoStop}%
\bibitem [{\citenamefont {Guttmann}(2009)}]{Guttmann2009}%
  \BibitemOpen
  \bibfield  {author} {\bibinfo {author} {\bibfnamefont {A.~J.}\ \bibnamefont
  {Guttmann}},\ }\href@noop {} {\emph {\bibinfo {title} {Polygons, Polyominoes
  and Polycubes}}},\ Lecture Notes in Physics 775\ (\bibinfo  {publisher}
  {Springer},\ \bibinfo {year} {2009})\BibitemShut {NoStop}%
\bibitem [{\citenamefont {Jensen}\ and\ \citenamefont
  {Guttmann}(2000)}]{Jensen2000}%
  \BibitemOpen
  \bibfield  {author} {\bibinfo {author} {\bibfnamefont {I.}~\bibnamefont
  {Jensen}}\ and\ \bibinfo {author} {\bibfnamefont {A.~J.}\ \bibnamefont
  {Guttmann}},\ }\href {https://doi.org/10.1088/0305-4470/33/29/102} {\bibfield
   {journal} {\bibinfo  {journal} {J. Phys. A Math. Theor.}\ }\textbf {\bibinfo
  {volume} {33}},\ \bibinfo {pages} {L257} (\bibinfo {year}
  {2000})}\BibitemShut {NoStop}%
\bibitem [{\citenamefont {Lov{\'a}sz}(1993)}]{Lovasz1993}%
  \BibitemOpen
  \bibfield  {author} {\bibinfo {author} {\bibfnamefont {L.}~\bibnamefont
  {Lov{\'a}sz}},\ }\href {https://web.cs.elte.hu/~lovasz/erdos.pdf} {\bibfield
  {journal} {\bibinfo  {journal} {Combinatorics, Paul Erd\H{o}s is eighty}\
  }\textbf {\bibinfo {volume} {2}},\ \bibinfo {pages} {1} (\bibinfo {year}
  {1993})}\BibitemShut {NoStop}%
\bibitem [{\citenamefont {Lin}\ \emph {et~al.}(2012)\citenamefont {Lin},
  \citenamefont {Julaiti},\ and\ \citenamefont {Zhang}}]{Lin2012}%
  \BibitemOpen
  \bibfield  {author} {\bibinfo {author} {\bibfnamefont {Y.}~\bibnamefont
  {Lin}}, \bibinfo {author} {\bibfnamefont {A.}~\bibnamefont {Julaiti}}, \ and\
  \bibinfo {author} {\bibfnamefont {Z.}~\bibnamefont {Zhang}},\ }\href
  {https://doi.org/10.1063/1.4754735} {\bibfield  {journal} {\bibinfo
  {journal} {J. Chem. Phys.}\ }\textbf {\bibinfo {volume} {137}},\ \bibinfo
  {pages} {124104} (\bibinfo {year} {2012})}\BibitemShut {NoStop}%
\bibitem [{\citenamefont {Tejedor}\ \emph {et~al.}(2009)\citenamefont
  {Tejedor}, \citenamefont {B\'enichou},\ and\ \citenamefont
  {Voituriez}}]{Tejedor2009}%
  \BibitemOpen
  \bibfield  {author} {\bibinfo {author} {\bibfnamefont {V.}~\bibnamefont
  {Tejedor}}, \bibinfo {author} {\bibfnamefont {O.}~\bibnamefont {B\'enichou}},
  \ and\ \bibinfo {author} {\bibfnamefont {R.}~\bibnamefont {Voituriez}},\
  }\href {https://link.aps.org/doi/10.1103/PhysRevE.80.065104} {\bibfield
  {journal} {\bibinfo  {journal} {Phys. Rev. E}\ }\textbf {\bibinfo {volume}
  {80}},\ \bibinfo {pages} {065104} (\bibinfo {year} {2009})}\BibitemShut
  {NoStop}%
\bibitem [{\citenamefont {Baronchelli}\ and\ \citenamefont
  {Loreto}(2006)}]{Baronchelli2006}%
  \BibitemOpen
  \bibfield  {author} {\bibinfo {author} {\bibfnamefont {A.}~\bibnamefont
  {Baronchelli}}\ and\ \bibinfo {author} {\bibfnamefont {V.}~\bibnamefont
  {Loreto}},\ }\href {https://link.aps.org/doi/10.1103/PhysRevE.73.026103}
  {\bibfield  {journal} {\bibinfo  {journal} {Phys. Rev. E}\ }\textbf {\bibinfo
  {volume} {73}},\ \bibinfo {pages} {026103} (\bibinfo {year}
  {2006})}\BibitemShut {NoStop}%
\bibitem [{\citenamefont {Noh}\ and\ \citenamefont {Rieger}(2004)}]{Noh2004}%
  \BibitemOpen
  \bibfield  {author} {\bibinfo {author} {\bibfnamefont {J.~D.}\ \bibnamefont
  {Noh}}\ and\ \bibinfo {author} {\bibfnamefont {H.}~\bibnamefont {Rieger}},\
  }\href {https://link.aps.org/doi/10.1103/PhysRevLett.92.118701} {\bibfield
  {journal} {\bibinfo  {journal} {Phys. Rev. Lett.}\ }\textbf {\bibinfo
  {volume} {92}},\ \bibinfo {pages} {118701} (\bibinfo {year}
  {2004})}\BibitemShut {NoStop}%
\bibitem [{\citenamefont {Ferrando}\ and\ \citenamefont
  {Tr{\'{e}}glia}(1994)}]{Ferrando1994}%
  \BibitemOpen
  \bibfield  {author} {\bibinfo {author} {\bibfnamefont {R.}~\bibnamefont
  {Ferrando}}\ and\ \bibinfo {author} {\bibfnamefont {G.}~\bibnamefont
  {Tr{\'{e}}glia}},\ }\href {https://doi.org/10.1103/physrevb.50.12104}
  {\bibfield  {journal} {\bibinfo  {journal} {Phys. Rev. B}\ }\textbf {\bibinfo
  {volume} {50}},\ \bibinfo {pages} {12104} (\bibinfo {year}
  {1994})}\BibitemShut {NoStop}%
\bibitem [{\citenamefont {Yu}\ and\ \citenamefont {Scheffler}(1997)}]{Yu1997}%
  \BibitemOpen
  \bibfield  {author} {\bibinfo {author} {\bibfnamefont {B.~D.}\ \bibnamefont
  {Yu}}\ and\ \bibinfo {author} {\bibfnamefont {M.}~\bibnamefont {Scheffler}},\
  }\href {https://doi.org/10.1103/physrevb.55.13916} {\bibfield  {journal}
  {\bibinfo  {journal} {Phys. Rev. B}\ }\textbf {\bibinfo {volume} {55}},\
  \bibinfo {pages} {13916} (\bibinfo {year} {1997})}\BibitemShut {NoStop}%
\bibitem [{\citenamefont {Mehl}\ \emph {et~al.}(1999)\citenamefont {Mehl},
  \citenamefont {Biham}, \citenamefont {Furman},\ and\ \citenamefont
  {Karimi}}]{Mehl1999}%
  \BibitemOpen
  \bibfield  {author} {\bibinfo {author} {\bibfnamefont {H.}~\bibnamefont
  {Mehl}}, \bibinfo {author} {\bibfnamefont {O.}~\bibnamefont {Biham}},
  \bibinfo {author} {\bibfnamefont {I.}~\bibnamefont {Furman}}, \ and\ \bibinfo
  {author} {\bibfnamefont {M.}~\bibnamefont {Karimi}},\ }\href
  {https://doi.org/10.1103/physrevb.60.2106} {\bibfield  {journal} {\bibinfo
  {journal} {Phys. Rev. B}\ }\textbf {\bibinfo {volume} {60}},\ \bibinfo
  {pages} {2106} (\bibinfo {year} {1999})}\BibitemShut {NoStop}%
\bibitem [{\citenamefont {Nelson}\ \emph {et~al.}(1993)\citenamefont {Nelson},
  \citenamefont {Einstein}, \citenamefont {Khare},\ and\ \citenamefont
  {Rous}}]{Nelson1993}%
  \BibitemOpen
  \bibfield  {author} {\bibinfo {author} {\bibfnamefont {R.}~\bibnamefont
  {Nelson}}, \bibinfo {author} {\bibfnamefont {T.}~\bibnamefont {Einstein}},
  \bibinfo {author} {\bibfnamefont {S.}~\bibnamefont {Khare}}, \ and\ \bibinfo
  {author} {\bibfnamefont {P.}~\bibnamefont {Rous}},\ }\href
  {https://doi.org/10.1016/0039-6028(93)90293-s} {\bibfield  {journal}
  {\bibinfo  {journal} {Surf. Sci.}\ }\textbf {\bibinfo {volume} {295}},\
  \bibinfo {pages} {462} (\bibinfo {year} {1993})}\BibitemShut {NoStop}%
\bibitem [{\citenamefont {Nozawa}\ \emph {et~al.}(2018)\citenamefont {Nozawa},
  \citenamefont {Uda}, \citenamefont {Guo}, \citenamefont {Toyotama},
  \citenamefont {Yamanaka}, \citenamefont {Ihara},\ and\ \citenamefont
  {Okada}}]{Nozawa2018}%
  \BibitemOpen
  \bibfield  {author} {\bibinfo {author} {\bibfnamefont {J.}~\bibnamefont
  {Nozawa}}, \bibinfo {author} {\bibfnamefont {S.}~\bibnamefont {Uda}},
  \bibinfo {author} {\bibfnamefont {S.}~\bibnamefont {Guo}}, \bibinfo {author}
  {\bibfnamefont {A.}~\bibnamefont {Toyotama}}, \bibinfo {author}
  {\bibfnamefont {J.}~\bibnamefont {Yamanaka}}, \bibinfo {author}
  {\bibfnamefont {N.}~\bibnamefont {Ihara}}, \ and\ \bibinfo {author}
  {\bibfnamefont {J.}~\bibnamefont {Okada}},\ }\href
  {https://doi.org/10.1021/acs.cgd.8b00942} {\bibfield  {journal} {\bibinfo
  {journal} {Cryst. Growth Des.}\ }\textbf {\bibinfo {volume} {18}},\ \bibinfo
  {pages} {6078} (\bibinfo {year} {2018})}\BibitemShut {NoStop}%
\bibitem [{\citenamefont {Helden}\ \emph {et~al.}(2015)\citenamefont {Helden},
  \citenamefont {Eichhorn},\ and\ \citenamefont {Bechinger}}]{Helden2015}%
  \BibitemOpen
  \bibfield  {author} {\bibinfo {author} {\bibfnamefont {L.}~\bibnamefont
  {Helden}}, \bibinfo {author} {\bibfnamefont {R.}~\bibnamefont {Eichhorn}}, \
  and\ \bibinfo {author} {\bibfnamefont {C.}~\bibnamefont {Bechinger}},\ }\href
  {https://doi.org/10.1039/c4sm02833c} {\bibfield  {journal} {\bibinfo
  {journal} {Soft Matter}\ }\textbf {\bibinfo {volume} {11}},\ \bibinfo {pages}
  {2379} (\bibinfo {year} {2015})}\BibitemShut {NoStop}%
\bibitem [{\citenamefont {Braibanti}\ \emph {et~al.}(2008)\citenamefont
  {Braibanti}, \citenamefont {Vigolo},\ and\ \citenamefont
  {Piazza}}]{Braibanti2008}%
  \BibitemOpen
  \bibfield  {author} {\bibinfo {author} {\bibfnamefont {M.}~\bibnamefont
  {Braibanti}}, \bibinfo {author} {\bibfnamefont {D.}~\bibnamefont {Vigolo}}, \
  and\ \bibinfo {author} {\bibfnamefont {R.}~\bibnamefont {Piazza}},\ }\href
  {https://doi.org/10.1103/physrevlett.100.108303} {\bibfield  {journal}
  {\bibinfo  {journal} {Phys. Rev. Lett.}\ }\textbf {\bibinfo {volume} {100}}
  (\bibinfo {year} {2008})}\BibitemShut {NoStop}%
\bibitem [{\citenamefont {W\"{u}rger}(2010)}]{Wrger2010}%
  \BibitemOpen
  \bibfield  {author} {\bibinfo {author} {\bibfnamefont {A.}~\bibnamefont
  {W\"{u}rger}},\ }\href {https://doi.org/10.1088/0034-4885/73/12/126601}
  {\bibfield  {journal} {\bibinfo  {journal} {Rep. Prog. Phys.}\ }\textbf
  {\bibinfo {volume} {73}},\ \bibinfo {pages} {126601} (\bibinfo {year}
  {2010})}\BibitemShut {NoStop}%
\bibitem [{\citenamefont {Ganapathy}\ \emph {et~al.}(2010)\citenamefont
  {Ganapathy}, \citenamefont {Buckley}, \citenamefont {Gerbode},\ and\
  \citenamefont {Cohen}}]{Ganapathy2010}%
  \BibitemOpen
  \bibfield  {author} {\bibinfo {author} {\bibfnamefont {R.}~\bibnamefont
  {Ganapathy}}, \bibinfo {author} {\bibfnamefont {M.~R.}\ \bibnamefont
  {Buckley}}, \bibinfo {author} {\bibfnamefont {S.~J.}\ \bibnamefont
  {Gerbode}}, \ and\ \bibinfo {author} {\bibfnamefont {I.}~\bibnamefont
  {Cohen}},\ }\href {https://doi.org/10.1126/science.1179947} {\bibfield
  {journal} {\bibinfo  {journal} {Science}\ }\textbf {\bibinfo {volume}
  {327}},\ \bibinfo {pages} {445} (\bibinfo {year} {2010})}\BibitemShut
  {NoStop}%
\bibitem [{\citenamefont {Garc\'ia}\ \emph {et~al.}(2007)\citenamefont
  {Garc\'ia}, \citenamefont {Sapienza}, \citenamefont {Blanco},\ and\
  \citenamefont {L\'opez}}]{Garcia2007}%
  \BibitemOpen
  \bibfield  {author} {\bibinfo {author} {\bibfnamefont {P.}~\bibnamefont
  {Garc\'ia}}, \bibinfo {author} {\bibfnamefont {R.}~\bibnamefont {Sapienza}},
  \bibinfo {author} {\bibfnamefont {A.}~\bibnamefont {Blanco}}, \ and\ \bibinfo
  {author} {\bibfnamefont {C.}~\bibnamefont {L\'opez}},\ }\href
  {https://onlinelibrary.wiley.com/doi/abs/10.1002/adma.200602426} {\bibfield
  {journal} {\bibinfo  {journal} {Adv. Mater.}\ }\textbf {\bibinfo {volume}
  {19}},\ \bibinfo {pages} {2597} (\bibinfo {year} {2007})}\BibitemShut
  {NoStop}%
\bibitem [{\citenamefont {Pariente}\ \emph {et~al.}(2020)\citenamefont
  {Pariente}, \citenamefont {Caselli}, \citenamefont {Pecharromán},
  \citenamefont {Blanco},\ and\ \citenamefont {López}}]{Pariente2020}%
  \BibitemOpen
  \bibfield  {author} {\bibinfo {author} {\bibfnamefont {J.~A.}\ \bibnamefont
  {Pariente}}, \bibinfo {author} {\bibfnamefont {N.}~\bibnamefont {Caselli}},
  \bibinfo {author} {\bibfnamefont {C.}~\bibnamefont {Pecharromán}}, \bibinfo
  {author} {\bibfnamefont {A.}~\bibnamefont {Blanco}}, \ and\ \bibinfo {author}
  {\bibfnamefont {C.}~\bibnamefont {López}},\ }\href
  {https://onlinelibrary.wiley.com/doi/abs/10.1002/smll.202002735} {\bibfield
  {journal} {\bibinfo  {journal} {Small}\ }\textbf {\bibinfo {volume} {16}},\
  \bibinfo {pages} {2002735} (\bibinfo {year} {2020})}\BibitemShut {NoStop}%
\end{thebibliography}%


\begin{thebibliography}{5}%
\makeatletter
\providecommand \@ifxundefined [1]{%
 \@ifx{#1\undefined}
}%
\providecommand \@ifnum [1]{%
 \ifnum #1\expandafter \@firstoftwo
 \else \expandafter \@secondoftwo
 \fi
}%
\providecommand \@ifx [1]{%
 \ifx #1\expandafter \@firstoftwo
 \else \expandafter \@secondoftwo
 \fi
}%
\providecommand \natexlab [1]{#1}%
\providecommand \enquote  [1]{``#1''}%
\providecommand \bibnamefont  [1]{#1}%
\providecommand \bibfnamefont [1]{#1}%
\providecommand \citenamefont [1]{#1}%
\providecommand \href@noop [0]{\@secondoftwo}%
\providecommand \href [0]{\begingroup \@sanitize@url \@href}%
\providecommand \@href[1]{\@@startlink{#1}\@@href}%
\providecommand \@@href[1]{\endgroup#1\@@endlink}%
\providecommand \@sanitize@url [0]{\catcode `\\12\catcode `\$12\catcode
  `\&12\catcode `\#12\catcode `\^12\catcode `\_12\catcode `\%12\relax}%
\providecommand \@@startlink[1]{}%
\providecommand \@@endlink[0]{}%
\providecommand \url  [0]{\begingroup\@sanitize@url \@url }%
\providecommand \@url [1]{\endgroup\@href {#1}{\urlprefix }}%
\providecommand \urlprefix  [0]{URL }%
\providecommand \Eprint [0]{\href }%
\providecommand \doibase [0]{http://dx.doi.org/}%
\providecommand \selectlanguage [0]{\@gobble}%
\providecommand \bibinfo  [0]{\@secondoftwo}%
\providecommand \bibfield  [0]{\@secondoftwo}%
\providecommand \translation [1]{[#1]}%
\providecommand \BibitemOpen [0]{}%
\providecommand \bibitemStop [0]{}%
\providecommand \bibitemNoStop [0]{.\EOS\space}%
\providecommand \EOS [0]{\spacefactor3000\relax}%
\providecommand \BibitemShut  [1]{\csname bibitem#1\endcsname}%
\let\auto@bib@innerbib\@empty
%</preamble>
\bibitem [{\citenamefont {Noh}\ and\ \citenamefont {Rieger}(2004)}]{Noh2004}%
  \BibitemOpen
  \bibfield  {author} {\bibinfo {author} {\bibfnamefont {J.~D.}\ \bibnamefont
  {Noh}}\ and\ \bibinfo {author} {\bibfnamefont {H.}~\bibnamefont {Rieger}},\
  }\href {https://link.aps.org/doi/10.1103/PhysRevLett.92.118701} {\bibfield
  {journal} {\bibinfo  {journal} {Phys. Rev. Lett.}\ }\textbf {\bibinfo
  {volume} {92}},\ \bibinfo {pages} {118701} (\bibinfo {year}
  {2004})}\BibitemShut {NoStop}%
\bibitem [{\citenamefont {Baronchelli}\ and\ \citenamefont
  {Loreto}(2006)}]{Baronchelli2006}%
  \BibitemOpen
  \bibfield  {author} {\bibinfo {author} {\bibfnamefont {A.}~\bibnamefont
  {Baronchelli}}\ and\ \bibinfo {author} {\bibfnamefont {V.}~\bibnamefont
  {Loreto}},\ }\href {https://link.aps.org/doi/10.1103/PhysRevE.73.026103}
  {\bibfield  {journal} {\bibinfo  {journal} {Phys. Rev. E}\ }\textbf {\bibinfo
  {volume} {73}},\ \bibinfo {pages} {026103} (\bibinfo {year}
  {2006})}\BibitemShut {NoStop}%
\bibitem [{\citenamefont {Lin}\ \emph {et~al.}(2012)\citenamefont {Lin},
  \citenamefont {Julaiti},\ and\ \citenamefont {Zhang}}]{Lin2012}%
  \BibitemOpen
  \bibfield  {author} {\bibinfo {author} {\bibfnamefont {Y.}~\bibnamefont
  {Lin}}, \bibinfo {author} {\bibfnamefont {A.}~\bibnamefont {Julaiti}}, \ and\
  \bibinfo {author} {\bibfnamefont {Z.}~\bibnamefont {Zhang}},\ }\href
  {https://doi.org/10.1063/1.4754735} {\bibfield  {journal} {\bibinfo
  {journal} {J. Chem. Phys.}\ }\textbf {\bibinfo {volume} {137}},\ \bibinfo
  {pages} {124104} (\bibinfo {year} {2012})}\BibitemShut {NoStop}%
\bibitem [{\citenamefont {Tejedor}\ \emph {et~al.}(2009)\citenamefont
  {Tejedor}, \citenamefont {B\'enichou},\ and\ \citenamefont
  {Voituriez}}]{Tejedor2009}%
  \BibitemOpen
  \bibfield  {author} {\bibinfo {author} {\bibfnamefont {V.}~\bibnamefont
  {Tejedor}}, \bibinfo {author} {\bibfnamefont {O.}~\bibnamefont {B\'enichou}},
  \ and\ \bibinfo {author} {\bibfnamefont {R.}~\bibnamefont {Voituriez}},\
  }\href {https://link.aps.org/doi/10.1103/PhysRevE.80.065104} {\bibfield
  {journal} {\bibinfo  {journal} {Phys. Rev. E}\ }\textbf {\bibinfo {volume}
  {80}},\ \bibinfo {pages} {065104} (\bibinfo {year} {2009})}\BibitemShut
  {NoStop}%
\bibitem [{\citenamefont {Mosteller}\ and\ \citenamefont
  {Thomas}(1961)}]{Mosteller1961}%
  \BibitemOpen
  \bibfield  {author} {\bibinfo {author} {\bibfnamefont {R.~E.~K.}\
  \bibnamefont {Mosteller}, \bibfnamefont {F.;~Rourke}}\ and\ \bibinfo {author}
  {\bibfnamefont {G.~B.}\ \bibnamefont {Thomas}},\ }\href@noop {} {\emph
  {\bibinfo {title} {Probability and Statistics.}}}\ (\bibinfo  {publisher}
  {Addison-Wesley},\ \bibinfo {year} {1961})\BibitemShut {NoStop}%
\end{thebibliography}%

\end{document}

\title{Controlling the shape of small clusters with and without macroscopic fields:\\
Supplemental Material}

\author{Francesco Boccardo}
\email{francesco.boccardo@univ-lyon1.fr}
\author{Olivier Pierre-Louis}
\email{olivier.pierre-louis@univ-lyon1.fr}
\affiliation{Institut Lumi\`ere Mati\`ere, UMR5306 Universit\'e Lyon 1 - CNRS, 69622 Villeurbanne, France}
\date{\today}

\maketitle

%%%%%%%%%%%%%%%%%%%%%%%%%%%%%%%%%%%%%%%%%%%%%%%%%%%%%%%%%%%%
%%%%%%%%%%%%%%%%%%%%%%%%%%%%%%%%%%%%%%%%%%%%%%%%%%%%%%%%%%%%
\section{Expected return time to target vs time to target from other states}
%%%%%%%%%%%%%%%%%%%%%%%%%%%%%%%%%%%%%%%%%%%%%%%%%%%%%%%%%%%%
%%%%%%%%%%%%%%%%%%%%%%%%%%%%%%%%%%%%%%%%%%%%%%%%%%%%%%%%%%%%

\begin{figure}[ht]
	\includegraphics[width=0.9\linewidth]{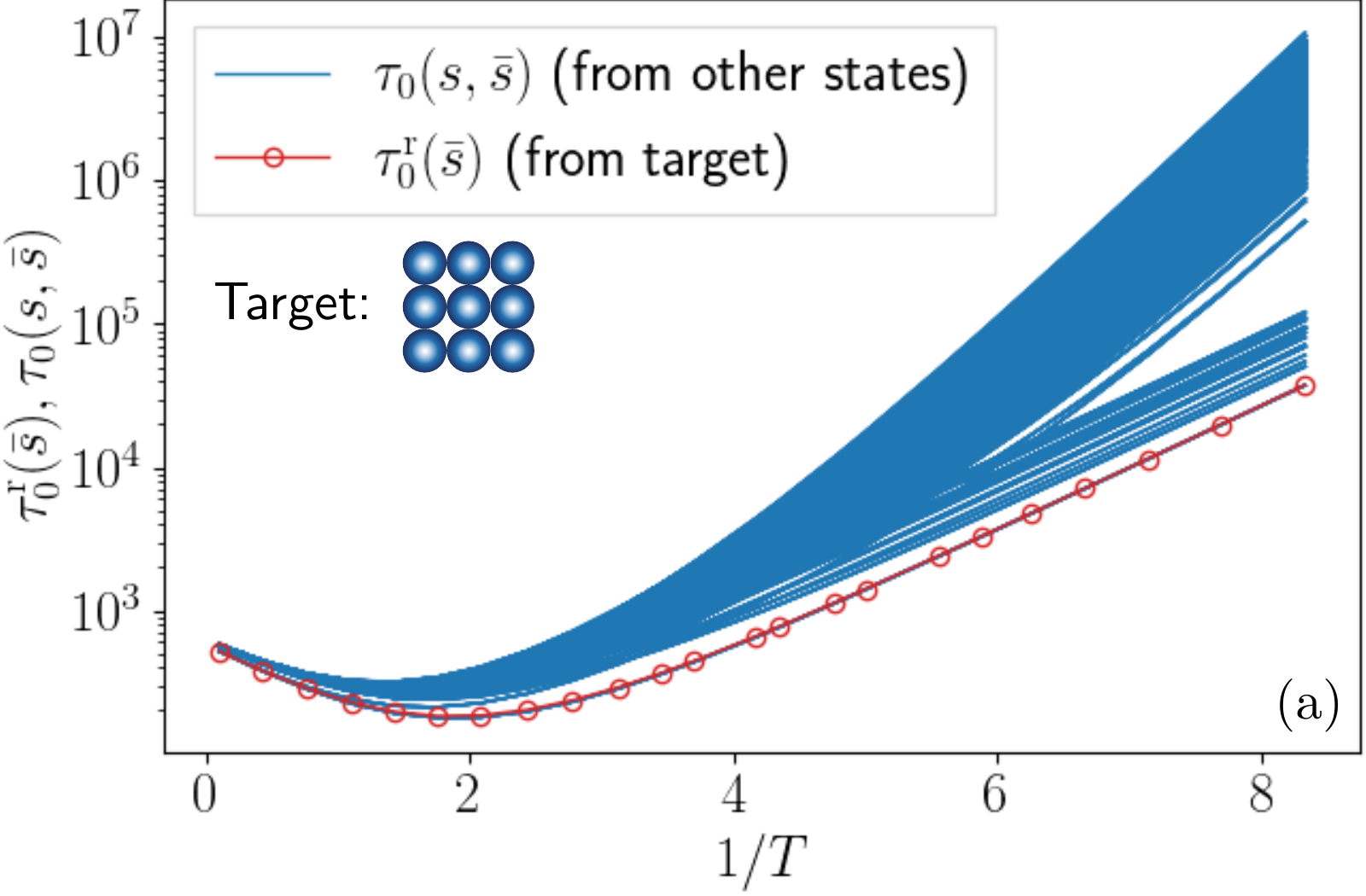}\\
	\includegraphics[width=0.9\linewidth]{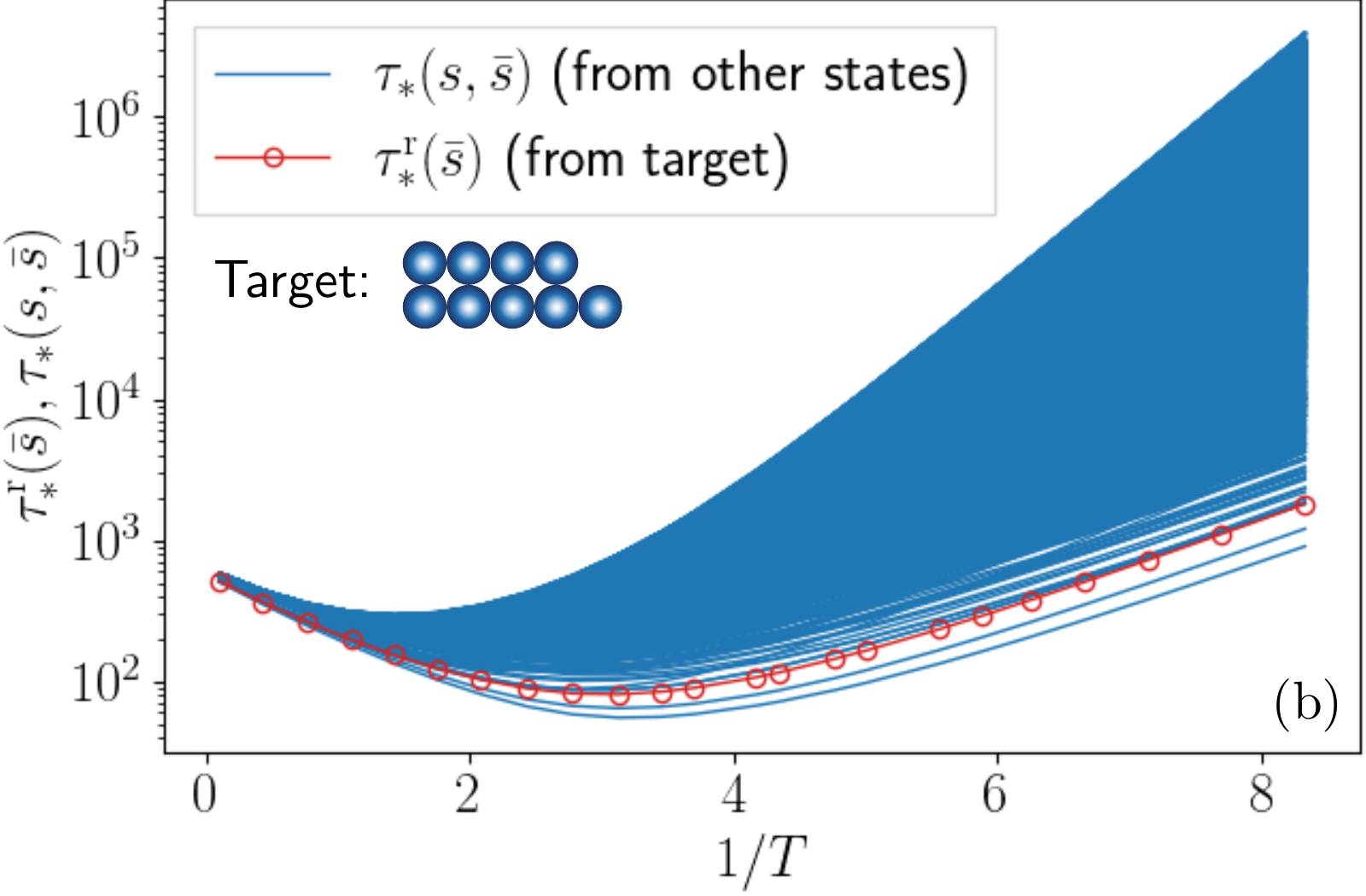}
	\caption{Expected time to reach the target as a function of $1/T$, starting from the target state itself (return time), and from all the other states in the system. (a) Zero-force case and (b) under an optimal policy ${\boldsymbol \phi}_*$. The number of blue curves in these graphs is therefore equal to the number of states minus one (the target), i.e. $S_9-1=9909$.}
	\label{fig:returntime_other_states}
\end{figure}

In the main text, we state that  when  the expected return time to target $\tau_{\boldsymbol \phi}^\mathrm{r}({\bar{s}})$ 
exhibits a minimum, 
the expected times to reach the target $\tau_{{\boldsymbol \phi}}(s,\bar{s})$ starting from any state $s$
usually also exhibit a minimum.
This is shown in \cref{fig:returntime_other_states}
for two 9-atom targets, for the zero-force and under an optimal policy ${\boldsymbol \phi}_*$
with $F_0=0.4$. In both cases, the presence of a minimum both in $\tau_{{\boldsymbol \phi}}^\mathrm{r}({\bar{s}})$ and $\tau_{{\boldsymbol \phi}}(s,\bar{s})$ is clearly visible.

%%%%%%%%%%%%%%%%%%%%%%%%%%%%%%%%%%%%%%%%%%%%%%%%%%%%%%%%%%%%
%%%%%%%%%%%%%%%%%%%%%%%%%%%%%%%%%%%%%%%%%%%%%%%%%%%%%%%%%%%%
\section{High temperature expansion of the expected return time to target}
%%%%%%%%%%%%%%%%%%%%%%%%%%%%%%%%%%%%%%%%%%%%%%%%%%%%%%%%%%%%
%%%%%%%%%%%%%%%%%%%%%%%%%%%%%%%%%%%%%%%%%%%%%%%%%%%%%%%%%%%%

%%%%%%%%%%%%%%%%%%%%%%%%%%%%%%%%%%%%%%%%%%%%%%%%%%%%%%%%%%%%
\subsection{Expression of the expected return time}
%%%%%%%%%%%%%%%%%%%%%%%%%%%%%%%%%%%%%%%%%%%%%%%%%%%%%%%%%%%%

Dividing the recursion relation Eq.~(\ref{MAIN-eq:recursion}) of the main text by $t_{\boldsymbol \phi}(s)$
and summing over all states but the target state, we obtain
\begin{align}
\label{aeq:1}
\bar{\sum_{s}}\frac{\tau_{{\boldsymbol \phi}}(s,\bar{s}) }{t_{\boldsymbol \phi}(s)} 
=S_N-1+ \bar{\sum_{s}}\sum_{s'\in {\cal B}_s}\frac{p_{\boldsymbol \phi}(s,s')}{t_{\boldsymbol \phi}(s)}\tau_{{\boldsymbol \phi}}(s',\bar{s})\,,
\end{align}
where $\bar\sum_{s}$ indicates the sum over all states in the system except for
the target state $\bar{s}$, and ${\cal B}_s$ is the set of bordering states of $s$, i.e. the states
that can be reached from $s$ in one physical move.

All physical moves except for those coming from the target
are summed over  in the last term of the right hand side of Eq.~(\ref{aeq:1}).
Hence, we can invert the indices $s$ and $s'$
in the sum if we subtract the contribution of the moves starting from the target, leading to
\begin{align}
\label{aeq:2}
\bar{\sum_{s}}\frac{\tau_{{\boldsymbol \phi}}(s,\bar{s}) }{t_{\boldsymbol \phi}(s)} 
=\,&S_N-1+ \bar{\sum_{s}}\sum_{s'\in {\cal B}_s}\frac{p_{{\boldsymbol \phi}}(s',s)}{t_{{\boldsymbol \phi}}(s')}\tau_{{\boldsymbol \phi}}(s,\bar{s})
\nonumber \\
%&-\bar{\sum_{s}}\frac{p_{{\boldsymbol \phi}}(\bar{s},s)}{t_{{\boldsymbol \phi}}(\bar{s})}\tau_{{\boldsymbol \phi}}(s,\bar{s})\delta_{{\cal B}_{\bar{s}}}(s)\,,
&-\sum_{s\in {\cal B}_{\bar{s}}} \frac{p_{{\boldsymbol \phi}}(\bar{s},s)}{t_{{\boldsymbol \phi}}(\bar{s})}\tau_{{\boldsymbol \phi}}(s,\bar{s})\,.
\end{align}
%where $\delta_{{\cal B}_s}(s')$ the Kronecker-delta
%function (or indicator function) which is $1$ when $s'$ belongs to ${\cal B}_s$ and zero otherwise.
The last term is proportional to the expected return time to target (Eq.(\ref{MAIN-eq:mean_return_time}) of the main text)
\begin{align}
\label{aeq:3}
\tau_{\boldsymbol \phi}^{\mathrm r}(\bar{s}) 
&= \sum_{s\in {\cal B}_{\bar{s}}}p_{{\boldsymbol \phi}}(\bar{s},s)\tau_{{\boldsymbol \phi}}(s,\bar{s})\, .
%&=\bar{\sum_{s}}p_{{\boldsymbol \phi}}(\bar{s},s)\tau_{{\boldsymbol \phi}}(s,\bar{s})\delta_{{\cal B}_{\bar{s}}}(s)\,,
\end{align}
We can therefore re-write Eq.~(\ref{aeq:2}) as
\begin{align}
% \label{aeq:4}
\tau_{\boldsymbol \phi}^{\mathrm r}(&\bar{s}) 
=t_{{\boldsymbol \phi}}(\bar{s})(S_N-1)
\nonumber \\
&+t_{{\boldsymbol \phi}}(\bar{s})\bar{\sum_{s}}\tau_{{\boldsymbol \phi}}(s,\bar{s})
\left\{\left(\,\sum_{s'\in {\cal B}_s}\frac{p_{{\boldsymbol \phi}}(s',s)}{t_{{\boldsymbol \phi}}(s')}\right)
-\frac{1}{t_{\boldsymbol \phi}(s)}\right\}\, .
\end{align}
Then,  writing $t_{\boldsymbol \phi}$ and $p_{\boldsymbol \phi}$ as a function of $\gamma_{{\boldsymbol \phi}}$,
we obtain an expression that is convenient for the study of the high temperature regime
\begin{align}
\label{aeq:4}
\tau_{\boldsymbol \phi}^{\mathrm r}(&\bar{s}) 
=t_{{\boldsymbol \phi}}(\bar{s})(S_N-1)
\nonumber \\
&+t_{{\boldsymbol \phi}}(\bar{s})\sum_{s}\tau_{{\boldsymbol \phi}}(s,\bar{s})
\sum_{s'\in {\cal B}_s}(\gamma_{{\boldsymbol \phi}}(s',s)-\gamma_{{\boldsymbol \phi}}(s,s'))\, ,
\end{align}
where we have dropped the exclusion of the target state (indicated by the bar)
in the sum since $\tau_{\boldsymbol \phi}(\bar{s},\bar{s})=0$.

%%%%%%%%%%%%%%%%%%%%%%%%%%%%%%%%%%%%%%%%%%%%%%%%%%%%%%%%%%%%
\subsection{Infinite temperature limit}
%%%%%%%%%%%%%%%%%%%%%%%%%%%%%%%%%%%%%%%%%%%%%%%%%%%%%%%%%%%%

In the high temperature limit $k_BT\rightarrow+\infty$, the rates 
given by Eq.(\ref{MAIN-eq:TST_hopping}) of the main text all converge to the same value 
$\gamma_{{\boldsymbol \phi}}(s,s')\rightarrow 1$, and therefore the residence time
$t_{\boldsymbol \phi}(s)=1/\!\sum_{s'\in {\cal B}_s}\!\gamma_{\boldsymbol \phi}(s,s')$ reads
\begin{align}
\label{aeq:5}
t_{\boldsymbol \phi}(s)\rightarrow \frac{1}{d_s}\, ,
\end{align}
where $d_s$, called the degree of state $s$,
is the number of possible moves from $s$.
Note that $d_s$  is the cardinal of ${\cal B}_s$. 
As a consequence, Eq.~(\ref{aeq:4}) implies 
\begin{align}
\label{aeq:6}
&\tau_{\boldsymbol \phi}^{\mathrm r}(\bar{s}) 
\rightarrow \tau_\infty^{\mathrm r}(\bar{s}) =\frac{S_N-1}{d_{\bar{s}}}
\end{align}
in the high temperature limit.

In addition, the general recursion relation Eq.~(\ref{MAIN-eq:recursion}) of the main text 
can be re-written as
\begin{align}
    1=\sum_{s'\in {\cal B}_s}\gamma_{{\boldsymbol \phi}}(s,s')(\tau_{{\boldsymbol \phi}}(s,\bar{s})-\tau_{{\boldsymbol \phi}}(s',\bar{s}))\, .
\end{align}
In the limit of infinite temperatures, this expression takes the form of the usual discretized Laplacian
for the expected first passage times to the target  $\tau_\infty(s,\bar{s})$
\begin{align}
\label{eq:Laplacian_HT}
    1=\sum_{s'\in {\cal B}_s}(\tau_\infty(s,\bar{s})-\tau_\infty(s',\bar{s})) \,.
\end{align}
This relation is valid everywhere but at the target state, where $\tau_\infty(\bar{s})=0$.
The value of $\tau_\infty(s,\bar{s})$ depends in general on the precise 
structure of the graph.

A vast literature is dedicated to the study of the first passage times on random graphs \cite{Noh2004,Baronchelli2006}. 
However, most studies are devoted to the case where $t_{\boldsymbol \phi}(s)=1$,
which is different from our infinite temperature limit.
The difference lies only in the factor $1/d_s$ in $t_{\boldsymbol \phi}(s)$.
A trace of this simple difference is that the mean first passage time (MFPT),
which is the average of the first passage time over all initial states for a fixed target state,
discussed in the literature with $t_{\boldsymbol \phi}(s)=1$, exhibits a lower bound~\cite{Lin2012,Tejedor2009} 
MFPT$\geq S_N \langle d_s\rangle/d_s$, where $\langle d_s\rangle$
is the average of $d_s$ over all states. Up to the factor $1/\langle d_s\rangle$,
this is identical to the expression of the expected return time to the target
$\tau_{\boldsymbol \phi}^{\mathrm r}(\bar{s}) $ in Eq.~(\ref{aeq:6}).

%%%%%%%%%%%%%%%%%%%%%%%%%%%%%%%%%%%%%%%%%%%%%%%%%%%%%%%%%%%%
\subsection{High temperature expansion}
%%%%%%%%%%%%%%%%%%%%%%%%%%%%%%%%%%%%%%%%%%%%%%%%%%%%%%%%%%%%

We  perform an expansion to linear order in $1/T$ to obtain the first correction to Eq.~(\ref{aeq:6}).
We start with the expansion of the rates, defined in Eqs.~(1,2) of the main text
\begin{align}
\label{eq:expansion_gamma_HT}
    \gamma_{{\boldsymbol \phi}}(s,s')=1+\frac{1}{T}(-n_{ss'}+\varphi(s)\, u_{ss'}),
\end{align}
where we recall that $n_{ss'}$ is the initial number of bonds of the atom 
which moves in the transition from state $s$ to state $s'$.
In addition, we have  $u_{ss'}=1/2$, $0$, $0$ or $-1/2$
when the move is in the direction $+\hat {\bm x}$,
$-\hat {\bm y}$, $+\hat {\bm y}$ or
$-\hat {\bm x}$ respectively.
The $+\hat {\bm x}$ and $+\hat {\bm y}$ axes correspond to the (10) and (01) lattice directions respectively. 
Remark the symmetry relation $u_{ss'}=-u_{s's}$.
Moreover, $\varphi(s)\,=-F_0$, $0$, or $F_0$ is the scalar force in state $s$
given by the policy ${\boldsymbol \phi}$, so that ${\boldsymbol \phi}(s)=\varphi(s)\,\hat {\bm x}$.

Inserting this expression in the expected return time Eq.~(\ref{aeq:4}), we obtain
to linear order in $1/T$
\begin{align}
\label{aeq:7}
\tau_{\boldsymbol \phi}^{\mathrm r}(\bar{s}) &=\tau_\infty^{\mathrm r}(\bar{s}) \left(1+ \frac{M_{\boldsymbol \phi}}{T}\right) 
\nonumber \\
M_{\boldsymbol \phi}&=\frac{1}{d_{\bar{s}}}\sum_{s\in {\cal B}_{\bar{s}}}(n_{\bar{s}s}-\varphi(\bar{s})\,u_{\bar{s}s})
\nonumber \\&
+\frac{1}{S_N-1}\sum_{s}\tau_\infty(s,\bar{s})
%\times\nonumber \\ &\hspace{1 cm}
\sum_{s'\in {\cal B}_s}(n_{ss'}-n_{s's})
\nonumber \\ &
+\frac{1}{S_N-1}\sum_{s}\!\tau_\infty(s,\bar{s})\!
\sum_{s'\in {\cal B}_s}\!(\varphi(s')u_{s's}\!-\!\varphi(s)\,u_{ss'})
\, ,
\end{align}
where the first passage time to target at infinite temperature $\tau_\infty(s,\bar{s})$
is independent on the policy ${\boldsymbol \phi}$.
The normalized slope $M_{\boldsymbol \phi}$ obeys
\begin{align}
\label{aeq:return_time_HT}
(1-&\frac{1}{S_N})M_{\boldsymbol \phi}=
\Bigl\langle 
(1-\delta_{s{\bar s}})\langle n_{ss'}\rangle_{s'\in {\cal B}_{s}} -d_s g_n(s,\bar{s})
\Bigr\rangle_{s\in {\cal S}}
% \Bigl\langle 
% (1-\delta_{s{\bar s}}) \Bigl( \langle n_{ss'}\rangle_{s'\in {\cal B}_{s}} -d_s g_n(s,\bar{s}) \Bigr)
% \Bigr\rangle_{s\in {\cal S}}
\nonumber \\
&\quad-\Bigl\langle 
 \varphi(s)\, \Bigl((1-\delta_{s{\bar s}})\langle u_{ss'}\rangle_{s'\in {\cal B}_{s}} -d_s g_u(s,\bar{s})\Bigr)
\Bigr\rangle_{s\in {\cal S}}\,,
% &-\Bigl\langle 
%  \varphi(s)\, (1-\delta_{s{\bar s}}) \Bigl(\langle u_{ss'}\rangle_{s'\in {\cal B}_{s}} -d_s g_u(s,\bar{s})\Bigr)
% \Bigr\rangle_{s\in {\cal S}}
%M_{\boldsymbol \phi}&=-\frac{1}{\tau_\infty^{\mathrm r}(\bar{s})}g_n(\bar{s})
%+\Bigl\langle 
%\langle n_{ss'}\rangle_{s'\in {\cal B}_{s}} -d_s g_n(s,\bar{s})
%\Bigr\rangle_{s\in \bar{\cal S}}
%\nonumber \\
%&+\frac{\varphi(\bar{s})\,}{\tau_\infty^{\mathrm r}(\bar{s})}g_u(\bar{s})
%-\Bigl\langle 
% \varphi(s)\, \Bigl(\langle u_{ss'}\rangle_{s'\in {\cal B}_{s}} -d_s g_u(s,\bar{s})\Bigr)
%\Bigr\rangle_{s\in \bar{\cal S}}
\end{align}
where $\delta_{ss'}$ is the Kronecker symbol and the notation 
\begin{align}
\langle q_s \rangle_{s\in{\cal Z}} =\frac{1}{|{\cal Z}|}\sum_{s\in{\cal Z}}q_s 
\end{align}
indicates the average of $q_s$
taken over the states $s$ belonging to the set of states ${\cal Z}$
with cardinal $|{\cal Z}|$. 
% We have also defined the ensemble $\bar{\cal S}$ of all states except the target state $\bar{s}$ \com{$\bar{\cal S}$ not needed anymore?}.
In addition, we have defined
the local covariances
%~\footnote{Strictly speaking, $g_n(s,\bar{s})$ and $g_u(s,\bar{s})$
%are not covariances because the neighborhood ${\cal B}_{s}$ is not a probability space. However,
%their definition is similar to that of a covariance.}
\begin{align}
\label{aeq:cov_n}
&g_n(s,\bar{s})=
\nonumber\\
&\!\Bigl\langle\! 
\Bigl(n_{ss'}\!-\!\langle n_{ss''}\rangle_{s''\in {\cal B}_{s}} \Bigr) \!
\Bigl(\tau_\infty(s',\bar{s})\!-\!\langle \tau_\infty(s'', \bar{s})\rangle_{s''\in {\cal B}_{s}} \Bigr)\!
\Bigr\rangle_{\!s'\in {\cal B}_{s}}
\\
\label{aeq:cov_u}
&g_u(s,\bar{s})=\nonumber\\
&\!\Bigl\langle \!
\Bigl(u_{ss'}\!-\!\langle u_{ss''}\rangle_{s''\in {\cal B}_{s}}\Bigr) \!
\Bigl(\tau_\infty(s',\bar{s})\!-\!\langle \tau_\infty(s'', \bar{s})\rangle_{s''\in {\cal B}_{s}}\Bigr)\!
\Bigr\rangle_{\!s'\in {\cal B}_{s}}
\end{align}
that are averaged over the bordering states of $s$.

In the case of zero force,
we set $\varphi(s)\,=0$ in Eq.~(\ref{aeq:return_time_HT})
and  readily obtain Eq.~(7) of the main text.

We now consider the case of non-zero forces.
Since $\tau_{\boldsymbol \phi}^{\mathrm r}(\bar{s})$ is linear in the forces $\varphi(s)\,$
from Eq.~(\ref{aeq:return_time_HT}), we obtain the optimal
policy ${\boldsymbol \phi}_*(s)=\varphi_*(s)\hat{\bm{x}}$ from the sign of the prefactor of $\varphi(s)\,$
in Eq.~(\ref{aeq:return_time_HT}):
\begin{align}
\label{eq:optimal_forces_HT}
\varphi_*(s)&=-F_0\;\mathrm{sign} \left[
    d_s g_u(s,\bar{s})-\langle u_{ss'}\rangle_{s'\in {\cal B}_{s}}
    \right] \, ,
\nonumber \\
\nonumber \\
    \varphi_*(\bar{s})&=-F_0\;\mathrm{sign} \left[
    g_u(\bar{s})
    \right] \, .
\end{align}
If the terms in the squared brackets vanish, then
the contribution of the force term to $\tau_{\boldsymbol \phi}^{\mathrm r}(\bar{s})$ is at least second order 
in $1/T$, and should therefore be negligible for high enough temperatures.

Using Eq.~(\ref{eq:optimal_forces_HT}) in Eq.~(\ref{aeq:return_time_HT}), we obtain the high temperature correction to the optimal expected return time to target
%\begin{align}
%\label{aeq:optimal_return_time_HT}
%M_*&=-\frac{1}{\tau_\infty^{\mathrm r}(\bar{s})}g_n(\bar{s})
%%\nonumber \\&
%+\Bigl\langle 
%\langle n_{ss'}\rangle_{s'\in {\cal B}_{s}} -d_s g_n(s,\bar{s})
%\Bigr\rangle_{s\in \bar{\cal S}}
%\nonumber \\
%&-\frac{F_0}{\tau_\infty^{\mathrm r}(\bar{s})}|g_u(\bar{s})|
%-F_0\Bigl\langle 
%\Bigl|\langle u_{ss'}\rangle_{s'\in {\cal B}_{s}} -d_s g_u(s,\bar{s})\Bigr|
%\Bigr\rangle_{s\in \bar{\cal S}}
%\end{align}
%Another writing of the same expression is
\begin{align}
\label{aeq:optimal_return_time_HT}
\left(1-\frac{1}{S_N}\right)&M_*=\Bigl\langle 
 (1-\delta_{s\bar{s}}) \langle n_{ss'}\rangle_{s'\in {\cal B}_{s}} -d_s g_n(s,\bar{s})
\Bigr\rangle_{s\in {\cal S}}
\nonumber \\
&\!\!-F_0\Bigl\langle 
 \Bigl| (1-\delta_{s\bar{s}}) \langle u_{ss'}\rangle_{s'\in {\cal B}_{s}} -d_s g_u(s,\bar{s})\Bigr|
\Bigr\rangle_{s\in {\cal S}},
\end{align}
which is identical to Eq.~(\ref{MAIN-aeq:M_HT_*_approx}) of the main text.

%%%%%%%%%%%%%%%%%%%%%%%%%%%%%%%%%%%%%%%%%%%%%%%%%%%%%%%%%%%%
\subsection{Approximate expression of $M_0$}
%%%%%%%%%%%%%%%%%%%%%%%%%%%%%%%%%%%%%%%%%%%%%%%%%%%%%%%%%%%%

\begin{figure}[ht]
	\includegraphics[width=\linewidth]{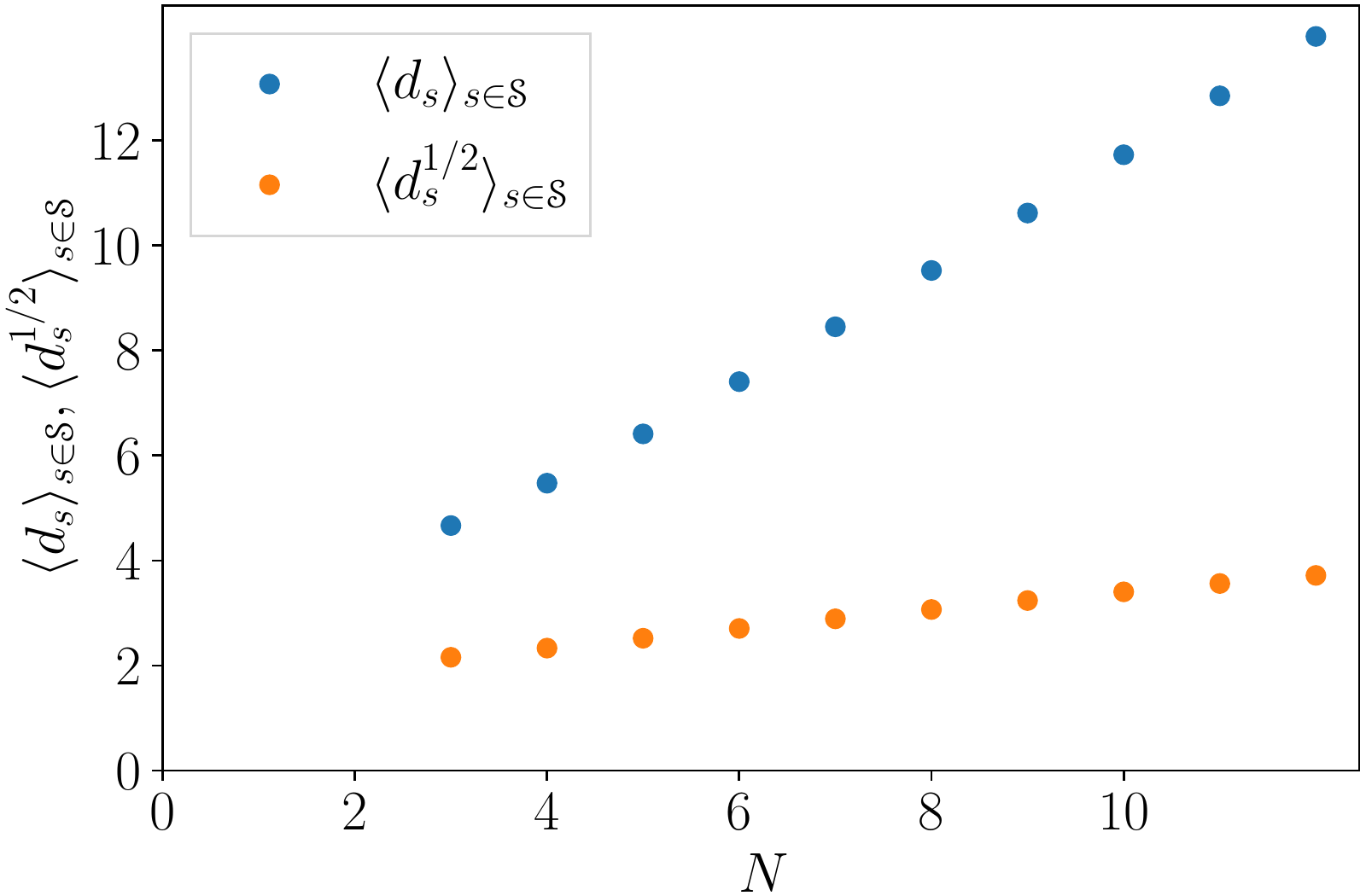}
	\caption{Mean of the degrees of the states $d_s$ and their square roots over all states, for clusters of increasing size.}
	\label{fig:ds_vs_N}
\end{figure}

\begin{figure}[ht]
	\includegraphics[width=\linewidth]{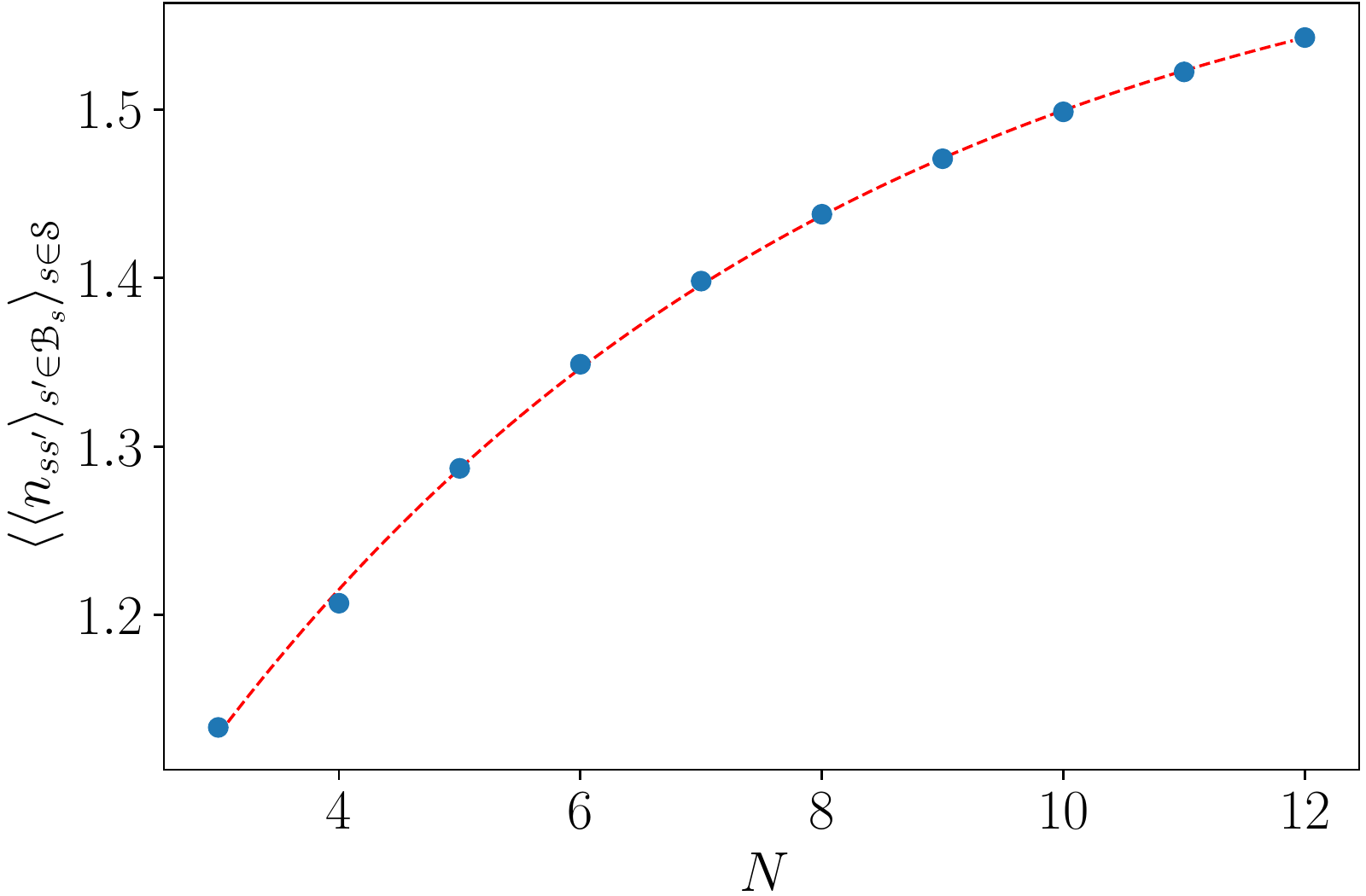}
	\caption{Mean of $\langle n_{ss'} \rangle_{s' \in {\cal B}_s}$ over all states, for clusters of increasing size. The points are fitted with a decaying exponential, which gives an asymptotic value $\rho_0(\infty) \approx 1.64$.}
	\label{fig:nss_vs_N}
\end{figure}

In Eq.~(\ref{MAIN-eq:M_0}) of the main text, the first term  $\langle (1-\delta_{s\bar{s}}) \langle n_{ss'}\rangle_{s'\in {\cal B}_{s}}\rangle_{s\in {\cal S}}$ is the average 
over all possible states $s$ except the target $\bar{s}$
of the average number of bonds $n_{ss'}$ that is broken to perform a move from $s$.
This term essentially accounts for the global slowing down of the dynamics
when the temperature is decreased.
The number $d_s$ of possible moves from a given state $s$ grows moderately
(like a power law) with $N$, as shown in Fig.\ref{fig:ds_vs_N}, 
while the total number of states $S_N$ grows exponentially with $N$ as discussed in the main text.
As a consequence, the contribution of the moves that start from the target is negligible 
when $N$ is large enough. Hence, the first term in Eq.~(\ref{MAIN-eq:M_0}) of the main text can be approximated
with the average $\langle  \langle n_{ss'}\rangle_{s'\in {\cal B}_{s}}\rangle_{s\in {\cal S}}$ of $n_{ss'}$ over all states. 
Note that $n_{ss'}$ is bounded ($1\leq n_{ss'} \leq 4$),
and this average converges quickly to a value $\rho_0$ around 1.64. 
A fit of $\rho_0(N)=\langle  \langle n_{ss'}\rangle_{s'\in {\cal B}_{s}}\rangle_{s\in {\cal S}}$ as for $N>2$, 
provided in \cref{fig:nss_vs_N}, shows that this convergence is exponential
$\rho_0(N)=\rho_0(\infty)-\tilde\rho_{0}\exp(-N/N_0)$ with  $\tilde\rho_0\approx0.89$ and $N_0 \approx 5.4$.

\begin{figure}[ht]
	\includegraphics[width=\linewidth]{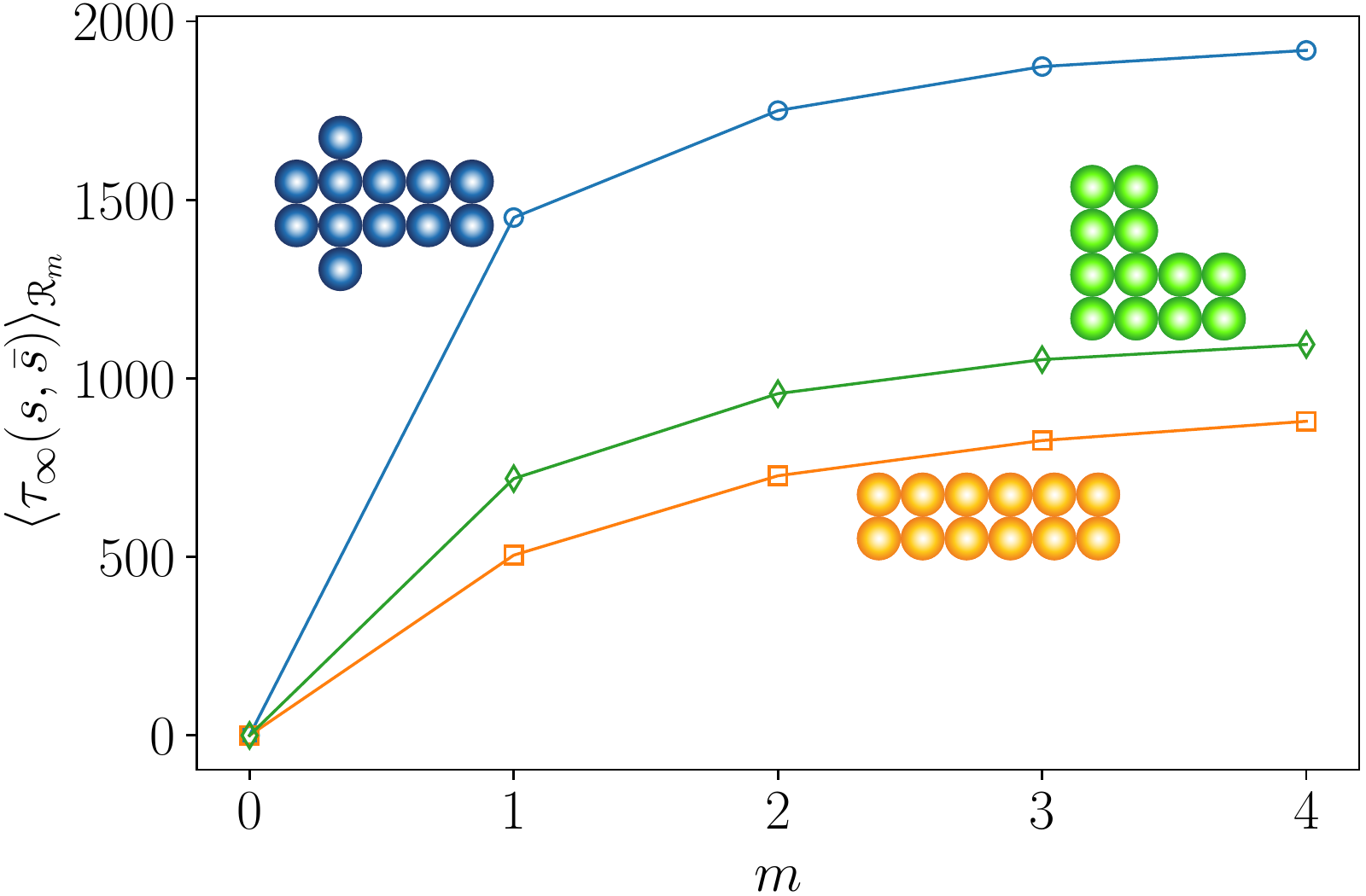}
	\caption{Mean on rings of the first passage time to target at infinite temperature as a function of the ring index $m$ for three different targets with $N=12$.}
	\label{fig:tau_inf_rings}
\end{figure}

The second term $\langle d_s g_n(s,{\bar{s}})\rangle_{s\in {\cal S}}$ in Eq.~(\ref{MAIN-eq:M_0}) of the main text
accounts for the correlations between the number of bonds 
that are broken and the first passage time to target 
at infinite temperature when starting from a given state $s$. 
This second term accounts for the relative trend for the system
to go faster towards low energy states.
In order to analyze this expression, we introduce the concept of rings
following Ref.~\cite{Baronchelli2006}.
The index of the rings $m$ is defined as the minimum number
of moves to reach the target.  A ring ${\cal R}_m$ is the ensemble of states with the same 
ring index $m$. 
Note that our definitions are such that ${\cal R}_1={\cal B}_{\bar s}$.
The first passage time to target at infinite temperature $\tau_\infty(s,{\bar{s}})$
is known to increase quickly up to an asymptotic value as $m$ increases~\cite{Baronchelli2006}.
The asymptotic value is reached after a few rings.
A plot of the mean first passage time to target as a function of the ring
index is reported in \cref{fig:tau_inf_rings}.
We therefore simply assume that the second term is dominated by the contributions
related the large variation of $\tau_\infty(s,{\bar{s}})$
between the target state where $\tau_\infty(\bar{s},{\bar{s}})=0$ and the first ring.

%[[not needed!]]
%Furthermore, we assume that the value of $\tau_\infty(s,\bar{s})$
%in the first ring is similar to the asymptotic value.
%This asymptotic value is assumed to be similar to the
%Mean First Passage Time (MFPT), defined as the average of the
%first passage time over all states $\langle \tau_\infty(s,\bar{s})\rangle_{\cal S}$.
%The MFPT has been studied in the literature, and bounds
%are known for its possible value. 
When evaluating the term $\langle d_s g_n(s,\bar{s})\rangle_{s\in {\cal S}}$ 
in Eq.~(\ref{aeq:optimal_return_time_HT}) we therefore need to sum only over the moves
from states in ${\cal B}_{\bar s}$,
% ${\cal R}_1$ to the target, 
which leads to
\begin{align}
\nonumber
\langle d_s &g_n(s,\bar{s})\rangle_{s\in {\cal S}} \approx \\
&-\frac{d_{\bar{s}}}{S_N} \langle 
(n_{s\bar{s}}-\langle n_{ss''}\rangle_{s''\in {\cal B}_{s}} ) 
\tau_\infty(s,\bar{s}) \rangle_{s\in {\cal B}_{\bar s}} .
\label{aeq:dsgn_approx_1}
\end{align}
To obtain this relation, we have used the recursion relation Eq.~(\ref{eq:Laplacian_HT}),
so that for the term with $s'=\bar{s}$ and $s\in{\cal B}_{\bar s}$ in Eq.~(\ref{aeq:cov_n}), we have
\begin{align}
\tau_\infty(s',\bar{s})-&\langle \tau_\infty(s'')\rangle_{s''\in {\cal B}_{s}} =
\nonumber\\
&\tau_\infty(\bar{s},\bar{s})-\left(\tau_\infty(s,\bar{s})-\frac{1}{d_s}\right) =
\nonumber\\
&-\left(\tau_\infty(s,\bar{s})-\frac{1}{d_s}\right)\approx-\tau_\infty(s,\bar{s})\,,
\nonumber
\end{align}
where in the last line, we use $1/d_s\ll \tau_\infty(s,\bar{s})$, which is valid
for $N$ large enough.
Finally, we simply assume the statistical independence of the two terms
in the product in the brackets of Eq.(\ref{aeq:dsgn_approx_1}), leading to 
\begin{align}
\nonumber
\langle d_s &g_n(s,\bar{s})\rangle_{s\in {\cal S}}\approx\\
&-\frac{d_{\bar{s}}}{S_N} \langle 
n_{s\bar{s}}-\langle n_{ss''}\rangle_{s''\in {\cal B}_{s}}  \rangle_{s\in {\cal B}_{\bar s}}\langle 
\tau_\infty(s,\bar{s}) \rangle_{s\in {\cal B}_{\bar s}}\,.
\end{align}
Since $\langle\tau_\infty(s,\bar{s}) \rangle_{s\in {\cal B}_{\bar s}}=\tau^{\mathrm r}_\infty(\bar{s})=(S_N-1)/d_{\bar{s}}$
and $S_N\gg 1$,
we finally obtain 
\begin{align}
\label{aeq:dsgn_approx}
\langle d_s g_n(s,\bar{s})\rangle_{s\in {\cal S}}\approx -\langle 
n_{s\bar{s}}-\langle n_{ss''}\rangle_{s''\in {\cal B}_{s}}  \rangle_{s\in {\cal B}_{\bar s}}.
\end{align}

\begin{figure}[ht]
	\includegraphics[width=\linewidth]{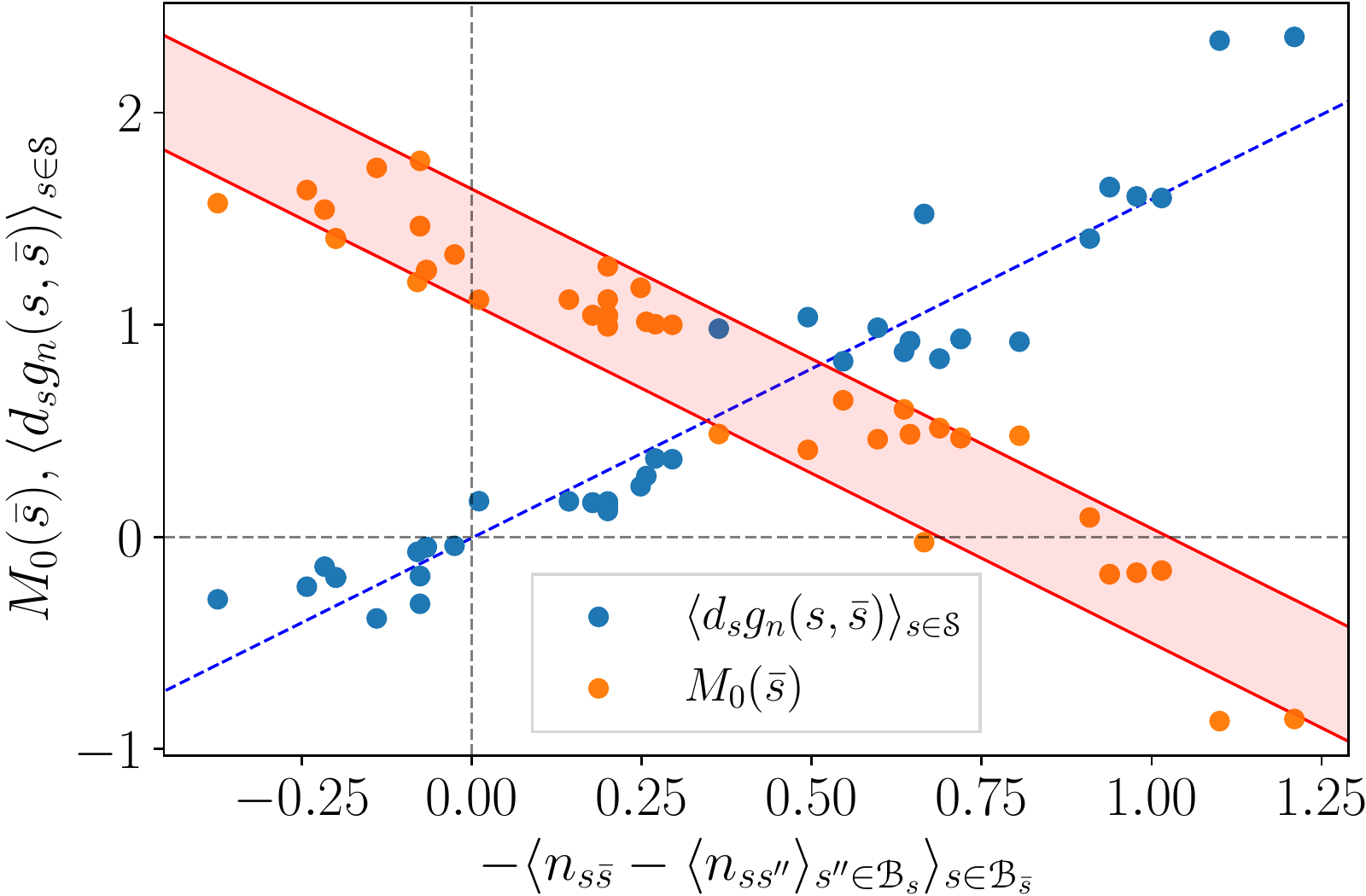}
	\caption{$M_0(\bar s)$ and mean of $d_sg_n(s,\bar s)$ over all states as a function of the quantity $-\langle 
n_{s\bar{s}}-\langle n_{ss''}\rangle_{s''\in {\cal B}_{s}}  \rangle_{s\in {\cal B}_{\bar s}}$ (the mean is taken over the bordering states of the target $\bar s$).
% ${\cal B}_{\bar s}$, or ${\cal R}_1$
The blue dashed line is a fit of $\langle d_sg_n(s,\bar s) \rangle_{s \in {\cal S}}$, that gives a slope $\rho_1 \approx 1.60$. This value is used to draw the two red lines, which correspond to the approximate relation $M_0^{\mathrm{appr}}(\bar s)= \rho_0(N) -\rho_1\langle n_{s\bar{s}}-\langle n_{ss''}\rangle_{s''\in {\cal B}_{s}}  \rangle_{s \in {\cal B}_{\bar s}}$,
% of Eq.~(\ref{eq:approximate_R0})
with $\rho_0(N=3)=1.1$ and $\rho_0(N=\infty)=1.64$ (the minimum and the asymptotic value of $\langle \langle n_{ss'} \rangle_{s' \in {\cal B}_s} \rangle_{s \in {\cal S}}$, see \cref{fig:nss_vs_N}).}
	\label{fig:dsgn_R0_first_ring}
\end{figure}

This approximate relation is shown to be accurate in \cref{fig:dsgn_R0_first_ring}
up to a multiplicative factor $\rho_1\approx1.60$ in the range 
of cluster sizes $N=2$ to $12$ that we have explored. 
When the target shape is compact,
then $n_{s\bar{s}}$ is smaller than the average of $n_{ss'}$ over the 
bordering states $s'$ of the states $s$ of the first ring.
Hence, compact shapes lead to a positive contribution to $\langle d_s g_n(s,\bar{s})\rangle_{s\in {\cal S}}$,
and a negative contribution to $M_0$.

Combining the approximations of both contributions to the first line
of Eq.~(\ref{aeq:optimal_return_time_HT}), we obtain Eq.~(\ref{MAIN-eq:approximate_R0})
of the main text. This expression provides a fair estimate of
$M_0$, as shown in \cref{fig:dsgn_R0_first_ring}.

%%%%%%%%%%%%%%%%%%%%%%%%%%%%%%%%%%%%%%%%%%%%%%%%%%%%%%%%%%%%
\subsection{Approximate expression of $M_*$}
%%%%%%%%%%%%%%%%%%%%%%%%%%%%%%%%%%%%%%%%%%%%%%%%%%%%%%%%%%%%

Assuming for the sake of simplicity
complete uncorrelation between $u_{ss'}$ and $\tau_{\infty}(s',{\bar{s}})$,
the covariance $|d_s g_u(s,{\bar{s}})|$ 
can be approximated as
the product of the standard deviations $\sigma_u$ and $\sigma_{\tau_\infty}({\bar{s}})$ of $u_{ss'}$ and $\tau_\infty(s,{\bar{s}})$ 
in ${\cal B}_s$, times the absolute value of a sum of $d_s$ random signs that account for the
possible signs of the products.
This sum is approximated by the standard formula for the 
expectation value of the absolute distance
after $d_s$ steps as $(2 d_s/\pi)^{1/2}$~\cite{Mosteller1961}.
%{\color{red}[[Francesco what is the right name for this quantity? reference?]]} \com{it's called the \q{expectation value of the absolute distance}. I found the formula here: Weisstein, Eric W. "Random Walk--1-Dimensional." From MathWorld--A Wolfram Web Resource. https://mathworld.wolfram.com/RandomWalk1-Dimensional.html  but in that page they cite three other references: Grünbaum (1960), Mosteller et al. (1961), and König et al. (1999). I don't know which is best to cite.}.
As a consequence, we have 
\begin{align}
\langle |d_s g_u(s,{\bar{s}})|\rangle_{s\in{\cal S}}\approx (2/{\pi})^{1/2}\langle d_s^{1/2}
\rangle_{s\in{\cal S}} \;\sigma_u\; \sigma_{\tau_\infty}(\bar s) \,.
\label{eq:approx_ds_gu}
\end{align}

\begin{figure}[ht]
	\includegraphics[width=\linewidth]{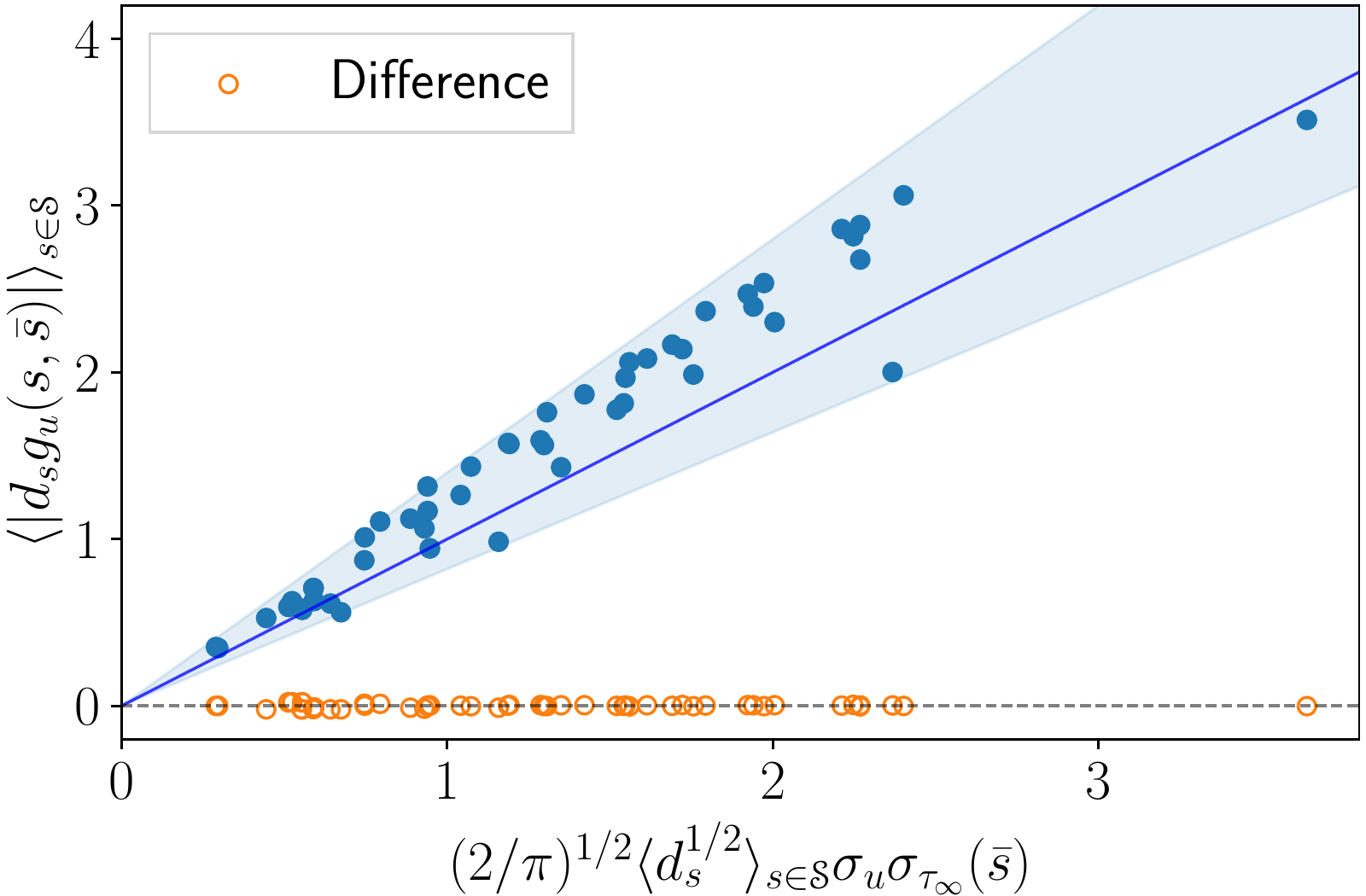}
	\caption{Blue dots: mean of the covariance $|d_s g_u(s,{\bar{s}})|$ over all states as a function of its approximation Eq.~(\ref{eq:approx_ds_gu}). The orange dots are the difference between $\langle |d_s g_u(s,{\bar{s}})|\rangle_{s\in{\cal S}}$ and $\langle |d_s g_u(s,{\bar{s}})-\langle u_{ss'}\rangle_{s'\in{\cal B}_s}|\rangle_{s\in{\cal S}}$.}
	\label{fig:dsgu_vs_2pids}
\end{figure}

As shown in \cref{fig:dsgu_vs_2pids}, this expression provides a fair approximation to $\langle |d_s g_u(s,{\bar{s}})|\rangle_{s\in{\cal S}}$. The averages $\langle u_{ss'}\rangle_{s'\in{\cal B}_s}$ are small and their contribution to the term $\langle 
 |(1-\delta_{s\bar{s}})  \langle u_{ss'}\rangle_{s'\in{\cal B}_s} -d_s g_u(s,{\bar{s}})|
\rangle_{s\in {\cal S}}$ is negligible. This is also shown in \cref{fig:dsgu_vs_2pids}, where the difference between $\langle |d_s g_u(s,{\bar{s}})|\rangle_{s\in{\cal S}}$ and $\langle |d_s g_u(s,{\bar{s}})-\langle u_{ss'}\rangle_{s'\in{\cal B}_s}|\rangle_{s\in{\cal S}}$ is 
seen to be small.

\begin{figure}[ht]
	\includegraphics[width=\linewidth]{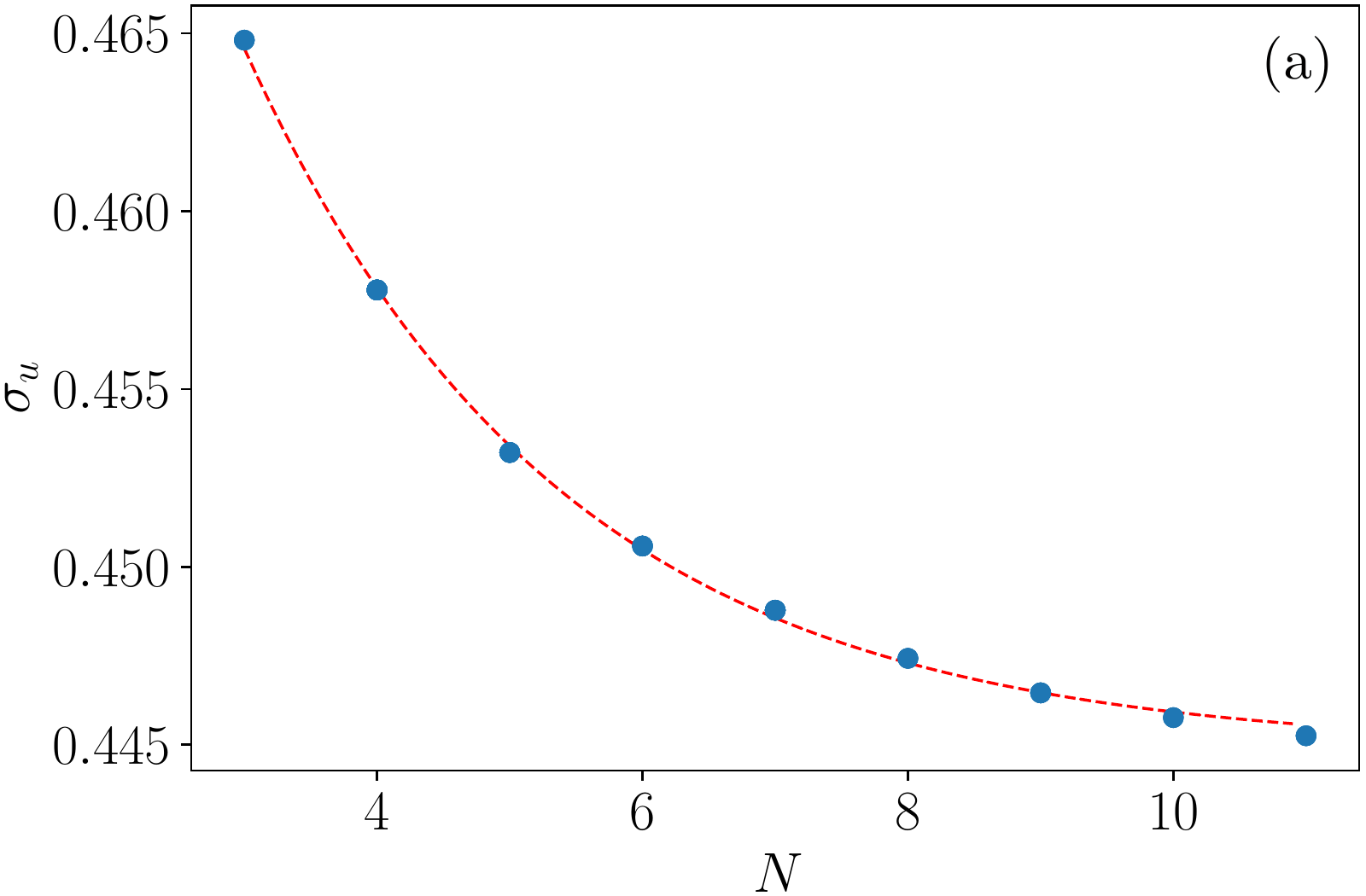}\\
	[10pt]
	\includegraphics[width=\linewidth]{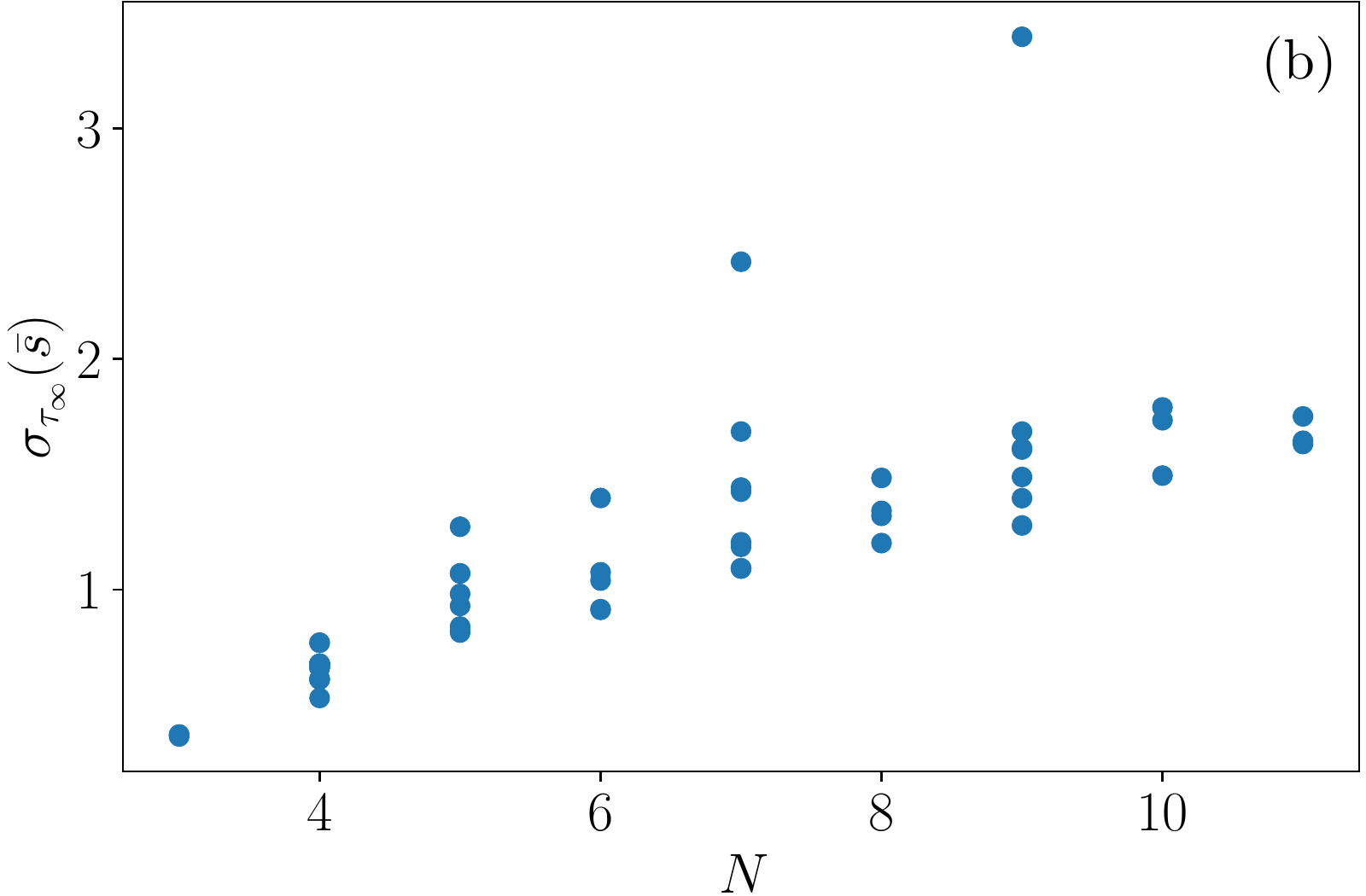}
	\caption{(a) Standard deviation $\sigma_u$ as a function of $N$. The points are fitted with a decreasing exponential, which gives an asymptotic value approximately equal to 0.44. (b) Standard deviation $\sigma_{\tau_{\infty}}(\bar s)$ for different targets as a function of $N$.}
	\label{fig:sigma_u_vs_N}
\end{figure}

Finally, in \cref{fig:sigma_u_vs_N,fig:ds_vs_N}, we show that the standard deviation $\sigma_u$, which is bounded, converges to a constant roughly equal to $0.44$, 
while the quantities $\sigma_{\tau_\infty}({\bar{s}})$ and $\langle d_s^{1/2}\rangle_{s\in{\cal S}}$ grow with $N$.

%\begin{figure}[ht]
%	\includegraphics[width=\linewidth]{sigma_tau_vs_N.pdf}
%	\caption{Standard deviation $\sigma_{\tau_{\infty}}(\bar s)$ for different targets as a function of $N$.}
%	\label{fig:sigma_tau_vs_N}
%\end{figure}

\section{Zero-force vs random policy}

\begin{figure}[ht]
	\includegraphics[width=\linewidth]{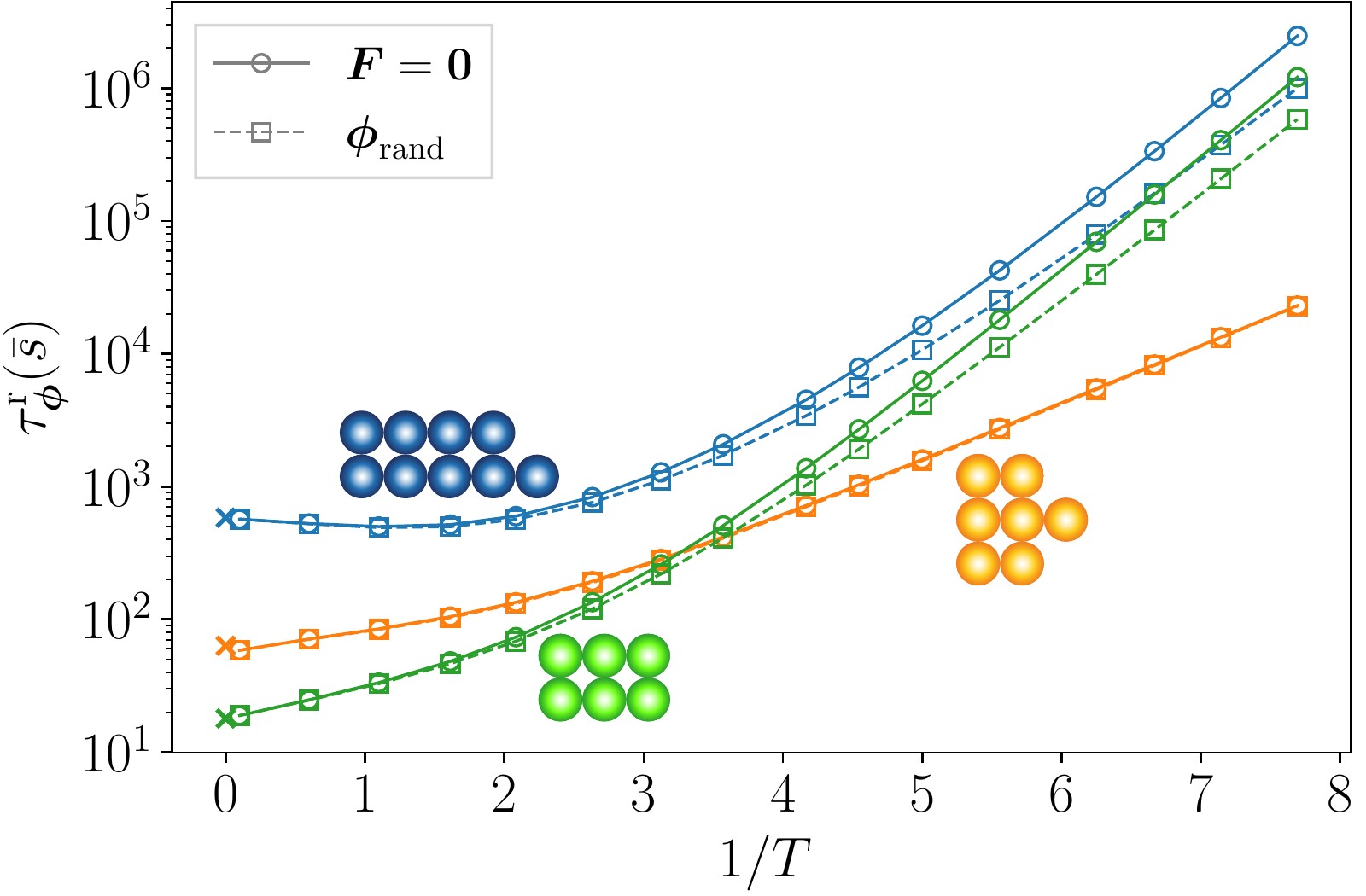}
	\caption{Expected return time to target $\tau^{\textrm{r}}_{{\boldsymbol \phi}}(\bar s)$ for the zero-force case and under a random policy ${\boldsymbol \phi}_{\textrm{rand}}$ (i.e. a policy that chooses a random force with equal probability in each state) for the three targets shown in the figure, as a function of $1/T$. 
% 	For the blue and green targets, there is a small gain at low temperatures when using ${\boldsymbol \phi}_{\textrm{rand}}$. This is because these shapes are elongated on the horizontal direction, and the allowed random actions of the policy are also along the same direction. This is not true for the yellow target, which is elongated on the vertical direction.
}
	\label{fig:F0_vs_rand}
\end{figure}

To calculate the gain in the expected return time to target due to the optimization, we use the zero-force policy as a reference. However, using a 
random-force policy leads to similar results: in \cref{fig:F0_vs_rand} the return time to target for the zero-force and the random policy for three different targets are seen to lead to similar values of $\tau^{\textrm{r}}_{{\boldsymbol \phi}}(\bar s)$.

\section{Optimal return time to target for big targets}

\begin{figure}[ht]
	\includegraphics[width=\linewidth]{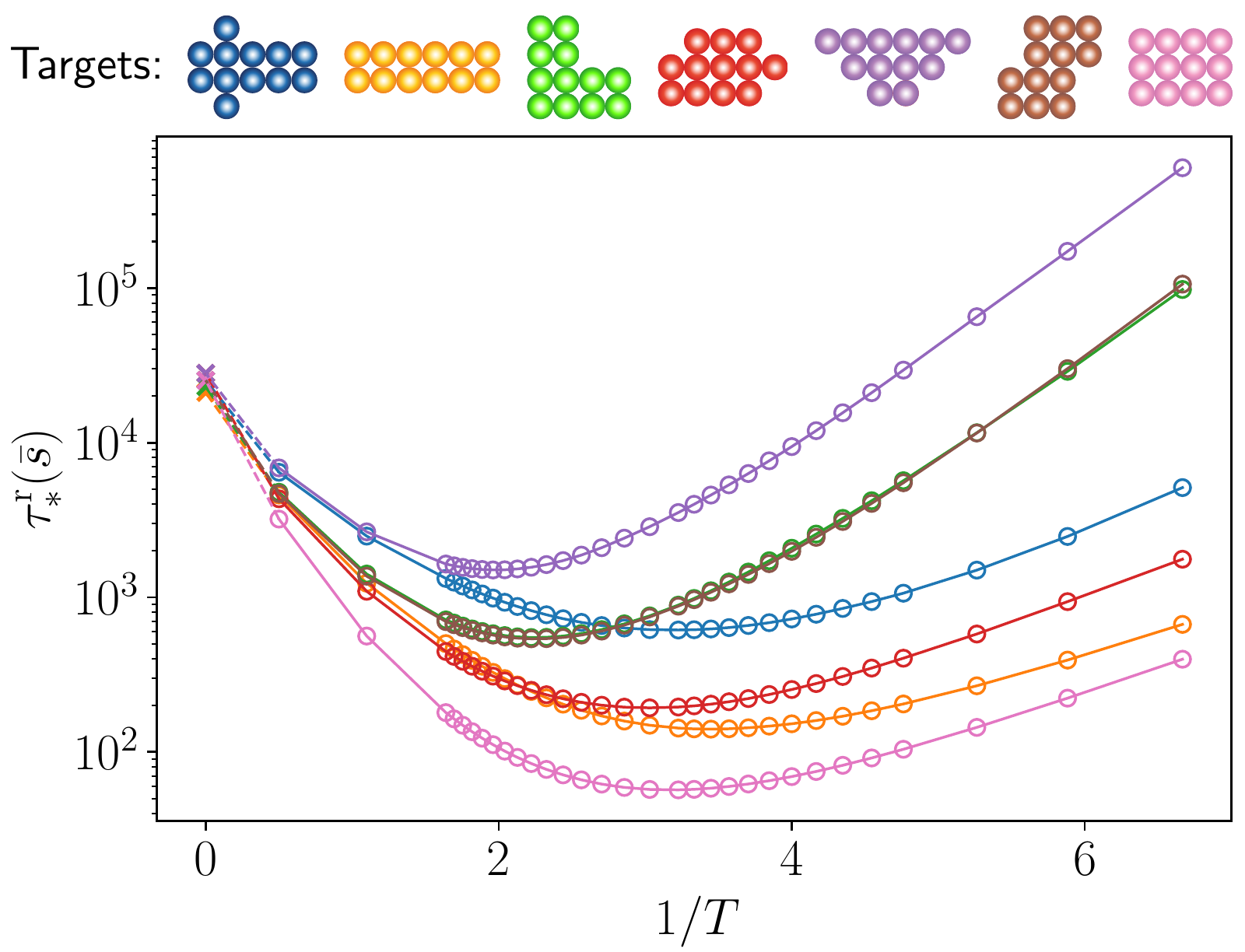}
	\caption{Optimal return time to target $\tau^{\mathrm{r}}(\bar s)$ as a function of $1/T$ for targets with $N=12$. All the curves exhibit a minimum, corresponding to an optimal temperature. The $\times$ correspond to the analytical expression of $\tau_{\infty}^\mathrm{r}(\bar{s})$ for $T \rightarrow \infty$.}
	\label{fig:returntime_big_targets}
\end{figure}

For big targets, we expect $M_*(\bar s)$ to be negative and thus to observe a minimum in the optimal return time to target $\tau^{\mathrm{r}}(\bar s)$. 
This is confirmed by \cref{fig:returntime_big_targets}, where  $\tau^{\mathrm{r}}(\bar s)$ 
is plotted as a function of $1/T$ for seven targets with $N=12$ and $F_0=0.4$.

\section{Computation time as a function of cluster size}

\begin{figure}[ht]
	\includegraphics[width=\linewidth]{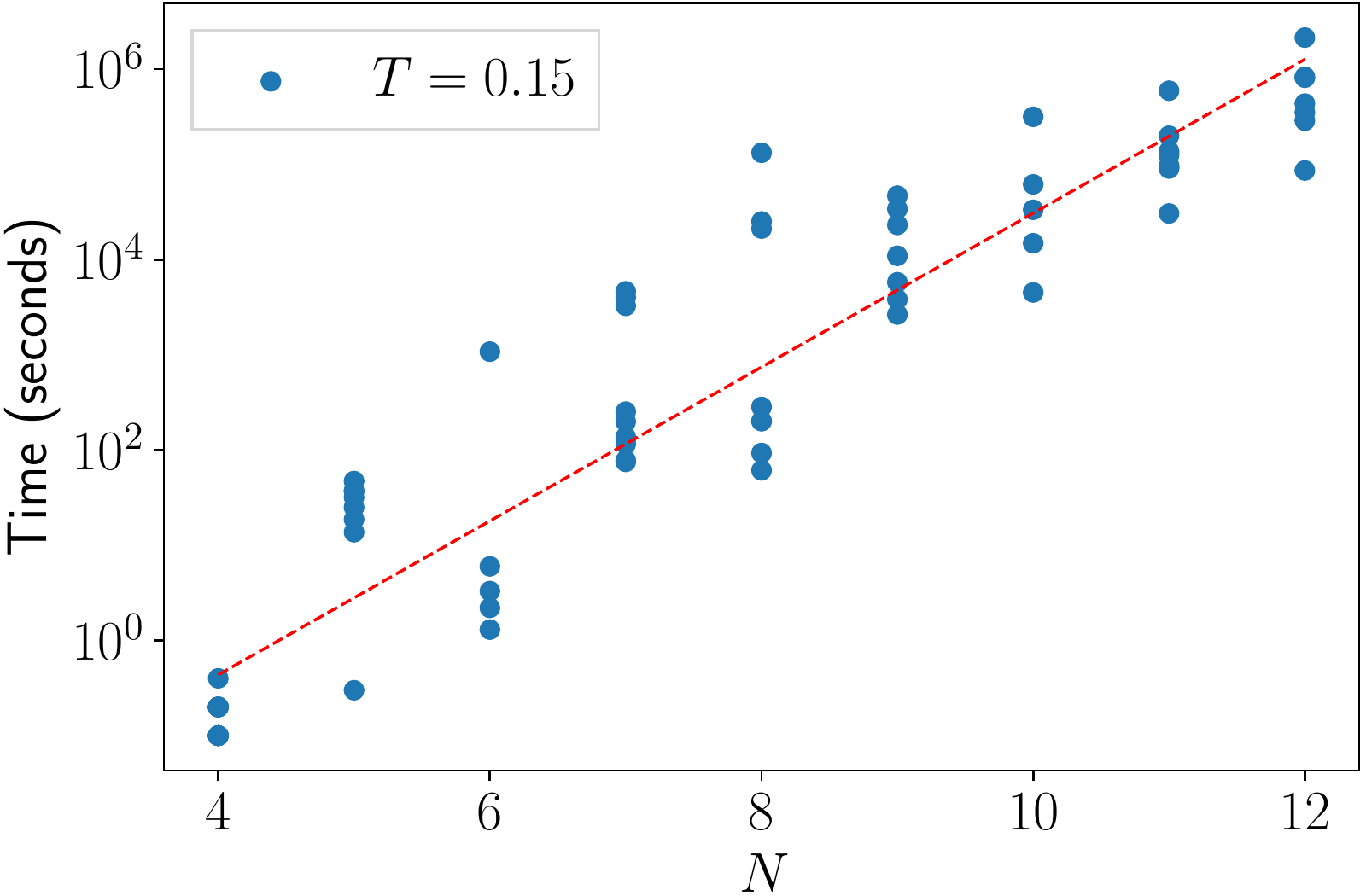}\\
	[10pt]
	\includegraphics[width=\linewidth]{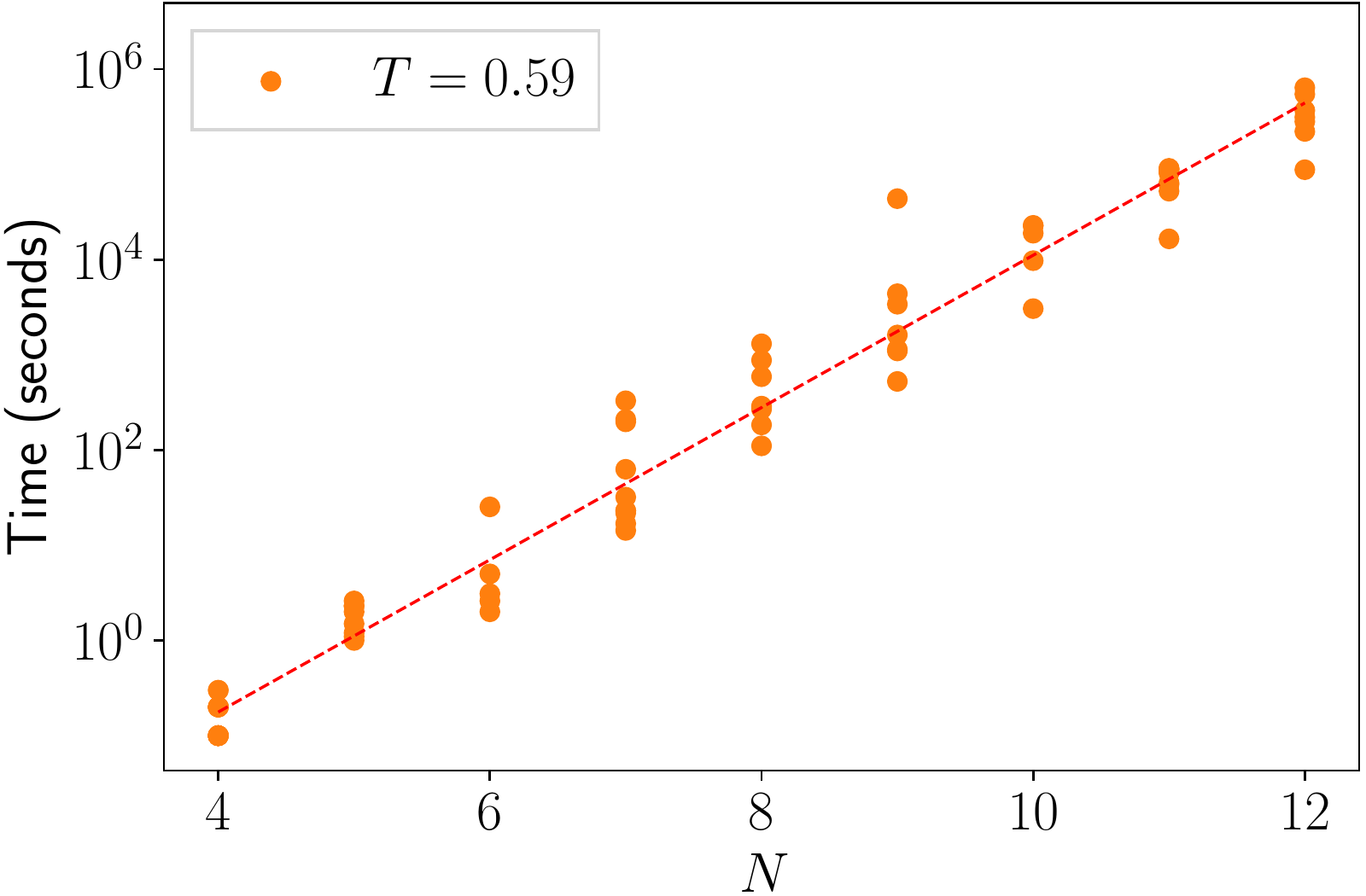}
	\caption{Computation time to obtain an optimal policy as a function of cluster size $N$, for several targets of size $4 \leq N \leq 12$, at $T=0.15$ and $T=0.59$.}
	\label{fig:comp_time}
\end{figure}

In \cref{fig:comp_time}, we report measurements of the computation time required to obtain an optimal policy with value iteration for several targets of size $4 \leq N \leq 12$, at two different temperatures $T=0.15$ and $T=0.59$.
The amplitude of the force is $F_0=0.4$. 
In both cases, the time increases exponentially as $\sim 6^N$. 
The computations have been performed with a sequential
code on two different computers, with processors Intel Xeon E5-1650 3.50GHz and Intel Xeon E5-2630 2.40GHz.

\section{Effect of freezing atoms with 4 nearest neighbors}

\begin{figure}[H]
    \vspace{15pt}
	\includegraphics[width=\linewidth]{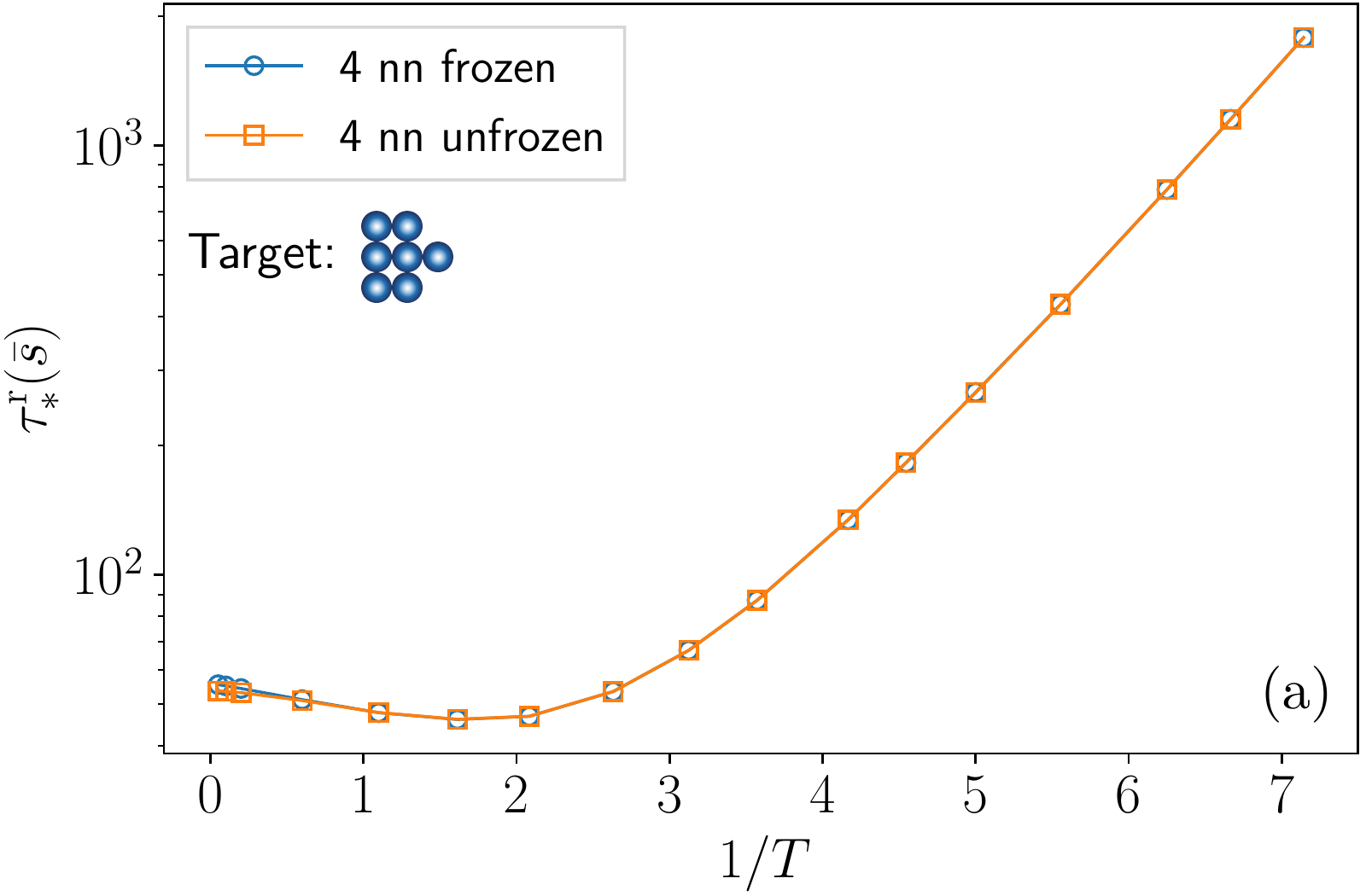}\\[10pt]
	\includegraphics[width=\linewidth]{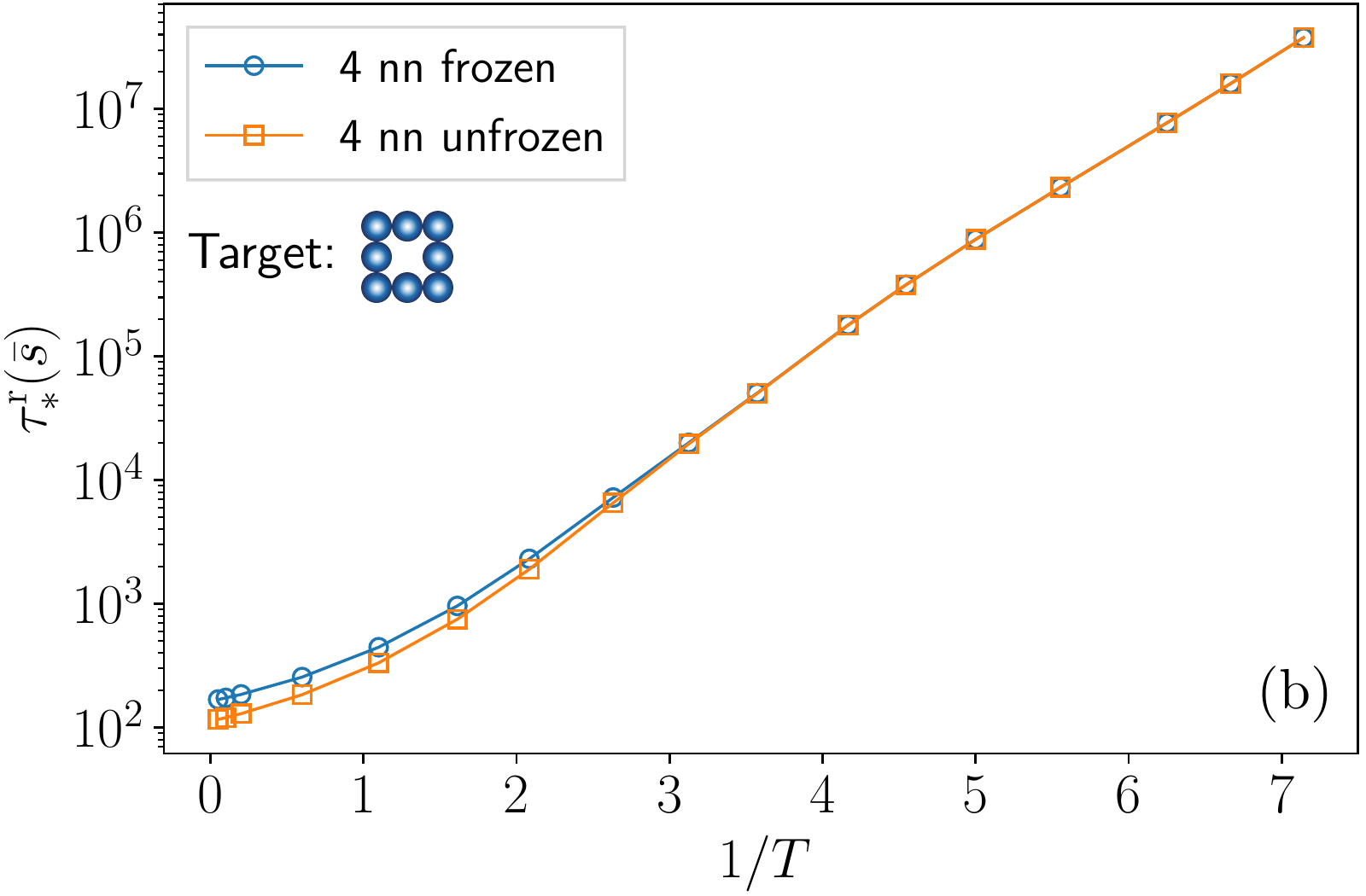}
	\caption{Comparison of the mean return time to target $\tau^\mathrm{r}_*(\bar s)$ as a function of temperature for the case where the atoms with 4 nearest neighbors are frozen and the case where they are unfrozen, for (a) a 7-atom and (b) a 8-atom target. }
	\label{fig:unfreeze}
\end{figure}

In \cref{fig:unfreeze} we show a comparison of the mean return time to target $\tau^\mathrm{r}_*(\bar s)$ calculated when freezing atoms with 4 nearest neighbors (as in the results of the main text), and when unfreezing them, for two targets with $N=7$ and $N=8$. Note that, for $N \leq 6$, there is no difference between the two cases, since there are no configurations that allow a displacement of an atom with 4 nearest neighbors, without leading to cluster breaking (which is forbidden in our model).

% The two targets considered in \cref{fig:unfreeze} are expected to be highly sensible to the freezing or unfreezing of atoms with 4 nearest neighbors, since they have several atomic moves

While the results are always accurate at low
temperature, some deviations are found at high temperatures.
The maximum deviation at high temperature is around 3$\%$
in \cref{fig:unfreeze}(a), and around 45$\%$ for the 8-particle target with a hole 
in \cref{fig:unfreeze}(b).
Indeed, in this latter target the reverse moves of the 8 moves 
bringing a particle to the empty center of the cluster are forbidden 
by the freezing of particles with 4 nearest neighbors.

\bibliography{references}